\documentclass[preprint,3p]{elsarticle}
\usepackage{graphicx,epsfig}
\usepackage{amssymb}
\usepackage{amsmath}
\usepackage{amsbsy}
\usepackage{slashed}
\usepackage{subfig}

\newcommand{\eg}[0]{\textit{e.g.}}
\newcommand{\ie}[0]{\textit{i.e.}}

\newcommand{\eq}[1]{Eq.~(\ref{#1})}
\newcommand{\eqs}[3]{Eqs~(\ref{#1}), (\ref{#2}), and (\ref{#3})}
\newcommand{\eqsfromto}[2]{Eq.s~(\ref{#1}) to (\ref{#2})}

\newcommand{\leer}[1]{}

\newcommand{\URxUBL}[0]{\ensuremath{U(1)_{R} \times U(1)_{B-L}}}
\newcommand{\UYxUBL}[0]{\ensuremath{U(1)_{Y} \times U(1)_{B-L}}}

\newcommand{\Da}[0]{\Delta m^2_{\textsc{A}}}
\newcommand{\seesaw}[0]{see-saw}
\newcommand{\Seesaw}[0]{See-saw}
\newcommand{\PDGIX}[0]{PDG.IX}
\newcommand{\RE}[0]{\text{Re}}
\newcommand{\IM}[0]{\text{Im}}

\begin{document}

\title{Proposal for generalised Supersymmetry Les Houches Accord \\for see-saw models and PDG numbering scheme}


\author[sou,oxon,fr]{L.~Basso}
\ead{lorenzo.basso@physik.uni-freiburg.de}

\author[sou,oxon]{A.~Belyaev}
\ead{a.belyaev@soton.ac.uk}

\author[ban]{D.~Chowdhury}
\ead{debtosh@cts.iisc.ernet.in}


\author[val]{M.~Hirsch}
\ead{mahirsch@ific.uv.es}

\author[sou,zew]{S.~Khalil}
\ead{skhalil@zewailcity.edu.eg}

\author[sou,oxon]{S.~Moretti}
\ead{S.Moretti@soton.ac.uk}

\author[wu]{B.~O'Leary}
\ead{boleary@physik.uni-wuerzburg.de}

\author[wu]{W.~Porod}
\ead{porod@physik.uni-wuerzburg.de}

\author[bonn,wu]{F.~Staub}
\ead{fnstaub@th.physik.uni-bonn.de}

\address[sou]{School of Physics \& Astronomy, University of Southampton,
 Highfield, Southampton SO17 1BJ, UK}
\address[oxon]{Particle Physics Department, Rutherford Appleton Laboratory, Chilton,
  Didcot, Oxon OX11 0QX, UK}
\address[fr]{Fakult\"at f\"ur Mathematik und Physik,
  Albert-Ludwigs-Universit\"at, D-79104 Freiburg i.Br., Germany}
\address[ban]{Centre for High Energy Physics, Indian Institute of Science,
  Bangalore 560 012, India}
\address[val]{Instituto de Fisica Corpuscular, CSIC, Universidad de Valencia, 46071
  Valencia, Spain}
\address[zew]{Centre for Theoretical Physics, Zewail City of Science and Technology,
 Sheikh Zayed, 12588, Giza, Egypt}
\address[wu]{Institut f\"ur Theoretische Physik und Astrophysik,
 Universit\"at  W\"urzburg, 97074  W\"urzburg, Germany}
\address[bonn]{Bethe Center for Theoretical Physics \& Physikalisches Institut der
Universit\"at Bonn, Nu{\ss}allee 12, 53115 Bonn, Germany}

\begin{flushright}
Bonn-TH-2012-05, FR-PHENO-2012-012, SHEP-12-09, IFIC/12-39
\end{flushright}

\begin{abstract}
The SUSY Les Houches Accord (SLHA) $2$ extended the first SLHA to include
 various generalisations of the Minimal Supersymmetric Standard Model (MSSM) as
 well as its simplest next-to-minimal version.
 Here, we propose further extensions to it, to include the most general and
 well-established \seesaw\ descriptions (types I/II/III, inverse, and linear) in
 both an effective and a simple gauged extension of the MSSM framework. In addition, we generalise the PDG numbering scheme to reflect the properties of the particles.

\end{abstract}

\maketitle

\section{INTRODUCTION}
If neutrinos are {\em Majorana} particles, their mass at low energy is
described by a unique dimension-5 operator \cite{Weinberg:1979sa}
\begin{equation}\label{eq:dim5}
m_{\nu} = \frac{f}{\Lambda} (H L) (H L).
\end{equation}
Using only renormalisable interactions, there are exactly three tree-level
models leading to this operator \cite{Ma:1998dn}. The first one is the
exchange of a heavy fermionic singlet, called the right-handed neutrino.
This is the celebrated \seesaw\
mechanism~\cite{Minkowski:1977sc,GellMann:1980vs,Yanagida:1979as,
S.L.Glashow,Mohapatra:1979ia}, nowadays called
\seesaw\ type I. The second possibility is the exchange of a scalar $SU(2)_L$
triplet~\cite{Schechter:1980gr,Cheng:1980qt}. This is commonly known
as \seesaw\ type II. And lastly, one could also add one (or more) fermionic
triplets to the field content of the SM~\cite{Foot:1988aq}. This is
known as \seesaw\ type III. The \seesaw\ mechanism provides a rationale for
the observed smallness of neutrino masses, by the introduction of the
inverse of some large scale $\Lambda$. In \seesaw\ type I, for example,
$\Lambda$ is equal to the mass(es) of the right-handed neutrinos. Since
these are $SU(2)_L$ singlets, their masses can take any value, and with
neutrino masses as indicated by the results from oscillation experiments
$m_{\nu} \sim \sqrt{\Da} \sim 0.05$ eV, where $\Da$ is the atmospheric
neutrino mass splitting, and couplings of order ${\cal O}(1)$, the scale
of the \seesaw\ is estimated to be very roughly $m_{SS} \sim 10^{15}$ GeV.
This value is close to, but slightly lower than, the scale of grand
unification. In addition there exist \seesaw\ models with large couplings at the
 electroweak scale, such as the linear
 \cite{Akhmedov:1995vm} and inverse \cite{Mohapatra:1986bd} \seesaw\ models.

In the MSSM the gauge couplings unify nearly perfectly at an
energy scale close to $m_{G} \simeq 2 \times 10^{16}$ GeV. Adding
new particles which are charged under the SM group at a scale below $m_G$
tends to destroy this attractive feature of the MSSM, unless the
new superfields come in complete $SU(5)$ multiplets. For this
reason, within supersymmetric \seesaw\ models, one usually realizes
the type II \seesaw\ by adding $\bf 15$-plets~\cite{Ma:1998dn,
Rossi:2002zb,Hirsch:2008gh,Borzumati:2009hu}
and the type III \seesaw\ by the addition of
$\bf 24$-plets \cite{Buckley:2006nv,Borzumati:2009hu,Esteves:2010ff}.

Just like with any other extensions of the SM based on supersymmetry (SUSY),
 those implementing \seesaw\ realisations have seen several different
 conventions used over the years, many of which have become widespread. Such a
 proliferation of conventions has some drawbacks though from a calculational
 point of view: results obtained by different authors or computing tools are not
 always directly comparable. Indeed, to enable this comparison, a consistency
 check of all the relevant conventions and the implementation of any necessary
 translations thereof must first be made. Needless to say, this is a
 time-consuming and rather error-prone task.

To remedy this problem, the original SUSY Les Houches Accord (SLHA1) was
 proposed \cite{Skands:2003cj}. The SLHA1 uniquely defined a set of conventions
 for SUSY models together with a common interface between codes. The latter can
 be broadly categorised in terms of four different kinds of tools:
 ($i$) spectrum calculators (which calculate the SUSY mass and coupling
 spectrum, assuming some SUSY-breaking terms and a matching of SM parameters to
 known data); ($ii$) observables calculators (packages which calculate one or
 more of the following: inclusive cross sections, decay partial widths, relic
 dark matter densities and  indirect/precision  observables); ($iii$) Monte
 Carlo (MC) event generators (which calculate exclusive cross sections through
 explicit  simulation of high-energy particle collisions, by including resonance
 decays, parton showering, hadronisation and underlying-event effects);
($iv$) SUSY fitting programs (which fit SUSY model parameters to data).
(See http://www.ippp.dur.ac.uk/montecarlo/BSM/ for an up-to-date collection and
 description of such tools.) Further, SLHA1 provided users with input and output
 in a common format, which is more readily comparable and transferable. In
 short, the basic philosophy was to specify a unique set of conventions for SUSY
 extensions of the SM together with generic file structures to be communicated
 across the four types of codes above, $(i)-(iv)$, based on the transfer of
 three different ASCII files: one for model input, one for spectrum calculator
 output and one for decay calculator output.

The original protocol, SLHA1, was strictly limited to the MSSM with real
 parameters and R-parity conservation neglecting generation mixing.  An expanded version was proposed in
 \cite{Allanach:2008qq} (see also \cite{Allanach:2005kk} in
 Ref.~\cite{Allanach:2006fy}), known as SLHA2, whereby various MSSM
 generalisations were included: \ie, those involving CP, $R$-parity and flavour
 violation as well as the simplest extension of the MSSM, the so-called
 next-to-MSSM (NMSSM). Herein, we follow and extend the development of this protocol started at Les Houches 2011~\cite{Brooijmans:2012yi}, by including
 the most general and well-established \seesaw\ descriptions in both an
 effective and a simple gauged extension of the MSSM. The new conventions and
 control switches described here comply with those of SLHA2 (and,
 retrospectively, also SLHA1) unless explicitly mentioned in the text.
Our effort here  is
parallelled by other generalisations of the previous accords
 documented in \cite{Brooijmans:2012yi}, altogether eventually contributing
 to the definition of a future release of the original protocol.

\section{THE SEE-SAW MECHANISM}
\label{sec:seesawmechanisms}

In this section, we discuss different implementations of the \seesaw\
 mechanism. As already stated in the introduction, the aim of the \seesaw\
 mechanism is to explain the neutrino masses and mixing angles. This is done by
 linking the tiny masses to other parameters which are of the naturally expected
 order. The general idea can be summarised by writing down the most general mass
 matrix combining left-handed neutrino ($L$), right-handed neutrino ($R$) and
 additional singlet fields carrying lepton number ($S$): 
\begin{equation}
 \left(\begin{array}{ccc}
m_{LL} & m_{LR} & m_{LS} \\
m^T_{LR} & m_{RR} & m_{RS} \\
m^T_{LS} & m^T_{RS} & m_{SS}
\end{array}  \right)
\end{equation}
Looking at specific limits of this matrix, we can recover the different \seesaw\
 realizations: $m_{LL} = m_{LS} = m_{RS} = 0$ leads to type I, type III is
 obtained in the same limit but with $m_{RR}$ stemming from $SU(2)_L$ triplets.
 $m_{LL} = m_{RR} = m_{LS} = 0$ is the characteristic matrix for inverse
 \seesaw\ , while $m_{LL} = m_{RR} = m_{SS} = 0$ is the standard parametrisation
 of the linear \seesaw\ . What most of these different  \seesaw\ models have in
 common is
the way the tiny neutrino masses are recovered, just by suppressing them with
very high scales for the new fields. This is strictly true for the type I/II/III
models. The linear and inverse \seesaw\ versions work slightly differently: the
 heaviness of the
 new fields is reduced at the price of introducing a relatively small
dimensionful parameter, usually connected to an explicit violation of the
lepton number.

In the following subsection we discuss models which can explain the origin of
 the distinct neutrino mass matrices.




\subsection{Type I/II/III}
The simplest see-saw models describe neutrino masses with an effective operator
arising after integrating out heavy superfields.
While one generation of $15$-plets
is sufficient to explain the entire neutrino data, this is not the case with
just one $24$-plet if $SU(5)$-invariant boundary conditions are assumed on the
new parameters and more generations have to be included. We will therefore treat
the number of generations of singlets, 15- and 24-plets as free parameter.
Bearing this in mind,
the models will be described with minimal addition of superfields, in the basis
after $SU(5)$ symmetry breaking (see table~\ref{eff_SS_new_chiral} \footnote{We use 
always the convention that all given 
$U(1)$  charges are those appearing in the covariant derivative, i.e. $\partial_\mu - i g Q A_\mu$}). 
All these
fields are integrated out during the RGE evaluation, such that, at the SUSY
scale, only the particle content of the MSSM remains. 
 
We give in the following the unified equations which would lead to a mixed
scenario of \seesaw\ types I, II, and III. The
 \eqs{eff_SS_T1_potential}{eff_SS_T1_softbreaking_W}{eff_SS_T1_softbreaking_phi}
 are specific to type I,
 \eqs{eff_SS_T2_potential}{eff_SS_T2_softbreaking_W}{eff_SS_T2_softbreaking_phi}
 refer to type II and
 \eqs{eff_SS_T3_potential}{eff_SS_T3_softbreaking_W}{eff_SS_T3_softbreaking_phi}
 refer to type III.

\begin{table}[!ht]
\begin{center}
\begin{tabular}{|c|c|c|c|c|c|} 
\hline \hline 
\multicolumn{6}{|c|}{Type I}\\
\hline \hline 
SF & Spin 0 & Spin \(\frac{1}{2}\) & Generations & \(U(1)\otimes\,
\text{SU}(2)\otimes\, \text{SU}(3)\) &  $R$-parity of fermion \\ 
\hline 
\(\hat{\nu}^c\) & \(\tilde{\nu}^c\) & \(\nu^c\) & $n_{1}$
 & \((0,{\bf 1},{\bf 1}) \) & $+$\\ 
\hline \hline
\multicolumn{6}{|c|}{Type II}\\
\hline \hline 
SF & Spin 0 & Spin \(\frac{1}{2}\) & Generations & \(U(1)\otimes\,
\text{SU}(2)\otimes\, \text{SU}(3)\) & $R$-parity of fermion \\ 
\hline 
\(\hat{T}\) & \(\tilde{T}\) & \(T\) & $n_{15}$ & \((1,{\bf 3},{\bf 1}) \) & $-$\\ 
\(\hat{\bar{T}}\) & \(\tilde{\bar{T}}\) & \(\bar{T}\) & $n_{15}$ & \((-1,{\bf 3},
{\bf 1}) \) & $-$\\ 
\(\hat{S}\) & \(\tilde{S}\) & \(S\) & $n_{15}$
 & \((-\frac{2}{3},{\bf 1},{\bf 6}) \) & $-$\\ 
\(\hat{\bar{S}}\) & \(\tilde{\bar{S}}^*\) & \(\bar{S}^*\) & $n_{15}$ &
\((\frac{2}{3},{\bf 1},{\bf \overline{6}}) \) & $-$\\ 
\(\hat{Z}\) & \(\tilde{Z}\) & \(Z\) & $n_{15}$
 & \((\frac{1}{6},{\bf 2},{\bf 3}) \) & $-$\\ 
\(\hat{\bar{Z}}\) & \(\tilde{\bar{Z}}\) & \(\bar{Z}\) & $n_{15}$
 & \((-\frac{1}{6},{\bf 2},{\bf \overline{3}}) \) & $-$\\ 
\hline \hline
\multicolumn{6}{|c|}{Type III}\\
\hline \hline 
SF & Spin 0 & Spin \(\frac{1}{2}\) & Generations & \(U(1)\otimes\,
\text{SU}(2)\otimes\, \text{SU}(3)\) &  $R$-parity of fermion \\ 
\hline 
\(\hat{W}_M\) & \(\tilde{W}_M\) & \(W_M\) & $n_{24}$ & \((0,{\bf 3},{\bf 1}) \)
 & $+$\\ 
\(\hat{G}_M\) & \(\tilde{G}_M\) & \(G_M\) & $n_{24}$ & \((0,{\bf 1},{\bf 8}) \)
 & $+$\\ 
\(\hat{B}_M\) & \(\tilde{B}_M\) & \(B_M\) & $n_{24}$ & \((0,{\bf 1},{\bf 1}) \)
 & $+$\\ 
\(\hat{X}_M\) & \(\tilde{X}_M\) & \(X_M\) & $n_{24}$ & \((\frac{5}{6},{\bf 2},
{\bf \overline{3}}) \) & $+$\\ 
\(\hat{\bar{X}}_M\) & \(\tilde{\bar{X}}_M\) & \(\bar{X}_M\) & $n_{24}$ &
\((-\frac{5}{6},{\bf 2},{\bf 3}) \) & $+$\\ 
\hline \hline
\end{tabular} 
\caption{\label{eff_SS_new_chiral} 
New chiral superfields appearing in the effective type I/II/III \seesaw\
models. While $n_{15} = 1$ is sufficient to explain neutrino data, 
$n_1$ and $n_{24}$ must be at least 2. }
\end{center} 
\end{table}

The combined superpotential of all three types can be written as
\begin{displaymath}
W -  W_{\text{MSSM}} = W_{I} + W_{II} + W_{III}
\end{displaymath}
where
\begin{align} \label{eff_SS_T1_potential}
W_{I} = & Y_{\nu} \, \hat{\nu}^c\,\hat{L}\,\hat{H}_u\,+\frac{1}{2} M_{\nu^c}
\,\hat{\nu}^c\,\hat{\nu}^c \\ 
W_{II} = & \frac{1}{\sqrt{2}} \nonumber
Y_T \,\hat{L}\,\hat{T}\,\hat{L}\,+\frac{1}{\sqrt{2}} Y_S
\,\hat{d}^c\,\hat{S}\,\hat{d}^c\,+Y_Z\,\hat{d}^c\,\hat{Z}\,\hat{L}\,\\ 
                 \label{eff_SS_T2_potential}
        & +\frac{1}{\sqrt{2}} \lambda_1
\,\hat{H}_d\,\hat{T}\,\hat{H}_d\,+\frac{1}{\sqrt{2}} \lambda_2
\,\hat{H}_u\,\hat{\bar{T}}\,\hat{H}_u\,+M_T\,\hat{T}\,\hat{\bar{T}}\,+M_Z\,
\hat{Z}\,\hat{\bar{Z}}\,+M_S\,\hat{S}\,\hat{\bar{S}}\, \\ \nonumber
 W_{III} = & \sqrt{\frac{3}{10}} Y_B
\,\hat{H}_u\,\hat{B}_M\,\hat{L}\,+Y_W\,\hat{H}_u\,\hat{W}_M\,\hat{L}\,+Y_X\,
\hat{H}_u\,\hat{\bar{X}}_M\,\hat{d}^c\,\\ \label{eff_SS_T3_potential}
 & +M_X\,\hat{X}_M\,\hat{\bar{X}}_M\,+\frac{1}{2} M_W
\,\hat{W}_M\,\hat{W}_M\,+\frac{1}{2} M_G \,\hat{G}_M\,\hat{G}_M\,+\frac{1}{2}
M_B \,\hat{B}_M\,\hat{B}_M\,
\end{align} 
For the MSSM part we use the conventions
\begin{align}
W_{MSSM} = & \,  Y_u\,\hat{u}^c\,\hat{Q}\,\hat{H}_u\,- Y_d \,\hat{d}^c\,\hat{Q}\,\hat{H}_d\,- 
Y_e \,\hat{e}^c\,\hat{L}\,\hat{H}_d\,+\mu\,\hat{H}_u\,\hat{H}_d\,
\end{align}

The soft-breaking terms can be split into three categories: terms stemming from
the superpotential couplings when replacing the fermions with their scalar
superpartners ($L_{SB,W}$), the scalar soft-breaking masses for each chiral
superfield ($L_{SB,\phi}$) and the soft-breaking masses for the gauginos
($L_{SB,\lambda}$). Since the gauge sector is not modified, $L_{SB,\lambda}$
reads as in the MSSM. The soft-breaking terms stemming from the superpotential
are

\begin{displaymath}
L_{\text{SB},W} - L_{\text{SB},W,\text{MSSM}}
= L_{\text{SB},W}^{I} + L_{\text{SB},W}^{II} + L_{\text{SB},W}^{III}
\end{displaymath}
where
\begin{align} \label{eff_SS_T1_softbreaking_W}
L_{\text{SB},W}^{I} = \, & \, T_{\nu}\,\tilde{\nu}^c\,\tilde{L}\,H_u\,
+\frac{1}{2} B_{\nu^c} \,\tilde{\nu}^c\,\tilde{\nu}^c\,+ \mbox{H.c.}\\ \nonumber
L_{\text{SB},W}^{II} = \, & \frac{1}{\sqrt{2}}
T_T \,\tilde{L}\,\tilde{T}\,\tilde{L}\,+\frac{1}{\sqrt{2}} T_S \,\tilde{d}^{c}\,
\tilde{S}\,\tilde{d}^{c}\,+T_Z\,\tilde{d}^{c}\,\tilde{Z}\,\tilde{L}\,
+\frac{1}{\sqrt{2}} T_1\,H_d\,\tilde{T}\,H_d\,\\
\label{eff_SS_T2_softbreaking_W}
 & +\frac{1}{\sqrt{2}} T_2 \,H_u\,\tilde{\bar{T}}\,H_u\,+ B_T\,\tilde{T}\,
\tilde{\bar{T}}\,+ B_Z\,\tilde{Z}\,\tilde{\bar{Z}}\,+ B_S\,\tilde{S}\,
\tilde{\bar{S}}\, + \mbox{h.c.}\\  \nonumber
L_{\text{SB},W}^{III} = \, & \sqrt{\frac{3}{10}} T_B \,H_u\,\tilde{B}_M\,
\tilde{L}\,+T_W\,H_u\,\tilde{W}_M\,\tilde{L}\,+T_X\,H_u\,\tilde{\bar{X}}_M\,
\tilde{d}^{c}\,+B_X\,\tilde{X}_M\,\tilde{\bar{X}}_M\\
\label{eff_SS_T3_softbreaking_W}
 & \,+\frac{1}{2} B_W \,\tilde{W}_M\,\tilde{W}_M\,+\frac{1}{2} B_G \,
\tilde{G}_M\,\tilde{G}_M\,+\frac{1}{2} B_B \,\tilde{B}_M\,\tilde{B}_M\,
+\mbox{H.c.} 
\end{align} 
while the soft-breaking scalar masses read
\begin{displaymath}
L_{\text{SB},\phi} - L_{\text{SB},\phi,MSSM} = L_{\text{SB},\phi}^{I}
+ L_{\text{SB},\phi}^{II} + L_{\text{SB},\phi}^{III}
\end{displaymath}
where
\begin{align} \label{eff_SS_T1_softbreaking_phi}
L_{\text{SB},\phi}^{I} = & - (\tilde{\nu}^c)^\dagger {m_{\nu^c}^{2}}
\tilde{\nu}^c \\ \label{eff_SS_T2_softbreaking_phi}
L_{\text{SB},\phi}^{II} = & -m_S^2 \tilde{S}^* \tilde{S} -m_{\bar{S}}^2
\tilde{\bar{S}}^* \tilde{\bar{S}} -m_T^2 \tilde{T}^* \tilde{T} -m_{\bar{T}}^2
\tilde{\bar{T}}^* \tilde{\bar{T}} -m_Z^2 \tilde{Z}^* \tilde{Z} -m_{\bar{Z}}^2
\tilde{\bar{Z}}^* \tilde{\bar{Z}} \\         \nonumber
L_{\text{SB},\phi}^{III} = \, & - \tilde{B}_M^\dagger m^2_B \tilde{B}_M
-\tilde{W}_M^\dagger m^2_W \tilde{W}_M  - \tilde{G}_M^\dagger m^2_G \tilde{G}_M
\\ \label{eff_SS_T3_softbreaking_phi}
& \, - \tilde{X}_M^\dagger m^2_X \tilde{X}_M  -  \tilde{\bar{X}}_M^\dagger
m^2_{\bar{X}} \tilde{\bar{X}}_M 
\end{align}

\paragraph{GUT conditions and free parameters}
Since the new interactions in
\eqsfromto{eff_SS_T2_potential}{eff_SS_T3_potential} are the result of
$SU(5)$-invariant terms, it is natural to assume a unification of the different
couplings at the GUT scale. 
\begin{align}
& M_T = M_Z = M_S \equiv M_{15}, \hspace{1cm} 
Y_S = Y_T = Y_Z \equiv Y_{15} \\
& Y_B = Y_W = Y_X  \equiv Y_{24}, \hspace{1cm}
 M_X = M_W = M_G = M_B  \equiv M_{24}
\end{align}
In the same way the bi- and trilinear soft-breaking terms unify and they are
connected to the superpotential parameters by
\begin{align}
& B_{\nu^c} \equiv B_0 M_{\nu^c}, \hspace{1cm} T_{\nu} \equiv A_0 Y_{\nu} \\
& B_{15} \equiv B_0 M_{15}, \hspace{1cm}  T_{15} \equiv A_0 Y_{15}\\
& B_{24} \equiv B_0 M_{24}, \hspace{1cm} T_{24} \equiv A_0 Y_{24} 
\end{align}
 In case of
CMSSM-like boundary conditions, this leads to the following free parameters
\begin{equation}
B_0, \thickspace M_{\nu^c}, \thickspace Y_{\nu} \thickspace M_{15}, \lambda_1,
 \lambda_2, \thickspace Y_{15}, \thickspace M_{24}, \thickspace Y_{24}
\end{equation}
in addition to the well-known MSSM parameters
\begin{equation}
 m_0, \thickspace M_{1/2}, \thickspace A_0, \thickspace \tan\beta, \thickspace
\text{sign}(\mu) \label{eq:CMSSM}
\end{equation}
In principle, this $B_{0}$ is not the same as the $B$ for the Higgs, though in a
minimal case they may be defined to be equal at the GUT scale. Furthermore,
 $T_i = A_0 Y_i$ holds at the GUT scale. 


\paragraph{Effective neutrino masses.}
The effective neutrino mass matrices appearing in type I/II/III at SUSY scale
are
\begin{align}
 \label{eq:mnuI}
m^I_\nu = &\, - \frac{v^2_u}{2} Y^T_\nu M^{-1}_R Y_\nu \\
\label{eq:ssII}
m^{II}_\nu = &\, \frac{v_u^2}{2} \frac{\lambda_2}{M_T}Y_T. \\
\label{eq:mnu_seesawIIIa}
m^{III}_\nu = &\,- \frac{v^2_u}{2} 
\left( \frac{3}{10} Y^T_B M^{-1}_B Y_B + \frac{1}{2} Y^T_W M^{-1}_W Y_W \right).
\end{align}


\subsection{Inverse and linear \seesaw}
\label{sec:inverse_linear}
The inverse and linear \seesaw\ realisations are obtained in models that provide
three generations of a further gauge singlet carrying lepton number in addition
to three generations of the well-known right-handed neutrino superfields, here
\(\hat{\nu}^c\) (see table~\ref{tab:chiral_lin_inv}).

\begin{table}[!ht]
\begin{center}
\begin{tabular}{|c|c|c|c|c|c|c|c|} 
\hline \hline 
SF & Spin 0 & Spin \(\frac{1}{2}\) & Generations & \(U(1)\otimes\,
\text{SU}(2)\otimes\, \text{SU}(3)\) &  lepton number & $R$-parity of fermion\\
\hline 
\(\hat{\nu}^c\) & \(\tilde{\nu}^c\) & \(\nu^c\) & $n_{\nu^c}$
 & \((0,{\bf 1},{\bf 1}) \) & $+1$ & $+$\\ 
\({\hat{N}}_{S}\) & \({\tilde{N}}_{S}\) & \(N_{S}\) & $n_{N_S}$
 & \((0,{\bf 1},{\bf 1}) \) & $-1$ & $+$\\ 
\hline \hline
\end{tabular} 
\caption{\label{tab:chiral_lin_inv}
New chiral superfields appearing in models with inverse and linear
\seesaw.}
\end{center} 
\end{table}

The only additional terms in the superpotential which are allowed by
conservation of gauge quantum numbers are 
\begin{equation}
\label{eq:W_lin_inv}
 W - W_{\text{MSSM}} = Y_{\nu}\,\hat{\nu}^c\,\hat{L}\,\hat{H}_u\,
+M_R\,\hat{\nu}^c\,{\hat{N}}_{S}\,
+ \left\{ \begin{array}{c c}
\frac{1}{2} {\mu}_{N}\,{\hat{N}}_{S}\,{\hat{N}}_{S} & \mbox{inverse \seesaw}\\
Y_{LN} \,{\hat{N}}_{S}\,\hat{L}\,\hat{H}_u\, & \mbox{linear \seesaw}
\end{array} \right. .
\end{equation}
It is important to note that the last term in each model breaks lepton number
explicitly, but are expected for different reasons.

The soft-breaking terms read
\begin{align}
L_{\text{SB},W} = \, & L_{\text{SB},W,\text{MSSM}}\, + T_{\nu}\,\tilde{\nu}^c\,
\tilde{L}\,H_u\,+B_R\,\tilde{\nu}^c\,{\tilde{N}}_{S}\,
+ \left\{ \begin{array}{c c}
\frac{1}{2} B_N \,{\tilde{N}}_{S}\,{\tilde{N}}_{S} & \mbox{inverse \seesaw}\\
T_{LN} \,{\tilde{N}}_{S}\,\tilde{L} H_u & \mbox{linear \seesaw}
\end{array} \right\} + \mbox{H.c.} \\
L_{\text{SB},\phi} = \, & L_{\text{SB},\phi,\text{MSSM}}
- (\tilde{\nu}^c)^\dagger {m_{\nu^c}^{2}} \tilde{\nu}^c - {\tilde{N}}_{S} m_N^2
{\tilde{N}}_{S}^{\ast} \, .
\end{align}
while $L_{\text{SB},\lambda}$ is again the same as for the MSSM. 
It is necessary to split the sneutrinos and the singlets into their scalar and
pseudoscalar components:
\begin{equation} 
\label{eq:decomposition_sneutrinos}
\tilde{\nu}_L =  \, \frac{1}{\sqrt{2}} \left(\sigma_L  + i \phi_L\right),
\thickspace 
\tilde{\nu}^c =  \, \frac{1}{\sqrt{2}} \left(\sigma_R  + i  \phi_R \right),
\thickspace 
{\tilde{N}}_{S} = \, \frac{1}{\sqrt{2}} \left(\sigma_S  + i  \phi_S \right)\, .
\end{equation} 
In comparison to the MSSM, additional mixings between fields take place: the
left- and right-handed scalar components mix with the scalar component of the
singlet fields. The same holds for the pseudoscalar components. Furthermore, the
neutrinos mix with the fermionic singlet fields to build up $9$ Majorana
fermions. All three appearing $9 \times 9$ mass matrices can be diagonalised by
unitary matrices. We define the basis for the mass matrices as
\begin{itemize}
 \item Scalar sneutrinos: \( \left(\sigma_{L}, \sigma_{R}, \sigma_{S}\right)^T\)
 \item Pseudoscalar sneutrinos:
       \( \left(\phi_{L}, \phi_{R}, \phi_{S}\right)^T\)
 \item Neutrinos: \( \left(\nu_L, \nu^c, N_{S}\right)^T\)
\end{itemize}
The neutrino mass matrix then reads
\begin{equation}
\label{eq:InvLinMM}
 \left(\begin{array}{ccc}
0 & \frac{v_u}{\sqrt{2}}Y_\nu & 0 \\
\frac{v_u}{\sqrt{2}}Y^T_\nu & 0 & M_R \\
0 & M^T_R &  \mu_N
       \end{array} \right)
\mbox{ (inverse) or }
 \left(\begin{array}{ccc}
0 & \frac{v_u}{\sqrt{2}}Y_\nu & \frac{v_u}{\sqrt{2}} Y_{LN} \\
\frac{v_u}{\sqrt{2}}Y^T_\nu & 0 & M_R \\
\frac{v_u}{\sqrt{2}} Y^T_{LN} & M^T_R & 0
       \end{array} \right)
\mbox{ (linear),}
\end{equation}
Note, the presence of $v_u$ in all terms of the first column and row is just
 coincidence caused by the given, minimal particle content. For more general
 models different VEVs can appear. 

\paragraph{Free parameters}
If CMSSM-like boundary conditions are assumed, the following new free parameters
arise:
\begin{equation}
M_R, \thickspace Y_{LN}, \thickspace \mu_N, \thickspace B_0
\end{equation}
in addition of those given in \eq{eq:CMSSM}.

Calculating the eigenvalues of the above mass matrices, it can be seen that the
 light neutrino masses are linear functions of $Y_{LN}$ in the linear
 \seesaw\ models, while the neutrino masses are linearly proportional to
 $\mu_N$, as in the inverse \seesaw\ models. The neutrino masses in the two
 models read
\begin{align}
m^{LS}_\nu \simeq &\, \frac{v_u^2}{2} \left(Y_\nu (Y_{LN} M_R^{-1})^T +
(Y_{LN} M_R^{-1})Y_\nu^T \right) \, ,\\
m^{IS}_\nu \simeq &\, \frac{v_u^2}{2} Y_\nu (M^T_R)^{-1} \mu_N M^{-1} Y^T_\nu \,.
\end{align}
Hence we propose that both models, and any combination of the two, be specified
 by extending $Y_{\nu}$ to a $3 \times ( n_{\nu^c} + n_{N_S} )$ Yukawa matrix,
 incorporating $Y_{LN}$ as $Y_{\nu}^{ij}$ with $i$ running from $n_{\nu^c} + 1$
 to $n_{\nu^c} + n_{N_S}$. Implementation of each model consists of zeros being
 specified in the appropriate entries in the relevant matrices (\textit{e.g.}
 specifying that the elements of $Y_{\nu}$ corresponding to $Y_{LN}$ are zero
 recovers the inverse \seesaw\ model, while specifying that ${\mu}_{N}$ is zero
 recovers the linear \seesaw\ model).


\subsection{\Seesaw\ in models with \URxUBL\ gauge sector}
\label{sec:RxB-L}
Models based on a $SO(10)$ GUT theory can lead to a gauge sector containing the
product group \URxUBL, through the breaking pattern 
\begin{align}
SO(10) & \rightarrow SU(3)_C \times SU(2)_L \times SU(2)_R \times U(1)_{B-L}
& \rightarrow SU(3)_C \times SU(2)_L \times U(1)_R \times U(1)_{B-L}\, .
\end{align}
The \URxUBL\ factors will be subsequently broken to the hypercharge $U(1)_Y$ of
the SM. However, it is possible that this final breaking scale is just around
the TeV scale without spoiling gauge unification \cite{Malinsky:2005bi}. This
can therefore lead to interesting phenomenology and can have important impact on
the Higgs sector \cite{Hirsch:2011hg}. The first version of these models
included a linear \seesaw\ mechanism, but it has been shown that also the
 inverse \seesaw\ can be included \cite{DeRomeri:2011ie}. Further, the minimal (type-I) \seesaw\ can also be included if the Higgs fields responsible for the \URxUBL\ breaking carry twice the traditional $U(1)$ charges.

Notice that, in general,
these models contain not only gauge couplings per each Abelian gauge group, but
also so-called `off-diagonal couplings', as discussed in~\ref{sec:kineticmixing}. The minimal particle content for such model
extending the MSSM, leading to the spontaneous breaking of \URxUBL\ and to
neutrino masses, is given in table~\ref{tab:chiral_RxBL}. This particle content
consists of $3$ generations of $16$-plets of $SO(10)$, $2$ additional Higgs
fields and $3$ generations of a singlet field. The vector superfields are given
in table \ref{tab:vector_RxBL}.
\begin{table}[!ht]
\begin{center} 
\begin{tabular}{|c|c|c|c|c|c|} 
\hline \hline 
SF & Spin 0 & Spin \(\frac{1}{2}\) & Generations & \( U(1)_{B-L}\otimes\,
\text{SU}(2)\otimes\, U(1)_R\otimes\, \text{SU}(3)\)\\ 
\hline 
\hline
\multicolumn{5}{|c|}{Matter fields (fermionic components have positive
$R$-parity)}
\\
\hline \hline
\(\hat{Q}\) & \(\tilde{Q}=\left(\begin{array}{c} \tilde{u}_L \\ \tilde{d}_L
\end{array} \right) \) & \(Q=\left(\begin{array}{c} u_L \\ d_L \end{array}
\right)\) & 3 & \((\frac{1}{6},{\bf 2},0,{\bf 3}) \) \\ 
\(\hat{L}\) & \(\tilde{L}=\left(\begin{array}{c} \tilde{\nu}_L \\ \tilde{e}_L
\end{array} \right)\) & \(L=\left(\begin{array}{c} \nu_L \\ e_L \end{array}
\right)\) & 3 & \((-\frac{1}{2},{\bf 2},0,{\bf 1}) \) \\ 
\(\hat{u^c}\) & \(\tilde{u}^c\) & \(u^c\) & 3 & \((-\frac{1}{6},{\bf 1},
-\frac{1}{2},{\bf \overline{3}}) \) \\ 
\(\hat{d^c}\) & \(\tilde{d}^c\) & \(d^c\) & 3 & \((-\frac{1}{6},{\bf 1},
\frac{1}{2},{\bf \overline{3}}) \) \\ 
\( {\hat \nu}^c \) & \(\tilde{\nu}^c\) & \( \nu^c\) & 3 & \((\frac{1}{2},
{\bf 1},-\frac{1}{2},{\bf 1}) \) \\ 
\(\hat{e^c}\) & \(\tilde{e}^c\) & \(e^c\) & 3 & \((\frac{1}{2},{\bf 1},
\frac{1}{2},{\bf 1}) \) \\ 
\( {\hat{N}}_{S}\) & \( {\tilde{N}}_{S} \) & \( N_{S} \) & $n_{N_S}$ &
\((0,{\bf 1},0,{\bf 1}) \) \\
\hline 
\hline
\multicolumn{5}{|c|}{Higgs fields (scalar components have positive
$R$-parity)}\\
\hline \hline
\(\hat{H}_d\) & \(H_d = \left(\begin{array}{c} H_d^0 \\ H_d^- \end{array}
\right)\) & \(\tilde{H}_d = \left(\begin{array}{c} \tilde{H}_d^0 \\
\tilde{H}_d^- \end{array} \right)\) & 1 & \((0,{\bf 2},-\frac{1}{2},{\bf 1}) \)
\\ 
\(\hat{H}_u\) & \(H_u= \left(\begin{array}{c} H_u^+ \\ H_u^0\end{array}
\right)\) & \(\tilde{H}_u= \left(\begin{array}{c} \tilde{H}_u^+ \\ \tilde{H}_u^0
\end{array} \right)\) & 1 & \((0,{\bf 2},\frac{1}{2},{\bf 1}) \) \\ 
\hline
\multicolumn{5}{|c|}{For minimal \seesaw}\\
\hline
\(\hat{\delta}_R\) & \(\delta_R^0\) & \(\tilde{\delta}_R^0\) & 1 & \((-1,
{\bf 1},1,{\bf 1}) \) \\
\(\hat{\bar{\delta}}_R\) & \(\bar{\delta}_R^0\) & \(\tilde{\bar{\delta}}_R^0\) & 1 &
\((1,{\bf 1},-1,{\bf 1}) \) \\ 
\hline
\multicolumn{5}{|c|}{For linear and inverse \seesaw}\\
\hline
\(\hat{\xi}_R\) & \(\xi_R^0\) & \(\tilde{\xi}_R^0\) & 1 & \((-\frac{1}{2},
{\bf 1},\frac{1}{2},{\bf 1}) \) \\
\(\hat{\bar{\xi}}_R\) & \(\bar{\xi}_R^0\) & \(\tilde{\bar{\xi}}_R^0\) & 1 &
\((\frac{1}{2},{\bf 1},-\frac{1}{2},{\bf 1}) \) \\ 
\hline \hline
\hline
\multicolumn{5}{|c|}{Fields integrated out (scalar components have positive
$R$-parity)}\\
\hline
\(\hat{\xi}_L\) & \(\xi_L^0\) & \(\tilde{\xi}_L^0\) & 1 & \((\frac{1}{2},
{\bf 2},0,{\bf 1}) \) \\
\(\hat{\bar{\xi}}_L\) & \(\bar{\xi}_L^0\) & \(\tilde{\bar{\xi}}_L^0\) & 1 &
\((-\frac{1}{2},{\bf 2},0,{\bf 1}) \) \\ 
\hline \hline
\end{tabular} 
\end{center} 
\caption{\label{tab:chiral_RxBL}
Chiral Superfields appearing in models with \URxUBL\ gauge sector which
incorporate minimal, linear and inverse \seesaw\ mechanisms.
}
\end{table}
\begin{table}[!ht]
\begin{center} 
\begin{tabular}{|c|c|c|c|c|c|} 
\hline \hline 
SF & Spin \(\frac{1}{2}\) & Spin 1 & \(SU(N)\) & Coupling & Name \\ 
 \hline 
\(\hat{B}'\) & \(\lambda_{\tilde{B}'}\) & \(B'\) & \(U(1)\) & \(g_{BL}\) &
\text{B-L}\\ 
\(\hat{W}_L\) & \(\lambda_{L}\) & \(W_L\) & \(\text{SU}(2)\) & \(g_L\) &
\text{left}\\ 
\({\hat B}_R \) &\(\lambda_{\tilde{B}_R}\) & \(B_R\) & \(U(1)\) & \(g_R\) &
\text{right}\\ 
\(\hat{g}\) & \(\lambda_{\tilde{g}}\) & \(g\) & \(\text{SU}(3)\) & \(g_s\) &
\text{color}\\ 
\hline \hline
\end{tabular} 
\end{center} 
\caption{\label{tab:vector_RxBL}
Vector superfields appearing in models with \URxUBL\ gauge sector.}
\end{table}
\paragraph{Superpotential}
We assume for the following discussion that the superpotential can contain the
following terms:
\begin{align} 
W -W_{MSSM} = & \, Y_{\nu}\,\hat{\nu^c}\,\hat{L}\,\hat{H}_u\, - \mu_{\delta} \,
\hat{\bar{\delta}}_R\,\hat{\delta}_R\,  + Y_M \hat{\nu}^c \hat{\delta}_R \hat{\nu}^c  
\label{eq:W_RxBL_min}
\end{align} 
for the realization of the minimal \seesaw\ and
\begin{align} 
\nonumber 
W -W_{MSSM} = & \, Y_{\nu}\,\hat{\nu^c}\,\hat{L}\,\hat{H}_u\, - \mu_{\xi} \,
\hat{\bar{\xi}}_R\,\hat{\xi}_R\, +Y_{N\nu^c} {\hat{N}}_{S}\hat{\nu}^c\hat{\xi}_R \\
& \hspace{0.5cm} + \left\{ \begin{array}{c c}
\frac{1}{2} \mu_N {\hat{N}}_{S}{\hat{N}}_{S} \, & \mbox{inverse \seesaw}\\
Y_{LS} \,{\hat{L}}\,\hat{\xi}_L\,\hat{N}_S\, + Y_{LR} \hat{\xi}_L \hat{\xi}_R \hat{H}_d  \ + \mu_L \hat{\xi}_L \hat{\bar{\xi}}_L  & \mbox{linear \seesaw}
\end{array} \right. 
\label{eq:W_RxBL}
\end{align} 
for linear and inverse \seesaw, respectively. Notice that, again, the ${\hat{N}}_{S}$ superfield carries lepton number and
that therefore the term $\mu_N$ provides its explicit violation. 
Since $\mu_L \gg m_{SUSY}$, the fields $\hat{\xi}_L$ and $\hat{\bar{\xi}}_L$ 
are integrated out and create an effective 
operator $\frac{Y_{LR} Y_{LS}}{\mu_L} \hat{L} \, \hat{N}_S \hat{H}_d \hat{\xi}_R$. 

\paragraph{Soft-breaking terms} 
The soft-breaking terms in the matter sector are 
\begin{align} 
\nonumber
L_{\text{SB},W} - L_{\text{SB},W,\text{MSSM}}
= \, & T_{\nu}\,\tilde{\nu}^c\,\tilde{L}\,H_u\,
- B_{\delta} \,\bar{\delta}_R\,\delta_R, +T_{M} \tilde{\nu}^c \delta_R^0 \tilde{\nu}^c  + \mbox{H.c.} \\ 
L_{\text{SB},\phi} - L_{\text{SB},\phi,\text{MSSM}}
= \,& - m_{\delta}^2 |\delta_R|^2 - m_{\bar{\delta}}^2 |\bar{\delta}_R|^2
- (\tilde{\nu}^c)^\dagger {m_{\nu^c}^{2}} \tilde{\nu}^c 
\end{align} 
respectively
\begin{align} 
\nonumber
L_{\text{SB},W} - L_{\text{SB},W,\text{MSSM}}
= \, & T_{\nu}\,\tilde{\nu}^c\,\tilde{L}\,H_u\,
- B_{\xi} \,\bar{\xi}_R\,\xi_R, +T_{N\nu^c} {\tilde{N}}_{S} \tilde{\nu}^c \xi_R \\
& \, + \left\{ \begin{array}{c c}
\frac{1}{2} B_N {\tilde{N}}_{S}{\tilde{N}}_{S} \, & \mbox{inverse \seesaw}\\
T_{LS} \,{\tilde{L}}\,{\xi}_L\,\tilde{N}_S\, + T_{LR} \xi_L \xi_R H_d  \ + B_L \xi_L \bar{\xi}_L  & \mbox{linear \seesaw}
\end{array} \right\}
 + \mbox{H.c.} \\ 
L_{\text{SB},\phi} - L_{\text{SB},\phi,\text{MSSM}}
= \,& - {\tilde{N}}_{S}^\dagger m_{N}^2 \tilde{S}
- m_{\xi}^2 |\xi_R|^2 - m_{\bar{\xi}}^2 |\bar{\xi}_R|^2
- (\tilde{\nu}^c)^\dagger {m_{\nu^c}^{2}} \tilde{\nu}^c - m_{\xi_L}^2 |\xi_L|^2 - m_{\bar{\xi}_L}^2 |\bar{\xi}_L|^2  
\end{align} 
While the soft-breaking gaugino sector reads
\begin{align}
L_{\text{SB},\lambda} = \, & \frac{1}{2}\left(- \lambda_{\tilde{B}'}^{2} M_{B-L}
- 2 \lambda_{\tilde{B}'} \lambda_{\tilde{B}_R} {M}_{R B}
- M_2 \lambda_{{\tilde{W}},{i}}^{2} - M_3 \lambda_{{\tilde{g}},{\alpha}}^{2}
- \lambda_{\tilde{B}_R}^{2} {M}_{R} + \mbox{H.c.}
\right) 
\end{align} 
The term $\lambda_{\tilde{B}'} \lambda_{\tilde{B}_R} {M}_{R B}$ is a consequence
of the presence of two Abelian gauge groups,
see \ref{sec:kineticmixing}.

\paragraph{Symmetry breaking}
Since it is assumed that in these models the scale of spontaneous symmetry
breaking to the SM gauge group is near the TeV scale, it is possible to restrict
ourselves to a direct one-step breaking pattern
(\textit{i.e.}, $SU(2)_L \times \URxUBL \rightarrow U(1)_{\text{EM}}$). This
breaking pattern takes place when the Higgs fields in the left and right sectors
receive VEVs. We can parametrise the scalar fields as follows: 
\begin{align} 
H^0_d = & \, \frac{1}{\sqrt{2}} \left(v_d  +  \sigma_d  +i \phi_d\right),
\thickspace 
H^0_u =  \, \frac{1}{\sqrt{2}} \left(v_u + \sigma_u  +  i \phi_u \right) \\
X_R = & \, \frac{1}{\sqrt{2}} \left(v_{X_R}  +  \sigma_{X_R}  +i \phi_{X_R}\right),
\thickspace 
\bar{X}_R =  \, \frac{1}{\sqrt{2}} \left(v_{\bar{X}_R} + \sigma_{\bar{X}_R}
+ i \phi_{\bar{X}_R} \right)\, .
\end{align}
with $X = \xi, \delta$
It is useful to define the quantities $v_R = v_{X_R}^2 + v_{{\bar{X}_R}}^2$
and $\tan\beta_R = \frac{v_{{\bar{X}_R}}}{v_{X_R}}$, in analogy to
$v^2 = v_d^2 + v_u^2$ and to $\tan\beta = \frac{v_u}{v_d}$.

\paragraph{Particle mixing}
Additional mixing effects take place in the gauge and Higgs sectors due to the
additional gauge fields considered, besides the neutrino and sneutrino cases.
The three neutral gauge bosons $B'$, $B_R$ and $W^3$ mix to form three mass
eigenstates: the massless photon, the well-known $Z$ boson, and a $Z'$ boson.
This mixing can be parameterised by a unitary $3 \times 3$ matrix which
diagonalises the mass matrix of the gauge bosons, such as
\begin{equation}
 (\gamma, Z, Z')^T = U^{{\gamma}ZZ^{\prime}} (B', B_R, W^3)^T\, .
\end{equation}
Similarly, this model contains $7$ neutralinos which are an admixture of the
three neutral gauginos, of the two neutral components of the Higgsino doublets
and of the two additional fermions coming from the right sector. The mass
matrix, written in the basis $\left(\lambda_{\tilde{B}'}, \tilde{W}^0,
\tilde{H}_d^0, \tilde{H}_u^0, \lambda_{\tilde{B}_R}, \tilde{X_R},
\tilde{\bar{X}}_R\right)$ (with $X=\xi,\delta$), can be diagonalised by a unitary matrix, here
denoted with $Z^N$. In the Higgs sector we choose the mixing basis and rotation
matrices to be, respectively,
\begin{itemize}
 \item Scalar Higgs fields: \( \left(\sigma_{d}, \sigma_{u}, \sigma_{X_R},
                            \sigma_{\bar{X}_R}\right)^T\) and $Z^H$
 \item Pseudoscalar Higgs fields: \( \left(\phi_{d}, \phi_{u}, \phi_{X},
                                  \phi_{\bar{X}_R}\right)^T\) and $Z^A$\, .
\end{itemize}
The neutrino and sneutrino sectors are similar to the case discussed in
section~\ref{sec:inverse_linear}: the scalar fields are decomposed into their
{\it{CP}}-even and odd components according to
eq.~(\ref{eq:decomposition_sneutrinos}). The mass matrices are defined in the
same basis. If all terms of \eq{eq:W_RxBL} are present, the resulting
masses of the light neutrinos are a result of
a mixed linear and
an inverse \seesaw. The mass matrix for inverse seesaw is analog to left matrix given in \eq{eq:InvLinMM}
with $M_R$ replaced by $\frac{1}{\sqrt{2}} Y_{N{\nu}^{c}} v_{\xi_R} \equiv \tilde{Y}$, while the 
mass matrix for linear seesaw is given by the right matrix in \eq{eq:InvLinMM} and the replacement
$Y_{LS} \to \frac{Y_{LS} Y_{LR}}{\sqrt{2}\mu_L} v_{\eta}$. For the minimal realization with the
superpotential given in \eq{eq:W_RxBL_min}, the neutrino mass matrix is given by 
\begin{equation}
\label{eq:RBLminMM}
 \left(\begin{array}{cc}
0 & \frac{v_u}{\sqrt{2}}Y_\nu  \\
\frac{v_u}{\sqrt{2}}Y^T_\nu & \frac{2 v_{\delta_R}}{\sqrt{2}} Y_M 
  \end{array} \right)
\end{equation}

\paragraph{Free parameters}
If CMSSM-like boundary conditions are assumed, the following new free parameters
arise in addition of those given in \eq{eq:CMSSM} in inverse seesaw
\begin{equation}
\thickspace Y_{N{\nu}^{c}},
 \thickspace \mu_N,\thickspace B_0, \thickspace \tan\beta_R, \thickspace
 \text{sign}(\mu_{\xi}) \thickspace M_{Z'}
\end{equation}
and in addition for linear seesaw
\begin{equation}
\thickspace Y_{LS}, \thickspace Y_{LR}, \thickspace \mu_L
\end{equation}

Here we have assumed that
the parameters $\mu_\xi$ and $B_\xi$ are fixed by the tadpole equations. The
 relationships of the soft trilinear terms to the Yukawa couplings are as
 before, and  $B_N = B_0 \mu_N$.


\subsection{\Seesaw\ in models with \UYxUBL\ gauge sector}
\label{sec:YxB-L}
The final category of models considered here includes an additional $B-L$ gauge
group tensored to the SM gauge groups, \textit{i.e.}
$SU(3)_C \times SU(2)_L \times \UYxUBL$. The corresponding vector superfields
are given in table~\ref{tab:vector_YxBL}. The minimal version of these models
\cite{Khalil:2007dr,FileviezPerez:2010ek,O'Leary:2011yq} extends the MSSM particle content with three generations of
right-handed superfields. Two additional scalars, singlets with respect to SM
gauge interactions but carrying $B-L$ charge, are added to break $U(1)_{B-L}$,
as well as allowing for a Majorana mass term for the right-handed neutrino
superfields. Furthermore, two new lepton fields per generation can be included to specifically implement the inverse \seesaw\ mechanism \cite{Elsayed:2011de}, as well as the linear \seesaw\ realisation if further two doublet fields ($\hat{\rho}$ and $\hat{\bar{\rho}}$) are considered, to be integrated out.
 All particles and their
quantum numbers are given in table~\ref{tab:chiral_YxBL}. This table contains
also the charge assignment under a $Z_2$ symmetry which is just present in the case of the inverse and linear \seesaw\ models\footnote{Notice that in comparison to Ref.~\cite{Elsayed:2011de}, the charge assignments of the new particles in the inverse \seesaw\ model, as well as the $Z_2$ symmetry, have been redefined for consistency with the similar minimal model of Ref.\cite{Khalil:2007dr}.}. 

\begin{table}[!ht]
\begin{center} 
\begin{tabular}{|c|c|c|c|c|c|} 
\hline \hline 
SF & Spin \(\frac{1}{2}\) & Spin 1 & \(SU(N)\) & Coupling & Name \\ 
 \hline 
\(\hat{B}\) & \(\lambda_{\tilde{B}}\) & \(B\) & \(U(1)\) & \(g_1\)
&\text{hypercharge}\\ 
\(\hat{W}\) & \(\lambda_{\tilde{W}}\) & \(W^-\) & \(\text{SU}(2)\) & \(g_2\)
&\text{left}\\ 
\(\hat{g}\) & \(\lambda_{\tilde{g}}\) & \(g\) & \(\text{SU}(3)\) & \(g_3\)
&\text{color}\\ 
\(\hat{B}'\) & \(\lambda_{\tilde{B}{}'}\) & \(B'\) & \(U(1)\) & \(g_{B}\)
&\text{B-L}\\ 
\hline \hline
\end{tabular} 
\end{center} 
\caption{\label{tab:vector_YxBL}
Vector superfields appearing in models with \UYxUBL\ gauge sector.}
\end{table}

\begin{table}[!ht]
\begin{center} 
\begin{tabular}{|c|c|c|c|c|c|c|c|} 
\hline \hline 
SF & Spin 0 & Spin \(\frac{1}{2}\) & Generations & \(U(1)_Y\otimes\,
\text{SU}(2)\otimes\, \text{SU}(3)\otimes\, U(1)_{B-L}\) & $Z_2$ inverse SS\\ 
\hline 
\hline
\multicolumn{6}{|c|}{Matter fields (fermionic components have positive
$R$-parity)}
\\
\hline \hline
\(\hat{Q}\) & \(\tilde{Q}=\left(\begin{array}{c} \tilde{u}_L \\ \tilde{d}_L
\end{array} \right) \) & \(Q=\left(\begin{array}{c} u_L \\ d_L \end{array}
\right)\) & 3 & \((\frac{1}{6},{\bf 2},{\bf
3},\frac{1}{6}) \)  & $+$\\ 
\(\hat{L}\) & \(\tilde{L}=\left(\begin{array}{c} \tilde{\nu}_L \\ \tilde{e}_L
\end{array} \right)\) & \(L=\left(\begin{array}{c} \nu_L \\ e_L \end{array}
\right)\) & 3 & \((-\frac{1}{2},{\bf 2},{\bf
1},-\frac{1}{2}) \)  & $+$\\ 
\(\hat{d}^c\) & \(\tilde{d}^c\) & \(d^c\) & 3 & \((\frac{1}{3},{\bf
1},{\bf \overline{3}},-\frac{1}{6}) \)  & $+$\\ 
\(\hat{u}^c\) & \(\tilde{u}^c\) & \(u^c\) & 3 & \((-\frac{2}{3},{\bf
1},{\bf \overline{3}},-\frac{1}{6}) \)  & $+$\\ 
\(\hat{e}^c\) & \(\tilde{e}^c\) & \(e^c\) & 3 & \((1,{\bf 1},{\bf
1},\frac{1}{2}) \)  & $+$\\ 
\(\hat{\nu}^c\) & \(\tilde{\nu}^c\) & \(\nu^c\) & 3 & \((0,{\bf 1},{\bf
1},\frac{1}{2}) \) & $+$\\ 
\hline 
\hline
\multicolumn{6}{|c|}{Higgs fields (scalar components have positive
$R$-parity)}\\
\hline \hline
\(\hat{\eta}\) & \(\eta\) & \(\tilde{\eta}\) & 1 & \((0,{\bf 1},{\bf
1},-1) \)  & $-$\\ 
\(\hat{\bar{\eta}}\) & \(\bar{\eta}\) & \(\tilde{\bar{\eta}}\) & 1 &
\((0,{\bf 1},{\bf 1},+1)\)  & $-$\\ 
\(\hat{H}_d\) & \(H_d\) & \(\tilde{H}_d\) & 1 & \((-\frac{1}{2},{\bf 2},
{\bf1},0) \)  & $+$\\ 
\(\hat{H}_u\) & \(H_u\) & \(\tilde{H}_u\) & 1 & \((\frac{1}{2},{\bf 2},
{\bf1},0) \) & $+$\\ 
\hline 
\hline
\multicolumn{6}{|c|}{Additional field for inverse \seesaw\ (fermionic components
have positive $R$-parity)}\\
\hline \hline
\({\hat{N}}_{S}\) & \({\tilde{N}}_{S}\) & \(N_{S}\) & $n_{N_S}$ &
\((0,{\bf 1},{\bf 1},-\frac{1}{2}) \) & $-$\\ 
\({\hat{N}}_{S}^{\prime}\) & \({\tilde{N}}_{S}^{\prime}\) & \(N_{S}^{\prime}\)
& $n_{N_S}$ & \((0,{\bf 1},{\bf 1},\frac{1}{2}) \) & $+$\\ 
\hline \hline
\multicolumn{6}{|c|}{Fields integrated out (for linear \seesaw)}\\
\hline 
\hline
\(\hat{\rho}\) & \(\rho\) & \(\tilde{\rho}\) & 1 & \((\frac{1}{2},{\bf 2},{\bf
1},1) \)  & $-$\\ 
\(\hat{\bar{\rho}}\) & \(\bar{\rho}\) & \(\tilde{\bar{\rho}}\) & 1 &
\((-\frac{1}{2},{\bf 2},{\bf 1},-1)\)  & $-$\\ 
\hline \hline
\end{tabular} 
\end{center} 
\caption{\label{tab:chiral_YxBL}
Chiral Superfields appearing in models with \UYxUBL\ gauge sector. The
minimal particle content is needed for \seesaw\ type I, the additional fields
 can be used to incorporate inverse or linear \seesaw. The $SU(2)_L$ doublets are named as
 in table~\ref{tab:chiral_RxBL}. The $Z_2$ in the last column is not present in
 the minimal model but just in the model with inverse and linear \seesaw. The number of
 generations of ${\hat{N}}_{S}^{\prime}$ must match those of ${\hat{N}}_{S}$ for
 anomaly cancellation.}
\end{table}

The additional terms for the superpotential in comparison to the MSSM read 
\begin{align} 
\label{eq:W_YxBL}
\nonumber W - W_{\text{MSSM}} =
& \, Y_{\nu}{\hat{\nu}}^{c}{\hat{l}}{\hat{H}}_{u}
+ \left\{ \begin{array}{c c}
Y_{{\eta}{\nu}^{c}}{\hat{\nu}}^{c}{\hat{\eta}}{\hat{\nu}}^{c}
- {\mu}_{{\eta}}{\hat{\eta}}{\hat{{\bar{\eta}}}} & \mbox{minimal \seesaw}\\
Y_{IS}{\hat{\nu}}^{c}{\hat{\eta}}{\hat{N}}_{S}
+ {\mu}_{N}{\hat{N}}_{S}{\hat{N}}_{S} & \mbox{inverse \seesaw} \\
Y_{IS}{\hat{\nu}}^{c}{\hat{\eta}}{\hat{N}}_{S} +
Y_{LS}\, \hat{L} \hat{\rho} \hat{N}_S + Y_{LR}\, \hat{\rho} \hat{\eta} \hat{H}_d + \overline{Y}_{LR}\, \hat{\bar{\rho}} \hat{\bar{\eta}} \hat{H}_u
+ {\mu}_{\rho}{\hat{\rho}}{\hat{\bar{\rho}}} & \mbox{linear \seesaw}
\end{array} \right. .
\end{align}

The extra parity is not required in the minimal case, thus the \seesaw\ term
$Y_{{\eta}{\nu}^{c}}{\hat{\nu}}^{c}{\hat{\eta}}{\hat{\nu}}^{c}$ is allowed,
whereas the latter is forbidden in the other \seesaw\ realisations precisely due to the $Z_2$ symmetry. In the inverse
\seesaw\ model the ${\mu}_{N}{\hat{N}}_{S}{\hat{N}}_{S}$ term plays an important
role, even though since ${\hat{N}}_{S}$ carries $B-L$ charge, this term
violates $B-L$ charge conservation similarly to the terms in
\eq{eq:W_lin_inv} and \eq{eq:W_RxBL} which break lepton number.
This term is assumed to come from higher-order effects. (A similar term for the
${\hat{N}}_{S}^{\prime}$ field is possible but not relevant, since the
$N_{S}^{\prime}$ does not take part in the mixing with the neutrinos.) A
possible bilinear term ${\hat{N}}_{S} {\hat{N}}_{S}^{\prime}$ is forbidden by
the $Z_2$ symmetry given in the last column of Table~\ref{tab:chiral_YxBL}. This
discrete symmetry also forbids terms like ${\hat{N}}_{S} {\hat{N}}_{S} \eta$ or
${\hat{N}}_{S}^{\prime} {\hat{N}}_{S}^{\prime} \bar{\eta}$ as well as the
${\mu}_{{\eta}}$-term that is necessary to obtain a pure minimal \seesaw\
scenario. In case of linear \seesaw\, heavy fields $\hat{\rho}$ and $\hat{\bar{\rho}}$ are
present which get integrated at $\mu_\rho \gg M_{SUSY}$ similar to the linear
\seesaw\ in \URxUBL. This creates an effective operator of the form 
$\frac{Y_{LR} Y_{LS}}{\mu_\rho} \hat{L} \, \hat{N}_S \hat{H}_d \hat{\eta}$. Notice that it is assumed that
in linear \seesaw\ the term $\mu_N$ is not generated at higher loop-level. \\
The
additional soft-breaking terms are written as follows:
\begin{align} 
\nonumber
L_{\text{SB},W} - &L_{\text{SB},W,\text{MSSM}}
 =  T_{\nu}{\tilde{\nu}}^{c}{\tilde{l}} H_{u}  \\
+ & \left\{ \begin{array}{c c}
T_{{\eta}{\nu}^{c}}{\tilde{\nu}}^{c}{\eta}{\tilde{\nu}}^{c}
- B_{{\eta}}{\eta}{\bar{{\eta}}} & \mbox{minimal see-saw}\\
T_{IS}{\tilde{\nu}}^{c}{\eta}{\tilde{N}}_{S}
+ B_{N}{\tilde{N}}_{S}{\tilde{N}}_{S} & \mbox{inverse see-saw} \\
T_{IS}{\tilde{\nu}}^{c}{\eta}{\tilde{N}}_{S}
+ T_{LS} \tilde{L} \rho \tilde{N}_S
+ T_{LR} \rho \eta H_d + \overline{T}_{LR} \bar{\rho} \bar{\eta} H_u
+ B_{\rho} \rho \bar{\rho} & \mbox{linear see-saw}
\end{array} \right\} + \mbox{H.c.}\\
L_{\text{SB},\phi} - &L_{\text{SB},\phi,\text{MSSM}} =  \, - m_{\eta}^2 |\eta|^2
- m_{\bar{\eta}}^2
|\bar{\eta}|^2 - ( {\tilde{N}}_{S}^{\ast} m_{N}^{2} {\tilde{N}}_{S}
+ {\tilde{N}}_{S^{\prime}}^{\ast} m_{N^{\prime}}^{2} {\tilde{N}}_{S}^{\prime}
+ m_{\rho}^2 |\rho|^2 + m_{\bar{\rho}}^2 |\bar{\rho}|^2 ) \\
L_{\text{SB},\lambda} = \, & \frac{1}{2}\left(- \lambda_{\tilde{B}}^{2} M_1
- \lambda_{\tilde{B}} \lambda_{\tilde{B}{}'} {M}_{B B'} - M_2
\lambda_{{\tilde{W}},{i}}^{2} - M_3 \lambda_{{\tilde{g}},{\alpha}}^{2}
- \lambda_{\tilde{B}{}'}^{2} {M}_{BL} + \mbox{H.c.} \right) 
\end{align}
The soft-breaking terms $m_{\rho}^2$ and $m_{\bar{\rho}}^2$ are just present in linear \seesaw,
while $m_{N}^{2}$ and $m_{N'}^{2}$ only exists for linear and inverse \seesaw.
To break $SU(2)_L \times \UYxUBL$ to $U(1)_{em}$, the neutral
MSSM Higgs fields and the new SM scalar singlets acquire VEVs:
\begin{align} 
H^0_d = & \, \frac{1}{\sqrt{2}} \left(v_d + \sigma_d + i \phi_d\right),
\thickspace 
H^0_u = \, \frac{1}{\sqrt{2}} \left(v_u + \sigma_u + i \phi_u \right) \\
\eta = & \, \frac{1}{\sqrt{2}} \left(\sigma_{\eta} + v_{\eta} + i \phi_{\eta}
\right), \thickspace 
\bar{\eta} =  \, \frac{1}{\sqrt{2}} \left( \sigma_{\bar{\eta}} + v_{\bar{\eta}}
+ i \phi_{\bar{\eta}} \right)
\end{align} 
We also define here 
\begin{equation}
 \label{eq:v_tb}
v^2 = v_d^2 + v_u^2, \hspace{1cm} \tan\beta = \frac{v_u}{v_d}
\end{equation}
as well as 
\begin{equation}
v_x = v_{\eta_R}^2 + v_{\bar{\eta}_R}^2, \hspace{1cm}
\tan\beta' = \frac{v_{\bar{\eta}}}{v_{\eta}}
\end{equation}
The left- and right-handed sneutrinos are decomposed into their scalar and
pseudoscalar components according to \eq{eq:decomposition_sneutrinos}.
Similarly, in the inverse \seesaw\ model, the scalar component of
 ${\hat{N}}_{S}$ reads
\begin{equation}
{\tilde{N}}_{S} = \frac{1}{\sqrt{2}} \left( {\sigma}_{S} + i {\phi}_{S} \right)
\end{equation}
The additional mixing effects which take place in this model are similar to the
case of \URxUBL\ discussed in section~\ref{sec:RxB-L}. In the
gauge sector three neutral gauge bosons appear which mix to give rise to the
massless photon, the $Z$ boson, and a $Z'$ boson:
\begin{equation}
 (\gamma, Z, Z')^T = U^{\gamma Z} (B,W^3,B')^T\, .
\end{equation}
Notice that when the kinetic mixing is neglected, the $B'$ field decouples and
the mass matrix of the gauge bosons becomes block diagonal, where the upper
$2\times 2$ block reads as in the SM. In the matter sector we choose the basis
and the mixing matrices which diagonalise the mass matrices respectively as
follows:
\begin{itemize}
 \item Neutralinos: $\left(\lambda_{\tilde{B}}, \tilde{W}^0, \tilde{H}_d^0,
                    \tilde{H}_u^0, \lambda_{\tilde{B}{}'}, \tilde{\eta},
                    \tilde{\bar{\eta}}\right)^T$ and $Z^N$
 \item Scalar Higgs fields: \( \left(\sigma_{d}, \sigma_{u}, \sigma_{\eta},
                            \sigma_{\bar{\eta}}\right)^T\) and $Z^H$
 \item Pseudoscalar Higgs fields: \( \left(\phi_{d}, \phi_{u}, \phi_{\eta},
                                  \phi_{\bar{\eta}}\right)^T\) and $Z^A$
 \item Scalar sneutrinos: $\left(\sigma_{L},\sigma_{R}, \sigma_{S}\right)^T$ and
                          $Z^{\sigma v}$
 \item Pseudoscalar sneutrinos: $\left(\phi_{L},\phi_{R}, \phi_{S}\right)^T$ and
                                $Z^{\phi v}$
 \item Neutrinos: $\left(\nu_L,\nu^c, N_{S}\right)^T$ and $U^V$
\end{itemize}
where the ${\sigma}_{S}, {\phi}_{S}$, and $N_{S}$ are only present in the
inverse \seesaw\ case. The neutrino mass matrices for the minimal, inverse and linear
\seesaw\ cases can be written as
\begin{equation}
 \left(\begin{array}{cc}
0 & \frac{v_u}{\sqrt{2}}Y_\nu \\
\frac{v_u}{\sqrt{2}}Y^T_\nu & \frac{2 v_{\eta}}{\sqrt{2}} Y_{{\eta}{\nu}^{c}}
\end{array} \right) \mbox{ or }
\left(\begin{array}{ccc}
0 & \frac{v_u}{\sqrt{2}}Y_\nu & 0 \\
\frac{v_u}{\sqrt{2}}Y^T_\nu & 0 & \frac{v_{\eta}}{\sqrt{2}} Y_{IS} \\
0 & \frac{v_{\eta}}{\sqrt{2}} Y^T_{IS} &  \mu_N \end{array} \right) \mbox{ or }
\left(\begin{array}{ccc}
0 & \frac{v_u}{\sqrt{2}}Y_\nu & 
\frac{\tilde{Y}}{2} v_d v_\eta \\
\frac{v_u}{\sqrt{2}}Y^T_\nu & 0 & \frac{v_{\eta}}{\sqrt{2}} Y_{IS} \\
\frac{\tilde{Y}^T}{2} v_d v_\eta & \frac{v_{\eta}}{\sqrt{2}} Y^T_{IS} &  0 \end{array} \right)
\end{equation}
respectively. $\tilde{Y}$ is the running value of the effective $\frac{Y_{LR} Y_{LS}}{\mu_\rho}$
caused by the heavy superfields $\rho$ and $\bar{\rho}$.
In the minimal case, the \seesaw\ of type I is recovered. The light neutrino
 mass matrix can be approximated in this case by
\begin{equation}
 m_\nu \simeq -\frac{v_u^2}{2 \sqrt{2}v_\eta} Y_\nu Y^{-1}_{{\eta}{\nu}^{c}} Y_\nu
\,.
\end{equation}
In the inverse \seesaw\ case, the light neutrino masses can be written as
\begin{equation}
 m_\nu \simeq -\frac{v_u^2}{2} Y^T_\nu (Y^T_{IS})^{-1} \mu_N Y_{IS} Y_\nu
\end{equation}

\paragraph{Free parameters}
If the minimum conditions for the vacuum are solved with respect to $\mu$,
$B_\mu$, ${\mu}_{{\eta}}$ and $B_{\eta}$, the following parameters can be
treated as free in addition of those given in \eq{eq:CMSSM}:
\begin{equation}
Y_{{\eta}{\nu}^{c}}, \thickspace Y_{IS}, \thickspace \mu_N, \thickspace B_0,
\thickspace \tan\beta', \thickspace \text{sign}({\mu}_{{\eta}}),
\thickspace M_{Z'}\, 
\end{equation}
where we have used again $B_N = B_0 \mu_N$.


\section{TOWARDS SLHAv3}

The purpose of this paper is to define the extensions of the SUSY Les Houches
Accords required to incorporate the models described above. In this context and
with respect to the implementation of further models in future, it is helpful to
propose some general rules for the naming of blocks and the allocation of PDG
particle codes, which we present below.

We would like to point out that the SLHA conventions allow for redundant
information to be contained in additional blocks. Hence, in the interests of
backward compatibility, we propose that if the following names for the blocks
are used, but in cases where the old SLHA1/2 blocks would also suffice, they
should also be written in addition, \eg\ if there are only four
 neutralinos, \texttt{NMIX} should be written as well as \texttt{NEUTRALINORM},
 both containing the same information.

\paragraph{Block names for input and output}
Blocks used to give parameters as input end with ``\texttt{IN}'', while the
corresponding values as output are written in blocks without the ending
``\texttt{IN}''. For example, the gauge couplings at the output scale are given
in the \texttt{GAUGE} block, while one could define them as input at the input
scale with the \texttt{GAUGEIN} block.

\paragraph{Block names for mixing matrices}
In SLHA 2 the neutralino and Higgs mixing matrices of the MSSM and NMSSM were
named differently. However, the information about the current model is already
given in the block {\tt MODSEL} and the renaming is thus redundant. Therefore,
in order to prevent a confusing amount of names for mixing matrices in different
models we propose to use always the same names for the following mixing matrices
regardless of their dimension:
\begin{itemize}
 \item Scalar Higgs mixing matrix: {\tt SCALARRM} 
 \item Pseudoscalar Higgs mixing matrix: {\tt PSEUDOSCALARRM} 
 \item Charged Higgs mixing matrix: {\tt CHARGEDSCALARRM} 
 \item Neutralino mixing matrix: {\tt NEUTRALINORM} (corresponds to {\tt NMIX}
       in SLHA1/2)
 \item Chargino mixing matrices: {\tt CHARGINOPLUSRM} and {\tt CHARGINOMINUSRM}
       (corresponds to {\tt VMIX} and {\tt UMIX} respectively in  SLHA1/2)
 \item Up-squark mixing matrix: {\tt UPSQUARKRM} (corresponds to {\tt USQMIX} in
        SLHA1/2)
 \item Down-squark mixing matrix: {\tt DOWNSQUARKRM} (corresponds to
       {\tt DSQMIX} in  SLHA1/2)
 \item Charged slepton mixing matrix: {\tt CHARGEDSLEPTONRM} (corresponds to
       {\tt SELMIX} in  SLHA1/2)
 \item Sneutrino mixing matrix: {\tt SNEUTRINORM} (corresponds to {\tt SNUMIX}
       in the MSSM). In the case of splitting the sneutrinos into real and
       imaginary parts, {\tt SNEUTRINOEVENRM} for the CP-even and
       {\tt SNEUTRINOODDRM} for the CP-odd states should be used
       (corresponding to {\tt SNSMIX} and {\tt SNAMIX} in SLHA1/2) 
\end{itemize}
To distinguish these names from the SLHA1/2-specific names we have always used
the suffix {\tt RM} for {\tt R}otation {\tt M}atrix.

\paragraph{Block names for couplings}
For the naming of blocks which correspond to couplings and soft-breaking terms
we propose to use names which already give information about the meaning of the
parameter: the new names contain abbreviations for the involved fields and start
with a prefix to assign the meaning of the parameters. The different prefixes
should be
 \begin{itemize}
   \item {\tt Y}: trilinear superpotential coupling
   \item {\tt T}: trilinear softbreaking coupling
   \item {\tt M}: bilinear superpotential coupling
   \item {\tt B}: bilinear softbreaking coupling
   \item {\tt L}: linear superpotential coupling
   \item {\tt S}: linear softbreaking coupling
   \item {\tt M2}: soft breaking scalar mass-squared
 \end{itemize}
Because of the prefixes which already give information about the spin of the
involved particles, it is not necessary to distinguish between fermions and
scalars when naming the blocks. However, this requires the convention that all
Majorana mass terms for fermions from chiral superfields appear in the
superpotential, leaving only bilinear mass terms for the scalars in the soft
SUSY-breaking Lagrangian. We propose the following names:
 \begin{itemize}
  \item MSSM: $H_u$: {\tt HU}, $H_d$: {\tt HD}, $d^c$: {\tt D}, $u^c$: {\tt U},
              $q$: {\tt Q}, $e^c$: {\tt E}, $l$: {\tt L}
  \item \Seesaw\ types I-III: $\nu^c$: {\tt NUR}, $S$: {\tt 15S},
                              $\bar{S}$: {\tt 15SB}, $Z$: {\tt 15Z},
                              $\bar{Z}$: {\tt 15ZB}, $T$: {\tt 15T},
                              $\bar{T}$: {\tt 15TB}, $G$: {\tt 24G},
                              $W$: {\tt 24W}, $X$: {\tt 24X},
                              $\bar{X}$: {\tt 24XB}, $B$: {\tt 24B}
  \item inverse and linear \seesaw: $S$: {\tt SL}
  \item \URxUBL\ models: $\xi_R$: {\tt CR}, $\bar{\xi}_R$: {\tt CRB}, $\xi_L$: {\tt CL}, $\bar{\xi}_L$: {\tt CLB}
  \item \UYxUBL\ models: $\eta$: {\tt BIL}, $\bar{\eta}$: {\tt BILB},
                         $N_{S}$: {\tt NS}, $N_{S}^{\prime}$: {\tt NSP}, $\rho$: {\tt RHO}, $\bar{\rho}$: {\tt RHOB},
 \end{itemize}
Using these conventions, the block {\tt YU} for the up-type Yukawa coupling in
the MSSM would be replaced by {\tt YHUQU} and the block name for the $\mu$-term
would be {\tt MHUHD}.

\paragraph{EXTPAR and MSOFT}
 In the MSSM the input parameters for the gaugino masses and the Higgs
soft-breaking masses are given in {\tt EXTPAR}. However, their output is given
in {\tt MSOFT}. Similarly, the singlet couplings in the NMSSM are defined by
{\tt EXTPAR} but their output is in {\tt NMSSMRUN}. This is in some conflict
with the general rule to use always the the input name plus {\tt IN} as output.
Therefore, we propose that all new one-dimensional soft-breaking parameters and
couplings involved in the \seesaw\ models presented here are not given in
{\tt EXTPAR} as input, but rather in the given output with the prefix {\tt IN}.

\paragraph{PDG particle numbering scheme}
We would like to introduce a consistent numbering scheme for the additional
elementary particles introduced by such models, such that further particles may
be added without worrying about accidentally using a pre-existing code, which
also allows one to gather some information about the particle. While the
proposed scheme would give new numbers to particles which already have codes,
we restrict ourselves just to giving our new particles unique codes, while
hoping that codes compliant with the new standard would be able to read either
the old code or the new code for particles. The proposal is a signed nine digit
integer for each mass eigenstate, which should easily fit in a 32-bit integer.

We note that the existing scheme is already mildly inconsistent, insofar as
adding a fourth generation already has a PDG numbering scheme (17 for the extra
neutrino and 18 for the $\tau^{\prime}$), yet SLHA2 uses 100017 and 100018 for
CP-odd sneutrinos which are degrees of freedom from the first three generations.
We note also that the Flavour Les Houches Accord \cite{Mahmoudi:2010iz} uses 17
and 18 for summing over the three Standard Model generations.

Since the particle code is a signed integer, pairs of conjugate particles are
 assigned the same code with different signs. We propose a convention for which
 particle gets the positive sign:
\begin{itemize}
\item If the particle is self-conjugate, there is only the positive sign. If the
      fermion is Majorana, it only has positive sign. If the scalar part of the
      superfield can be written in terms of self-conjugate CP eigenstates, it
      should be. If it cannot, the scalar field should be assigned a sign
      according to the rules below.

\item If the particle has non-zero electric charge, the state with positive
      electric charge is taken as the particle (hence the positron are taken as
      the particle).

\item If the particle is electrically neutral, but has baryon number $B$ or
      lepton number $L$, the state with positive $B - L$ is taken as the
      particle (hence antineutrinos are taken as the particle).

\item If the particle has $B - L = 0$ according to traditional assignment, but
      is still baryon- or lepton-like, a ``temporary'' $B - L$ is assigned. If
      the particle has colour charge a temporary $B$ is first assigned by finding
      the combination of triplets and antitriplets which could combine to form
      the particle's representation: adding $+1/3$ for each triplet and $-1/3$
      for each antitriplet, the combination which has the {\em lowest magnitude}
      of temporary $B$ is assigned; \eg\ an octet may be formed by a
      triplet with an antitriplet, giving $B = 0$, or by three triplets, giving
      $B = 1$, or three antitriplets, giving $B = -1$: in this case, $B = 0$ as
      the lowest $|B|$ is assigned. This temporary $B - L$ is only for the
      purposes of determining which state is taken to be the particle with
      positive code.

\item If the neutral, colourless particle has no natural assignment by
      (temporary) $B - L$, there are still a few cases:
 \begin{itemize}
 \item the fermion is massless: the left-handed fermion is given the positive
       code, thus in supersymmetric models, the scalars left-chiral superfields
       also are given positive codes if their fermions are massless.
 \item the fermion has a Dirac mass: this does not occur in the MSSM, NMSSM, or
       any of the extended models described here. However it is conceivable. We
       propose that the model builder is responsible for deciding to assign a
       temporary lepton number to one of the fields and thus fix the convention.
 \end{itemize}
\end{itemize}

Under this scheme, quarks and squarks would have the same signs as they already
 have in the PDG conventions, leptons would have the opposite sign (\eg\ the
 muon would have a negative code), and charginos and $W$ bosons would have the
 same signs.

Once the sign is fixed, the digits are as follows:
\begin{itemize}
\item[1st digit:] 1 if it is a mass eigenstate which has an admixture of a
                  Standard Model gauge eigenstate (including the Higgs doublet),
                  2 if it does not mix with the
                  Standard Model particles. 
\item[2nd digit:] twice the spin of the particle. 
\item[3rd digit:] the CP nature: complex bosons (spin 0, 1, or 2) and Dirac
                  fermions have 0, while (massless or massive) scalar bosons,
                  massless vector bosons, massless tensor bosons, and
                  Majorana fermions have 1. Pseudoscalar bosons (massless or
                  massive) and massive vector and tensor bosons have 2.
\item[4th and 5th digits:] a 2-digit number for the $SU(3)$ representation;
                           $SU(3)$ singlets have 00; representations up to
                           dimension 64 have either the Dynkin labels for an
                           unbarred representation, or 99 minus the Dynkin
                           labels for the barred representation. These
                           representations are enumerated in
                           \cite{Slansky:1981yr}. Any particles with $SU(3)$
                           representation that does not fit into this scheme are
                           assigned 99 (for instance those of dimension 65 or
                           greater).
\item[6th and 7th digits:] a 2-digit number for the electric charge; six times
                           the absolute value of the electric charge, relative
                           to the electron, up to a maximum of $98/6$. Any
                           particle with a charge that does not fit into
                           this scheme is assigned 99. For example, an electron
                           would have 06, while a down quark would have 02, and
                           a doubly-charged scalar would have 12.
\item[8th and 9th digits:] a generation number; 01 should be given to the
                           lightest particle of any group which share the same
                           first seven digits, 02 to the 2nd-lightest, and so
                           on.
\end{itemize}

Consequently what might be considered to be the same model with or without
$R$-parity will have different codes for some of its mass eigenstates. In the
MSSM, with $R$-parity there would be three charged antileptons 110000601,
110000602, and 110000603, and two charginos 210000601 and 210000602, while
without $R$-parity there would be five charged antileptons 110000601, 110000602,
110000603, 110000604 and 110000605, and no 210000601 or 210000602.

Additionally, the codes for the neutrinos will depend on whether they are Dirac
or Majorana in the considered model.

\section{EXTENSIONS TO SLHA}


In this section we describe the implementation of the models presented in 
section~\ref{sec:seesawmechanisms} using the conventions defined in the previous section. Note that all
parameters can be implemented in complex forms and the
corresponding information can be passed by using the corresponding blocks
starting with ``{\tt IM}'' \cite{Allanach:2008qq}.

\subsection{Block format}
\label{sec:blockformat}
In the following, all blocks with a single index are to be written in the
FORTRAN format\\ (1x,I5,3x,1P,E16.8,0P,3x,’\#’,1x,A) (the same format as the
 SLHA1 blocks \texttt{HMIX}, \texttt{GAUGE}, \textit{etc.}). These blocks will
 be denoted as ``rank one''. All blocks with two indices are to be written in
 the FORTRAN format (1x,I2,1x,I2,3x,1P,E16.8,0P,3x,’\#’,1x,A) (the same format
 as the SLHA1 blocks \texttt{NMIX}, \texttt{UMIX}, \textit{etc.}). These blocks
 will be denoted as ``rank two''. All blocks with two indices are to be written
 in the FORTRAN format (1x,I2,1x,I2,1x,I2,3x,1P,E16.8,0P,3x,’\#’,1x,A) (the same
 format as the SLHA2 blocks \texttt{RVLAMLLE}, \texttt{RVLAMLQD},
 \textit{etc.}). These blocks will be denoted as ``rank three''.

\subsection{Blocks required for each model}
\label{sec:blocksformodels}
Each model requires the presence of certain blocks. Some blocks are common to
several models. We summarise the blocks needed for each of the models
described in section \ref{sec:seesawmechanisms} under the entry for selecting
this model in the \texttt{MODSEL} block \ref{sec:blockmodsel}.

\subsection{Extra Flags In Existing Blocks}
\label{sec:extendingexistingblocks}

\subsubsection{Block \texttt{MODSEL}}
\label{sec:blockmodsel}
Flag 3 (particle content) has further switches, arranged in groups: 11X is for
 effective realizations, 12X non-effective models, 13X for \URxUBL\ models, and
 14X is for \UYxUBL\ models, the initial digit 1 indicating \seesaw. 2XY might
 be used for fourth-generation models, 3XY for another set of models, and so on.
 In addition to the sets of required blocks listed for each model below, the
 \texttt{SEESAWGENERATIONS} block must also be given, and the appropriate flags
 set.
\begin{itemize}
\item Effective models
 \begin{itemize}
 \item[110:] Most general $SU(5)$-invariant effective \seesaw\ (combination of
             types I-III); all the blocks required for types I-III below are
             required.
 \item[111:] type I \seesaw; the following blocks are required:
             \texttt{MNURNURIN, BNURNURIN, M2NURNURIN, YNURLHUIN, TNURLHUIN}.
 \item[112:] type II \seesaw\ ($SU(5)$ version); the following blocks are
             required:
             \texttt{M15S15SBIN, B15S15SBIN, M215SIN, M215SBIN,
             M15T15TBIN, B15T15TBIN, M215TIN, M215TBIN, M15Z15ZBIN,
             B15Z15ZBIN, M215ZIN, M215ZBIN, YD15SDIN, TD15SDIN, YL15TLIN,
             TL15TLIN, YD15ZLIN, TD15ZLIN, YHD15THDIN, THD15THDIN, YHU15TBHUIN,
             THU15TBHUIN}.
             Any or all of the blocks
             \texttt{M15S15SBIN, M15T15TBIN, M15Z15ZBIN}
             may be left absent if the block \texttt{M15IN} is present, and the
             missing terms are taken as copies of \texttt{M15IN}. Similarly for
             \texttt{YD15SDIN, YL15TLIN, YD15ZLIN} and \texttt{Y15IN}, and
             \texttt{TD15SDIN, TL15TLIN, TD15ZLIN} and \texttt{Y15IN} times
             $A_{0}$.
 \item[113:] type III \seesaw\ ($SU(5)$ version); the following blocks are
             required:
             \texttt{M24W24WIN, B24W24WIN, M224WIN, M24G24GIN, B24G24GIN,
             M224GIN, M24B24BIN, B24B24BIN, M224BIN, M24X24XBIN,
             B24X24XBIN, M224XIN, M224XBIN, YHU24BLIN, THU24BLN,
             YHU24WLIN, THU24WLN, YHU24BLIN, THU24BLN, YHU24XBDIN, THU24XBDIN}.
             Any or all of the blocks
             \texttt{M24W24WIN, M24G24GIN, M24B24BIN, M24X24XBIN} may be left
             absent if the block \texttt{M24IN} is present, and the missing terms
             are taken as copies of \texttt{M24IN}. Similarly for
             \texttt{B24W24WIN, B24G24GIN, B24B24BIN, B24X24XBIN} and
             \texttt{B24IN}; \texttt{YHU24BLIN, YHU24WLIN, YHU24XBDIN} and
             \texttt{Y24IN}; and \texttt{THU24BLIN, THU24WLIN, THU24XBDIN} and
             \texttt{Y24IN} times $A_{0}$.
 \item[114:] type II \seesaw\ (minimal version, only triplets); the following
             blocks are required:
             \texttt{M15T15TBIN, B15T15TBIN, M215TIN, M215TBIN, YL15TLIN,
             TL15TLIN, YHD15THDIN, THD15THDIN, YHU15TBHUIN, THU15TBHUIN}.
             If \texttt{TL15TLIN} is absent, \texttt{YL15TLIN} times $A_{0}$
             is used in its place.
 \item[115:] type III \seesaw\ (minimal version, only triplets); the following
             blocks are required:
             \texttt{M24W24WIN, B24W24WIN, M224WIN, YHU24WLIN, THU24WLN}.
             If \texttt{THU24WLN} is absent, \texttt{YHU24WLIN} times $A_{0}$
             is used in its place.
 \end{itemize}
\item Linear and inverse \seesaw:
 \begin{itemize}
  \item[120]: combined inverse and linear \seesaw; the following blocks are
              required: \texttt{MNURSIN, BNURSIN, MNSNSIN, BNSNSIN, M2NSIN,
              M2NURIN, YNURLHUIN, TNURLHUIN}.
              Purely inverse or linear \seesaw\ is specified by appropriate
              zeroes in the block entries.
 \end{itemize}
\item \URxUBL:
 \begin{itemize}
 \item[131]: \texttt{YNURLHUIN, MDRDRBIN, YNURDRNURIN, TNURLHUIN,  
              BDRDRBIN, TNURDRNURIN, M2NURNURIN, M2DRIN, M2DRBIN}
 \item[132]: inverse \seesaw; the following blocks are required in
             addition to those required by the \URxUBL\ minimal \seesaw:
             \texttt{MCRCRBIN, BCRCRBIN, YNSNURCRIN, TNSNURCRIN, M2CRCRIN,
              M2CRBCRBIN, MNSNSIN, BNSNSIN, M2NSIN, YNSNURCRIN, TNSNURCRIN}.
 \item[133]: linear \seesaw; the following blocks are required in
             addition to those required by the \URxUBL\ linear \seesaw:
             \texttt{YLCLNSIN, YCLCRHDIN, TLCLNSIN, TCLCRHDIN, MCLCLBIN, 
             BCLCLBIN, M2CLIN, M2LCBIN}.
 \end{itemize}
\item \UYxUBL:
 \begin{itemize}
 \item[141]: minimal \seesaw; the following blocks are required:
             \texttt{MBILBILBIN, BBILBILBIN, M2NURIN, M2BILIN,
             M2BILBIN, YNURLHUIN, TNURLHUIN, YNURBILNURIN, TNURBILNURIN}.
 \item[142]: inverse \seesaw; the following blocks are required in
             addition to those required by the \UYxUBL\ minimal \seesaw:
             \texttt{MNSNSIN, BNSNSIN, M2NSIN, YNSNURBILIN, TNSNURBILIN}.
 \item[143]: linear \seesaw; the following blocks are required in
             addition to those required by the \UYxUBL\ minimal \seesaw:
             \texttt{YNSNURBILIN, TNSNURBILIN, YLRHONSIN, TLRHINSIN,
              YRHOETAHDIN, TRHOETAHDIN, MRHORHOB, BRHORHOB, M2RHOIN, 
              M2RHOBIN, M2NSIN}.
 \end{itemize}
\end{itemize}
In addition, the \URxUBL\ and \UYxUBL\ models require that the appropriate
entries in \texttt{GAUGEIN, MSOFTIN, MINPAR}, and \texttt{EXTPAR} are set
correctly.

\subsubsection{Block \texttt{MINPAR}}
\label{sec:blockminpar}
We extend the \texttt{MINPAR} block to include the input parameters necessary
for the addition of an extra Abelian gauge group. We do not reproduce the
existing entries here since there are quite a few. We only show those that are
new and those that have new interpretations.
\begin{itemize}
\item[2:] The common soft mass term for all the the gauginos.

\item[6:] The cosine of the phase of the ${\mu}^{\prime}$ parameter.

\item[7:] The ratio of the two vacuum expectation values that either break
          \URxUBL to $U(1)_{Y}$ or break $U(1)_{B-L}$ leaving $U(1)_{Y}$ intact.

\item[8:] The mass of the $Z^{\prime}$ boson.

\item[9:] The common bilinear mass parameter $B_0$.
\end{itemize}

\subsubsection{Block \texttt{GAUGE}}
\label{sec:blockgauge}
We extend the \texttt{GAUGE} block to include the couplings necessary for the
addition of an extra Abelian gauge group. We reproduce the existing entries here
for completeness.
\begin{itemize}
\item[1:] The coupling $g_{1}(Q)$ (also known as $g^{\prime}(Q)$ in some
          conventions) of the $U(1)_{Y}$ gauge in models which have it, or the
          coupling $g_{R}(Q)$ of the $U(1)_{R}$ gauge in such models where the
          $U(1)_{R}$ in combination with another $U(1)$ gauge breaks down to
          $U(1)_{Y}$.

\item[2:] The coupling $g_{2}(Q)$ (also known as $g(Q)$ in some conventions) of
          the $SU(2)_{L}$ gauge.

\item[3:] The coupling $g_{3}(Q)$ of the $SU(3)_{c}$ gauge.

\item[4:] The coupling $g_{BL}(Q)$ of the $U(1)_{B-L}$ gauge.

\item[14:] The off-diagonal coupling $g_{T}(Q)$ of the two $U(1)$ gauges in the
           triangle basis.
\end{itemize}

\subsubsection{Block \texttt{MSOFT}}
\label{sec:blockmsoft}
We extend the \texttt{MSOFT} block to include the soft mass terms necessary for
the addition of an extra Abelian gauge group. We do not reproduce the existing
entries here since there are quite a few. We only show those that are new and
those that have new interpretations.
\begin{itemize}
\item[1:] The soft mass term for the gaugino of the $U(1)_{Y}$ gauge in models
          which have it, or the gaugino of the $U(1)_{R}$ gauge in such models
          where the $U(1)_{R}$ in combination with another $U(1)$ gauge breaks
          down to $U(1)_{Y}$.

\item[4:] The soft mass term for the gaugino of the $U(1)_{B-L}$ gauge.

\item[5:] The soft mass term mixing the gauginos of the two $U(1)$ gauges.
\end{itemize}

\subsubsection{Block \texttt{EXTPAR}}
\label{sec:blockextpar}
We extend the \texttt{EXTPAR} block to include the parameters necessary for
the addition of an extra Abelian gauge group. We do not reproduce the existing
entries here since there are quite a few. We only show those that are new and
those that have new interpretations. Bear in mind that just as in the MSSM,
 contradictory inputs should not be given.
\begin{itemize}
\item[124:] The tree-level mass-squared $m_{{\phi}_{R}}^{2}$ of the pseudoscalar
            formed by the doublets breaking $SU(2)_R$.
\item[126:] The pole mass $m_{{\phi}_{R}}$ of the pseudoscalar formed by the
            doublets breaking $SU(2)_R$.
\end{itemize}

\subsection{Extended blocks superseding existing blocks}
\label{sec:supersedingexistingblocks}

\subsubsection{Block \texttt{XPMNSRM} (rank two), superseding \texttt{UPMNS}
 (rank one)}
\label{sec:blockneutrinorm}
In the super-PMNS basis the mass matrices of the charged
leptons and the three light neutrinos are diagonal and the
relevant generation mixing information is given in the PMNS matrix.
In the SLHA2 the input block \texttt{UPMNSIN} was defined in terms
of three mixing angles and three phases \cite{Allanach:2008qq}
whereas for the output the complete $3\times 3$ mixing has to be
given in the block \texttt{UPMNS}. In general more than 3 neutrinos contribute
and thus we propose that similarly the generalised PMNS matrix, which
is rectangular, shall be given for both input and output.
In the basis where, as above, the charged lepton and the extended neutrino
mass matrices are diagonal, this matrix corresponds to the coupling
between the left-handed charged leptons with the $W$ boson and the neutrinos
divided by $g/\sqrt{2}$.

\subsection{New blocks}
\label{sec:newblocks}
Blocks used to give parameters as input end with ``\texttt{IN}'', while the
corresponding values as output are written in blocks without the ending
``\texttt{IN}''.

\subsubsection{Block \texttt{SEESAWGENERATIONS} (rank one)}
This block specifies the number of generations of the extra fields of the
 \seesaw\ models.
\begin{itemize}
\item[1:] The number $n_{{\nu}^{c}}$ of right-handed neutrino generations in the
          type I, inverse, linear, and \UYxUBL\ \seesaw\ models.
\item[2:] The number $n_{N_{S}}$ of neutrino-like singlet generations in the
          inverse, linear, and \UYxUBL\ \seesaw\ models, and also the number of
          parity-odd neutrino-like singlet generations in \UYxUBL\ inverse
          \seesaw\ models.
\item[15:] The number $n_{15}$ of 15-plet generations in the type II model.
\item[24:] The number $n_{24}$ of 24-plet generations in the type III model.
\end{itemize}

\subsubsection{Block \texttt{SCALARRM} (rank two)}
This block specifies the mixing matrix of the neutral scalar Higgs bosons. The
gauge eigenstates that are rotated into the mass-ordered mass eigenstates are
ordered as $H_{d}^{0}, H_{u}^{0}$, then a model-dependent ordering. For \seesaw\
types II and III, there are no further entries, because these fields are assumed
to be integrated out. For \URxUBL, they are
${\xi}_{R}^{0}, {\bar{\xi}}_{R}^{0}$. For \UYxUBL, they are
${\eta}, {\bar{\eta}}$.

\subsubsection{Block \texttt{PSEUDOSCALARRM} (rank two)}
This block specifies the mixing matrix of the neutral pseudoscalar Higgs bosons.
They are ordered analogously to how \texttt{SCALARRM} is ordered.


\subsubsection{Block \texttt{GAUGEIN} (rank one)}
This block specifies the gauge couplings at the input scale. The entries are
analogous to those of the \texttt{GAUGE} block.

\subsubsection{Block \texttt{MSOFTIN} (rank one)}
This block specifies the soft mass terms at the input scale. The entries are
analogous to those of the \texttt{MSOFT} block.

\subsubsection{Superpotential mass matrix blocks (rank two)}
The superpotential mass matrix blocks are \texttt{MNURNUR(IN)} for
$M_{{\nu}^{c}}$, \texttt{M15S15SB(IN)} for $M_{S}$, \texttt{M15T15TB(IN)} for
$M_{T}$, \texttt{M15Z15ZB(IN)} for $M_{Z}$, \texttt{M24W24W(IN)} for $M_{W}$,
\texttt{M24G24G(IN)} for $M_{G}$, \texttt{M24B24B(IN)} for $M_{B}$,
\texttt{M24X24XB(IN)} for $M_{X}$, \texttt{MNURS(IN)} for $M_{R}$,
\texttt{MNSNS(IN)} for ${\mu}_{N}$, \texttt{MCRCRB(IN)} for ${\mu}_{{\xi}}$, and
\texttt{MBILBILB(IN)} for ${\mu}_{{\eta}}$. \texttt{M15(IN)} for $M_{15}$ is
used when $M_{S}$, $M_{T}$, and $M_{Z}$ are set to a common value.
\texttt{M24(IN)} for $M_{24}$ is used when $M_{W}$, $M_{B}$, $M_{G}$, and
$M_{X}$ are set to a common value. Though some of the models are described with
only a single generation of some types of field, we allow for extra generations,
and thus define all the mass matrix blocks as being rank two, even though the
minimal case would use only the $( 1, 1 )$ entry of some of them.

\subsubsection{Soft SUSY-breaking mass matrix blocks (rank two)}
The soft SUSY-breaking mass matrix blocks are \texttt{M2NUR(IN)} for
$m_{{\tilde{{\nu}}}^{c}}^{2}$, \texttt{M215S(IN)} for $m_{S}^{2}$,
\texttt{M215SB(IN)} for $m_{\bar{S}}^{2}$, \texttt{M215T(IN)} for $m_{T}^{2}$,
\texttt{M215TB(IN)} for $m_{\bar{T}}^{2}$, \texttt{M215Z(IN)} for $m_{Z}^{2}$,
\texttt{M215ZB(IN)} for $m_{\bar{Z}}^{2}$, \texttt{M224W(IN)} for $m_{W}^{2}$,
\texttt{M224G(IN)} for $m_{G}^{2}$, \texttt{M224B(IN)} for $m_{B}^{2}$,
\texttt{M224X(IN)} for $m_{X}^{2}$, \texttt{M224XB(IN)} for $m_{{\bar{X}}}^{2}$,
\texttt{M2NS(IN)} for $m_{N}^{2}$, \texttt{M2CR(IN)} for $m_{{\xi}}^{2}$,
\texttt{M2CRB(IN)} for $m_{{\bar{{\xi}}}}^{2}$, \texttt{M2BIL(IN)} for
$m_{{\eta}}^{2}$, \texttt{M2BILB(IN)} for $m_{{\bar{{\eta}}}}^{2}$, and
\texttt{M2NSP(IN)} for $m_{N^{\prime}}^{2}$. Though some of the models are
described with only a single generation of some types of field, we allow for
extra generations, and thus define all the mass matrix blocks as being rank two,
even though the minimal case would use only the $( 1, 1 )$ entry of some of
them.

\subsubsection{Soft SUSY-breaking bilinear matrix blocks (rank two)}
The soft SUSY-breaking bilinear matrix blocks are \texttt{BNURNUR(IN)} for
$B_{{\tilde{{\nu}}}^{c}}$, \texttt{B15S15SB(IN)} for $B_{S}$,
\texttt{B15T15TB(IN)} for $B_{T}$, \texttt{B15Z15ZB(IN)} for $B_{Z}$,
\texttt{B24W24W(IN)} for $B_{W}$, \textit{B24G24G(IN)} for $B_{G}$,
\texttt{B24B24B(IN)} for $B_{B}$, \texttt{B24X24XB(IN)} for $B_{X}$,
\texttt{BNURS(IN)} for $B_{R}$, \texttt{BNSNS(IN)} for $B_{N}$,
\texttt{BCRCRB(IN)} for $B_{\xi}$, and \texttt{BBILBILB(IN)} for $B_{\eta}$.
\texttt{B15(IN)} for $B_{15}$ is used when $B_{S}$, $B_{T}$, and $B_{Z}$ are set
to a common value. \texttt{B24(IN)} for $B_{24}$ is used when $B_{W}$, $B_{B}$,
$B_{G}$, and $B_{X}$ are set to a common value.

\subsubsection{Yukawa coupling matrix blocks (rank three)}
The first digit is the generation index of the 15-plet or 24-plet field in type
II and III models respectively, and is 1 for the other models. The 2nd pair of
indices correspond to the usual indices of the minimal cases.
The Yukawa coupling matrix blocks are \texttt{YNURLHU(IN)} for $Y_{{\nu}}$,
\texttt{YL15TL(IN)} for $Y_{T}$, \texttt{YD15SD(IN)} for $Y_{S}$,
\texttt{YD15ZL(IN)} for $Y_{Z}$, \texttt{YHD15THD(IN)} for ${\lambda}_{1}$,
\texttt{YHU15TBHU(IN)} for ${\lambda}_{2}$, \texttt{YHU24BL(IN)} for $Y_{B}$,
\texttt{YHU24WL(IN)} for $Y_{W}$, \texttt{YHU24XBD(IN)} for $Y_{X}$,
\texttt{YNSLHU(IN)} for $Y_{LN}$, \texttt{YNSNURCR(IN)} for $Y_{N{\nu}^{c}}$,
\texttt{YNURBILNUR(IN)} for $Y_{{\eta}{\nu}^{c}}$, and \texttt{YNSNURBIL(IN)}
for $Y_{IS}$. \texttt{Y15(IN)} for $Y_{15}$ is used when $Y_{S}$, $Y_{T}$, and
$Y_{Z}$ are set to a common value. \texttt{Y24(IN)} for $Y_{24}$ is used when
$Y_{B}$, $Y_{W}$, and $Y_{X}$ are set to a common value.

\subsubsection{Soft SUSY-breaking trilinear matrix blocks (rank three)}
There is a soft SUSY-breaking trilinear matrix block for each Yukawa coupling
matrix block, with the same name but with the initial ``\texttt{Y}'' replaced by
 ``\texttt{T}'', corresponding to the ``$T_{\text{blah}}$'' associated with the
``$Y_{\text{blah}}$'' matrices. However, there is no \texttt{T15(IN)} for
$T_{15}$ or \texttt{T24(IN)} for $T_{24}$; instead, in such constrained
parameter sets, $A_{0} \times Y_{15/24}$ is used.

\subsubsection{Block \texttt{GAMZZPRM} (rank two)}
This is the gauge boson rotation matrix $U^{{\gamma}ZZ^{\prime}}$ given at the
SUSY scale.

\section{SOFTWARE TO STUDY SEE-SAW MODELS}

We give in the following a short overview of software which support at least some of the models presented here.
\begin{itemize}
 \item {\tt SPheno} \cite{Porod:2003um,Porod:2011nf}: \\
{\tt SPheno} is a spectrum calculator written in Fortran. It supports the MSSM with and without bilinear R-parity violation as well as high-scale extensions with the same particle content as the MSSM at the SUSY scale. So far, seesaw type I-III of the models discussed here are implemented. The For type-II it can be chosen between the variant with two $SU(2)_L$ 15-plets as well as the version with only one $SU(2)_L$ triplet. For all seesaw models the entire two-loop RGEs are included and the threshold corrections to gauge couplings and gaugino masses at the threshold scales are calculated. {\tt SPheno} performs also a calculation of flavour observables like $l \to l' \gamma$ and $l \to 3 l'$. 
\item {\tt SuSeFLAV} \cite{Chowdhury:2011zr}: \\
{\tt SuSeFLAV} is a spectrum calculator written in Fortran. It calculates SUSY spectrum for MSSM with conserved R-parity and its extension for inputs at high scale. It is specifically geared up studying type-I seesaw
in great detail. Various input options for neutrino Yukawa couplings,  CKM-like, PMNS-like and  R-parameterisation  are provided for the user. 
For MSSM and type I seesaw case it uses full two loop RGEs including flavour mixing  and also calculates the threshold corrections at one loop level. {\tt SuSeFLAV} also 
calculates flavour observables like $l \to l' \gamma$, $l \to 3 l'$ and $(g_{\mu} -2)$  at low energy.

\item {\tt SARAH} \cite{Staub:2008uz,Staub:2009bi,Staub:2010jh}: \\
{\tt SARAH} is a Mathematica package to derive analytical expressions for the masses, vertices, RGEs and one-loop corrections for a given SUSY model. This information can be used by {\tt SARAH} to write model files for {\tt CalcHep}/{\tt CompHep} \cite{Pukhov:2004ca,Boos:1994xb}, {\tt WHIZARD} \cite{Kilian:2007gr}, {\tt FeynArts/FormCalc} \cite{Hahn:1998yk,Hahn:2000kx} and {\tt Madgraph} \cite{Alwall:2011uj}. The model files for {\tt CalcHep} can also be used with {\tt MicrOmegas} \cite{Belanger:2006is} for relic density calculation. In addition, the output of Fortran source-code is possible which can be compiled with {\tt SPheno}. This gives the possibility to create a full-fledged spectrum calculator based on 2-loop RGEs and 1-loop mass corrections for any model implemented in {\tt SARAH}. Also routines for the calculation of flavour observables as well as for  decay widths and branching ratios are written. Also input files for {\tt HiggsBounds} \cite{Bechtle:2008jh,Bechtle:2011sb} are created by the {\tt SPheno} modules of {\tt SARAH}. So far, seesaw type I-III, inverse and linear seesaw as well as the models with \UYxUBL\ gauge sector a part of the public version of {\tt SARAH}. 
\item {\tt SUSY Toolbox} \cite{Staub:2011dp}: \\
The SUSY toolbox is  a collection of scripts to create an environment consisting of {\tt SPheno}, {\tt CalcHep}, {\tt SARAH}, {\tt SSP}, {\tt HiggsBounds}, {\tt MicrOmegas} and {\tt WHIZARD} to study extensions of the MSSM. These scripts give the possibility for an automatised implementation of a new model in all tools based on the implementation in {\tt SARAH}.
\end{itemize}

\section{CONCLUSIONS AND OUTLOOK}
In this contribution we propose an extension of the existing SLHA accords to
 include several \seesaw\ models.
Firstly, this requires new blocks to be defined 
for the additional couplings and masses needed within the various models. In
 this connection, we do not restrict ourselves to individual versions of the
 latter but also allow for combinations of such models in order to be as general
 as possible. 
Secondly, several new particles have to be postulated. For these, we propose
 a 9-digit scheme for the corresponding PDG-codes, which could in fact be of
 more general use than for \seesaw\ models only, yet it needs to be tested
 extensively against the properties of existing and possibly new SUSY models
 before widespread adoption. 
One issue that has to be addressed as next step is the proper definition of the
 additional parameters in the so-called super-PMNS basis. Moreover, also
 $SU(2)_R$ models are not yet covered in this proposal.

\section*{ACKNOWLEDGEMENTS}
We would like to thank the organisers of the Les Houches 2011
``Physics at TeV colliders'' workshop where this project has
started. We also want to thank Nazila Mahmoudi for the enlightening
comments about FLHA. L.B., A.B. and S.M. thank the NExT Institute
and Royal Society for partial financial support. L.B. has also
been partially supported by the Deutsche Forschungsgemeinschaft
through the Research Training Group GRK\,1102 \textit{Physics of
Hadron Accelerators}. S.K. acknowledges partial support from the
Leverhulme Trust under the grant VP2-2011-012. B.O.L and
W.P. are supported by the German Ministry of Education and
Research (BMBF) under contract
 no.\ 05H09WWEF.

\appendix
\section{Kinetic mixing}
\label{sec:kineticmixing}
It is well known that in models with several \(U(1)\) gauge groups,
kinetic mixing terms
\begin{equation}
\label{eq:offfieldstrength}
- \chi_{ab}  \hat{F}^{a, \mu \nu} \hat{F}^b_{\mu \nu}, \quad a \neq b
\end{equation}
between the field-strength tensors are allowed by gauge and Lorentz
invariance \cite{Holdom:1985ag}, as $\hat{F}^{a, \mu \nu}$ and 
$\hat{F}^{b, \mu \nu}$ are gauge invariant quantities by themselves,
see \textit{e.g.}\  \cite{Babu:1997st}. Even if these terms are absent at tree
level at a particular scale, they might be generated by RGE effects
\cite{delAguila:1988jz,delAguila:1987st}. This happens usually if the
two Abelian gauge groups cannot be embedded in a larger
gauge group simultaneously or if incomplete gauge multiplets of the
fundamental theory are integrated out.
It is easier to work with  non-canonical covariant
derivates instead of off-diagonal field-strength tensors such as in eq.~(\ref{eq:offfieldstrength}). 
The equivalence of both approaches has been shown in \cite{delAguila:1988jz,Fonseca:2011vn}. 
We show here the special case of two Abelian gauge groups $U(1)_A \times U(1)_B$. The 
covariant derivatives has the form
\begin{equation}
\label{eq:kovariantDerivative}
 D_\mu  = \partial_\mu - i Q_{\phi}^{T} G  A 
\end{equation}
where \(Q_{\phi}\) is a vector containing the charges of the field $\phi$ with
respect to the two Abelian gauge groups, $G$ is the gauge coupling matrix
\begin{equation}
\label{eq:Gmatrix}
 G = \left( \begin{array}{cc} g_{AA} & g_{AB} \\ g_{BA} & g_{BB} \end{array} \right)
\end{equation}
and $A$ contains the gauge bosons $A = ( A^A_\mu, A^B_\mu )^T$. As long as the two 
Abelian gauge groups are unbroken there is freedom to rotate the gauge bosons. It is convenient
to choose a basis in which $G$ gets a triangle form
\begin{equation}
\label{eq:Gmatrixtri}
 G' = \left( \begin{array}{cc} g & \tilde{g} \\ 0 & g' \end{array} \right)
\end{equation}
Mixing effects of Abelian gauge groups appear not only in the gauge sector but also for the gauginos 
because also terms of the form 
\begin{equation}
 M_{AB} \lambda_A \lambda_B
\end{equation}
are allowed by gauge and Lorentz invariance \cite{Fonseca:2011vn,Braam:2011xh}.

\section{PDG CODES AND EXTENSIONS}\label{SLHA_sect_PDG-tables}

We summarise here our codes for existing and for the extra particle content of the \seesaw\ models. In the first column of each table, the conventional PDG number is shown.

\begin{table}[hbt]
\begin{center}
\subfloat{
\begin{tabular}{|l|c|c|}
\hline
\multicolumn{3}{|c|}{\bf Coloured Fermions} \\
\hline
 PDG & Particle & \PDGIX \\
\hline
2      & $u$          & 110100401 \\
4      & $c$          & 110100402 \\
6      & $t$          & 110100403 \\
-1     & $\bar{d}$    & 110890201 \\
-3     & $\bar{s}$    & 110890202 \\
-5     & $\bar{b}$    & 110890203 \\
\hline
1000021 & $\tilde{g}$ & 211110001 \\
\hline
%

%

\multicolumn{3}{c}{(a)} \\
\multicolumn{3}{c}{} \\
\hline
\multicolumn{3}{|c|}{\bf Colourless, Charged Fermions} \\
\hline
 PDG & Particle & \PDGIX \\
\hline 
-11      & $e^+$                 & 110000601 \\
-13      & $\mu^+$               & 110000602 \\
-15      & $\tau^+$              & 110000603 \\
 1000024 & $\tilde{\chi}_1^+$    & 210000601 \\
 1000037 & $\tilde{\chi}_2^+$    & 210000602 \\
\hline
\multicolumn{3}{c}{(b)} \\
\end{tabular}
 }
 \hspace{1cm}
\subfloat{
\begin{tabular}{|l|c|c|}
\hline
\multicolumn{3}{|c|}{\bf Neutral Fermions} \\
\hline
 PDG & Particle & \PDGIX \\
\hline 
-12/12      & $\bar{\nu}^{D}_1$ / $\nu^{M}_1$ & 110000001 /111000001  \\
-14/14      & $\bar{\nu}^{D}_2$ / $\nu^{M}_2$ & 110000002 /111000002  \\
-16/16      & $\bar{\nu}^{D}_3$ / $\nu^{M}_3$ & 110000003 /111000003  \\
          & $\bar{\nu}^{D}_4$ / $\nu^{M}_4$ & 110000004 /111000004  \\
          & $\bar{\nu}^{D}_5$ / $\nu^{M}_5$ & 110000005 /111000005  \\
          & $\bar{\nu}^{D}_6$ / $\nu^{M}_6$ & 110000006 /111000006  \\
          & $\bar{\nu}^{D}_7$ / $\nu^{M}_7$ & 110000007 /111000007  \\
          & $\bar{\nu}^{D}_8$ / $\nu^{M}_8$ & 110000008 /111000008  \\
          & $\bar{\nu}^{D}_9$ / $\nu^{M}_9$ & 110000009 /111000009  \\
 1000022 & $\tilde{\chi}^0_1$       & 211000001  \\
 1000023 & $\tilde{\chi}^0_2$       & 211000002  \\
 1000025 & $\tilde{\chi}^0_3$       & 211000003  \\
 1000035 & $\tilde{\chi}^0_4$       & 211000004  \\
 1000045 & $\tilde{\chi}^0_5$       & 211000005  \\
         & $\tilde{\chi}^0_6$       & 211000006  \\
         & $\tilde{\chi}^0_7$       & 211000007  \\
\hline
\multicolumn{3}{c}{(c)} \\
\end{tabular}
}
\end{center}
\caption{ (a) The down-type quarks are considered antiparticles due to having
 negative electric charge, hence the anti-downs are the defining states, and
 since they are colour antitriplets, their colour digits are 89 ($99-10$).\\
(b) In the case of $R$-parity violation, these fields mix to eigenstates
 with codes {\tt 1100006XY}.\\
(c) $\nu^D$ are Dirac neutrinos while $\nu^M$ are Majorana neutrinos.
 In the case of $R$-parity violation, the fields mix forming (Majorana) mass
 eigenstates with \PDGIX\ codes {\tt 1110000XY}.}
\end{table}

\begin{table}[hbt]
\begin{center}
\subfloat{
\begin{tabular}{|l|c|c|}
\hline
\multicolumn{3}{|c|}{\bf Coloured Scalars} \\
\hline
 PDG & Particle & \PDGIX \\
\hline 
 1000002 & $\tilde{u}_1$   & 200100401  \\ 
 1000004 & $\tilde{u}_2$   & 200100402  \\
 1000006 & $\tilde{u}_3$   & 200100403  \\
 2000002 & $\tilde{u}_4$   & 200100404  \\
 2000004 & $\tilde{u}_5$   & 200100405  \\
 2000006 & $\tilde{u}_6$   & 200100406  \\
-1000001 & $\tilde{d}^*_1$ & 200890201  \\ 
-1000003 & $\tilde{d}^*_2$ & 200890202  \\
-1000005 & $\tilde{d}^*_3$ & 200890203  \\
-2000001 & $\tilde{d}^*_4$ & 200890204  \\
-2000003 & $\tilde{d}^*_5$ & 200890205  \\
-2000005 & $\tilde{d}^*_6$ & 200890206  \\
\hline
\multicolumn{3}{c}{(a)} \\
\end{tabular}}
 \hspace{1cm}
\subfloat{
\begin{tabular}{|l|c|c|}
\hline
\multicolumn{3}{|c|}{\bf Colourless, Charged Scalars} \\
\hline
 PDG & Particle & \PDGIX \\
\hline 
-1000011      & $\tilde{e}^+_1$              & 200000601 \\
-1000013      & $\tilde{e}^+_2$              & 200000602 \\
-1000015      & $\tilde{e}^+_3$              & 200000603 \\
-2000011      & $\tilde{e}^+_4$              & 200000604 \\
-2000013      & $\tilde{e}^+_5$              & 200000605 \\
-2000015      & $\tilde{e}^+_6$              & 200000606 \\
 37 & $H^+$    & 100000601 \\
\hline
\multicolumn{3}{c}{(b)} \\
\multicolumn{3}{c}{} \\
\hline
\multicolumn{3}{|c|}{\bf Vector Bosons} \\
\hline
 PDG & Particle & \PDGIX \\
\hline 
21  & $g$                &  121110001 \\
22  & $\gamma$           &  121000001 \\
23  & $Z$                &  122000001 \\
32  & $Z'$               &  122000002 \\
33  & $Z^{\prime\prime}$ &  122000003 \\
24  & $W^+$              &  120000601 \\
34  & ${W'}^+$           &  120000602 \\
\hline
\multicolumn{3}{c}{(c)} \\
\end{tabular}}
\end{center}
\caption{
(a) New PDG code for coloured scalars. \\
(b) In the case of $R$-parity violation, the fields mix to eigenstates with \PDGIX\ codes {\tt 1000006XY}.\\
(c) The $W^{+}$ is considered to be the particle (and hence the $W^{-}$ the
 antiparticle) since it has positive electric charge.}
\end{table}

\begin{table}[htb]
\begin{center}
\begin{tabular}{|l|c|c|}
\hline
\multicolumn{3}{|c|}{\bf Neutral Scalars} \\
\hline
 PDG & Particle & \PDGIX \\
\hline
25    & $h_1$   & 101000001 \\
35    & $h_2$   & 101000002 \\
45    & $h_3$   & 101000003 \\
      & $h_4$   & 101000004 \\
36    & $A^0_1$ & 102000001 \\
46    & $A^0_2$ & 102000002 \\
-1000012/1000012 & $\tilde{\nu}^*_1$ / $\RE(\tilde{\nu}_1)$&  200000001 /  201000001  \\
-1000014/1000014 & $\tilde{\nu}^*_2$ / $\RE(\tilde{\nu}_2)$&  200000002 /  201000002  \\
-1000016/1000016 & $\tilde{\nu}^*_3$ / $\RE(\tilde{\nu}_3)$&  200000003 /  201000003  \\
1000017 &  $\IM(\tilde{\nu}_1)$ & 202000001 \\
1000018 &  $\IM(\tilde{\nu}_2)$ & 202000002 \\
1000019 &  $\IM(\tilde{\nu}_3)$ & 202000003 \\
        & $\tilde{\nu}^*_4$ / $\RE(\tilde{\nu}_4)$ / $\IM(\tilde{\nu}_4)$ & 200000004 /  201000004 / 202000004 \\
        & $\tilde{\nu}^*_5$ / $\RE(\tilde{\nu}_5)$ / $\IM(\tilde{\nu}_5)$ & 200000005 /  201000005 / 202000005 \\
        & $\tilde{\nu}^*_6$ / $\RE(\tilde{\nu}_6)$ / $\IM(\tilde{\nu}_6)$ & 200000006 /  201000006 / 202000006 \\
        & $\tilde{\nu}^*_7$ / $\RE(\tilde{\nu}_7)$ / $\IM(\tilde{\nu}_7)$ & 200000007 /  201000007 / 202000007 \\
        & $\tilde{\nu}^*_8$ / $\RE(\tilde{\nu}_8)$ / $\IM(\tilde{\nu}_8)$ & 200000008 /  201000008 / 202000008 \\
        & $\tilde{\nu}^*_9$ / $\RE(\tilde{\nu}_9)$ / $\IM(\tilde{\nu}_9)$ & 200000009 /  201000009 / 202000009 \\
\hline
\end{tabular}
\caption{In the case of $R$-parity violation but CP conservation, the CP-even
 and CP-odd components mix separately to form eigenstates with \PDGIX\ codes
 {\tt 1010000XY} and {\tt 1020000XY}, respectively. If CP violation is present,
 the numbers are {\tt 1000000XY} and {\tt 2000000XY} respectively (no $R$-parity
 violation) or just {\tt 1000000XY} (with $R$-parity violation).}
 \end{center}
\end{table}

\section{Choice of basis}
We propose in this section a choice of basis to fix ambiguities in the different models. This is to be understood as a generalization of the SCKM and SPMNS basis of the MSSM \cite{Allanach:2005kk}. We focus here only on scenarios with three generations of right-handed superfields $\hat{\nu}^c$. In case of more or less generations some rules might have to be adjusted individually, see for instance \cite{Ibarra:2005qi}. Before we start with the new models presented here, we recapitulate first the SCKM and SPMNS basis of the MSSM.

\subsection{SCKM and SPMNS basis}
In the SCKM basis the quark Yukawa matrices in $W_{MSSM}$ are diagonal. 
\begin{equation}
(\hat{Y}_d)_{ii} = (U_d^\dagger Y_d^T V_d)_{ii} \, \hspace{0.5cm} (\hat{Y}_u)_{ii} = (U_u^\dagger Y_u^T V_u)_{ii}
\end{equation}
This is reached by a rotation of the quarks
\begin{equation}
d_L^o = V_d d_L\,, \hspace{0.5cm}
u_L^o = V_u u_L\,, \hspace{0.5cm}
d_R^o = U_d d_R\,, \hspace{0.5cm}
u_R^o = U_u u_R\,, 
\end{equation}
The entire flavor structure can be absorbed in the CKM matrix 
defined as
\begin{equation}
V_{\text{CKM}}=V_u^\dagger V_d~,
\end{equation}
In addition, the following re-parametrization of the soft-breaking squark masses takes place
\begin{equation}
{\hat m_{\tilde q}}^2 \equiv V^\dagger_d \,m^2_{\tilde q}\, V_d\,, \hspace{0.5cm}
{\hat m_{\tilde u}}^2 \equiv U^\dagger_u \,{m^2_{\tilde u}}^T\, U_u\,, \hspace{0.5cm}
{\hat m_{\tilde d}}^2 \equiv U^\dagger_d \,{m^2_{\tilde d}}^T\, U_d\,,
\end{equation}
and the trilinear soft-breaking masses are defined as
\begin{equation}
{\hat T_{U}} \equiv U^\dagger_u \,T_{U}^T\, V_u\,,\hspace{0.5cm}
{\hat T_{D}} \equiv U^\dagger_d \,T_{D}^T\, V_d\,,
\label{eq:that}
\end{equation}
In the SPMNS basis the effective neutrino mass term in \eq{eq:dim5} is diagonalized by a rotation of 
the neutrino fields
\begin{equation}
\label{eq:Vnu}
\nu^o = V_\nu \nu \,,
\end{equation}
\textit{i.e.}
\begin{equation}
  (\hat{m}_\nu)_{ii} = (V_\nu^T m_\nu V_\nu)_{ii}
\end{equation}
In addition, also the lepton Yukawa coupling is diagonalized by a rotation 
of the charged lepton fields
\begin{equation}
\label{eq:rotVe}
e_L^o = V_e e_L \hspace{0.5cm} \mbox{and} \hspace{0.5cm} e_R^o = U_e e_R \, .
\end{equation}
The equivalent diagonalised charged lepton Yukawa matrix is 
\begin{equation}
(\hat{Y}_e)_{ii} = (U_e^\dag Y_e^T V_e)_{ii} 
\end{equation}
The PMNS basis can be defined by using the mixing matrices $V_e$ and $V_\nu$ as
\begin{equation}
U_{PMNS}=V_e^\dag V_\nu~, 
\end{equation}
and contains the information about the neutrino mxing. 
It's standard parametrization is given by
\begin{eqnarray}
\label{def:unu}
U_{PMNS}=
\left(
\begin{array}{ccc}
 c_{12}c_{13} & s_{12}c_{13}  & s_{13}e^{-i\delta}  \\
-s_{12}c_{23}-c_{12}s_{23}s_{13}e^{i\delta}  & 
c_{12}c_{23}-s_{12}s_{23}s_{13}e^{i\delta}  & s_{23}c_{13}  \\
s_{12}s_{23}-c_{12}c_{23}s_{13}e^{i\delta}  & 
-c_{12}s_{23}-s_{12}c_{23}s_{13}e^{i\delta}  & c_{23}c_{13}  
\end{array}
\right) 
 \times
 \left(
 \begin{array}{ccc}
 e^{i\alpha_1/2} & 0 & 0 \\
 0 & e^{i\alpha_2/2}  & 0 \\
 0 & 0 & 1
 \end{array}
 \right)
\end{eqnarray}with $c_{ij} = \cos \theta_{ij}$ and $s_{ij} = \sin \theta_{ij}$. The angles $\theta_{12}$,
 $\theta_{13}$ and  $\theta_{23}$ are the solar neutrino angle, the reactor (or CHOOZ) angle
and the atmospheric neutrino mixing angle, respectively. $\delta$ is the Dirac phase
and $\alpha_i$ are Majorana phases. The recent experimental values of the angles 
are given in \cite{Schwetz:2011zk}.

After that change of basis, the leptonic soft-breaking terms are written in the new
basis as
\begin{eqnarray}
& {\hat{m}_{\tilde l}}^2 = V^\dagger_e \,m^2_{\tilde l}\, V_e\,, \hspace{0.5cm}
{\hat{m}_{\tilde e}}^2 = U^\dagger_e \,{m^2_{\tilde e}}^T\, U_e\,, & \\
& \hat{T}_e =  U^\dagger_e \, T_e^T \, V_e  &
\end{eqnarray}

\subsection{Choice of basis in seesaw type I -- III}
\paragraph{Type I} In seesaw type-I we fix the additional superfield by diagonalizing $M_{\nu^c}$ due to 
a rotation of $\nu^c$
\begin{equation}
U^\dagger_{\nu^c} M_{\nu^c} U_{\nu^c} = \hat{M}_{\nu^c}^{ii} 
\end{equation}
with
\begin{equation}
\nu^{c,0} = U_{\nu^c} \nu^c \, . 
\end{equation}
The neutrino Yukawa coupling can be re-expressed by
\begin{eqnarray}
 & \hat{Y}_\nu =  V^\dagger_\nu  Y^T_\nu U_{\nu^c}
\end{eqnarray}
with the running $\overline{DR}$ value of the rotation matrix $V_\nu$ defined in  \eq{eq:Vnu}. 
To fulfill neutrino data, $\hat{Y}_\nu$ can be calculated in this basis using the approach by Casas-Ibarra \cite{Casas:2001sr}
\begin{equation}
\label{eq:Ynu}
\hat{Y}_{\nu} =\sqrt{2}\frac{i}{v_u}\sqrt{\hat M_{\nu^c}}\cdot R \cdot \sqrt{{\hat m_{\nu}}} \cdot U_{PMNS}^{\dagger},
\end{equation}
where the $\hat m_{\nu}$ is a diagonal matrices containing the neutrino masses. 
$R$ is  in general a complex orthogonal matrix.

\paragraph{Type II} In type II with one generation of 15-plets no new freedom in the field definition arises. 
The neutrino data is mostly given by the matrix $Y_T$ after $SU(5)$ breaking. 
This matrix is diagonalized by the same matrix as $m_{\nu}$. 
If all neutrino eigenvalues, angles and phases were known, $Y_T$ 
would be fixed up to an overall constant which can be easily 
estimated to be 
\begin{equation}\label{est}
\frac{M_T}{\lambda_2} \simeq 10^{15} {\rm GeV} \hskip2mm 
\Big(\frac{0.05 \hskip1mm {\rm eV}}{m_{\nu}}\Big).
\end{equation}

If a superposition of type-I and type-II is present, it is possible to find valid value for the superpotential
parameters to get correct neutrino data as discussed in Ref.~\cite{Akhmedov:2006de}.

If several generations of 15-plets are added we propose that the corresponding bilinear term 
has to be diagonalized to fix the basis
\begin{equation}
U^\dagger_{ \overline{15}} M_{15} V_{15} = \hat{M}_{15}^{ii} 
\end{equation}
with 
\begin{equation}
{\bf \overline{15}}^0 = U_{{ \overline{15}}} {\bf \overline{15}} \, \hspace{1cm} {\bf 15}^0 = V_{15} {\bf 15} . 
\end{equation}
If no GUT unifcation for the components of the 15-plets is assumed this diagonalization procedure has to 
be performed for the different components separately.

\paragraph{Type-III} In case of several generations of 24-plets their phases are fixed by diagonalizing the bilinear term $M_{24}$
\begin{equation}
U^\dagger_{24} M_{24} U_{24} = \hat{M}_{24}^{ii}
\end{equation}
with
\begin{equation}
{\bf 24}^0 = U_{24} {\bf 24} \, .
\end{equation}

\subsection{Choice of basis in linear and inverse seesaw}
We define here the basis for three generations of right-handed sneutrino superfields $\hat{\nu}^c$ and three generations of additional singlet superfields $\hat{N}_S$ by demanding $M_R$ to be diagonal
\begin{equation}
U^\dagger_N M^T_R U_{\nu^c} = \hat{M}_R^{ii} 
\end{equation}
with
\begin{equation}
\nu^{c,0} = U_{\nu^c} \nu^c \, \hspace{0.5cm} N_S^0 = U_N N_S
\end{equation}
The other superpotential parameters are rotated as
\begin{eqnarray}
 & \hat{Y}_\nu =  V^\dagger_\nu Y^T_\nu U_{\nu^c} \, \hspace{0.5cm} \hat{\mu}_N = U^\dagger_N \mu_N U_N \, \hspace{0.5cm} \hat{Y}_{LS} = V^\dagger_\nu Y^T_{LS} U_N & 
\end{eqnarray}
while the soft-breaking terms read in the new basis
\begin{eqnarray}
 & \hat{B}_R =  U^\dagger_N B^T_R U_{\nu^c} \, \hspace{0.5cm} \hat{B}_{N} = U^\dagger_N B_N U_N \, \\
& \hat{T}_\nu = V^\dagger_\nu T^T_\nu U_{\nu^c} \,  \hspace{0.5cm} \hat{T}_{LS} = V^\dagger_\nu T^T_{LS} U_N & \\
 & \hat{m}_{\nu^c}^2 =  U^\dagger_{\nu^c} m^{2,T}_{\nu^c} U_{\nu^c} \, \hspace{0.5cm}  \hat{m}_{N}^2 =  U^\dagger_{N} m^{2,T}_{N} U_{N} &
\end{eqnarray}
To get correct neutrino data $Y_\nu$ has to be choosen apropiatly. This can be done by using the 
parametrization given in Ref.~\cite{Forero:2011pc}. The formula in case of inverse seesaw reads
\begin{equation}
\label{eq:YnuInv}
Y_\nu = \frac{\sqrt{2}}{v_u} U_{PMNS} \sqrt{\hat{m}_\nu} R^T  \sqrt{\hat{\mu}}^{-1} \hat{M}_R
\end{equation}
Here, $\hat{m}_\nu$ and $R$ are the matrices
already introduced in \eq{eq:Ynu}. In case of linear seesaw $Y_\nu$ can be calculated by 
\begin{equation}
\label{eq:YnuLin}
Y_\nu = \frac{2}{v^2_u} U_{PMNS} \sqrt{\hat{m}_\nu} A^T \sqrt{\hat{m}_\nu} U_{PMNS}^T  \hat{Y}^{-1}_{LN} \hat{M}_R  
\end{equation}
with
\begin{equation}
A = \left( \begin{array}{ccc} \frac{1}{2} & a & b \\
  -a & \frac{1}{2} & c \\ 
  -b & -c & \frac{1}{2} \end{array} \right) 
\end{equation}
and real numbers $a$, $b$, $c$.

\subsection{Choice of basis in seesaw scenearios in  $U(1)_R \times U(1)_{B-L}$ gauge sector}
In case of minimal seesaw we fix the basis by demanding
\begin{equation}
U^\dagger_{\nu^c} Y_M U_{\nu^c} = \hat{Y}_M^{ii} \,
\end{equation}
while for the linear and inverse realization of the seesaw, $Y_{N \nu^c}$ has to be diagonal
\begin{equation}
U^\dagger_N Y^T_{N \nu^c} U_{\nu^c} = \hat{Y}_{N \nu^c}^{ii}  \, .
\end{equation}
Both conditions can be fullfilled by a rotation of $\nu^c$ respectively $\nu^c$ and $N_S$
\begin{equation}
\nu^{c,0} = U_{\nu^c} \nu^c \, \hspace{0.5cm} N_S^0 = U_N N_S \, . 
\end{equation}
The other superpotential parameters are rotated as
\begin{eqnarray}
 & \hat{Y}_\nu = V^\dagger_\nu Y^T_\nu U_{\nu^c} \, \hspace{0.5cm} \hat{\mu}_N = U^\dagger_N \mu_N U_N \, \hspace{0.5cm} \hat{Y}_{LS} = V^\dagger_\nu Y^T_{LS} U_N & 
\end{eqnarray}
while the soft-breaking terms read in the new basis
\begin{eqnarray}
 & \hat{T}_M = U^\dagger_{\nu^c} T_M U_{\nu^c} \, \hspace{1cm} \hat{T}_{N \nu^c}  =  U^\dagger_N T^T_{N \nu^c} U_{\nu^c}   & \\
& \hat{T}_\nu = V^\dagger_\nu T^T_\nu U_{\nu^c} \,  \hspace{0.5cm} \hat{T}_{LS} = V^\dagger_\nu T^T_{LS} V_N & \\
 & \hat{B}_{N} = U^\dagger_N B_N U_N \, \\
 & \hat{m}_{\nu^c}^2 =  U^\dagger_{\nu^c} m^{2,T}_{\nu^c} U_{\nu^c} \, \hspace{0.5cm}  \hat{m}_{N}^2 =  U^\dagger_{N} m^{2,T}_{N} U_{N} &
\end{eqnarray}
The neutrino Yukawa couplings can be choose in analogy to \eq{eq:Ynu},\eq{eq:YnuInv} and \eq{eq:YnuLin}.

\subsection{Choice of basis in seesaw scenearios in  $U(1)_Y \times U(1)_{B-L}$ gauge sector}
In case of minimal seesaw we fix the basis by demanding
\begin{equation}
U^\dagger_{\nu^c} Y_{\eta \nu^c} U_{\nu^c} = \hat{Y}_{\eta \nu^c}^{ii} \,
\end{equation}
while for the linear and inverse realization of the seesaw, $Y_{N \nu^c}$ has to be diagonal
\begin{equation}
U^\dagger_N Y^T_{IS} U_{\nu^c} = \hat{Y}_{IS}^{ii}  \, .
\end{equation}
Both conditions can be fullfilled by a rotation of $\nu^c$ respectively $\nu^c$ and $N_S$
\begin{equation}
\nu^{c,0} = U_{\nu^c} \nu^c \, \hspace{0.5cm} N_S^0 = U_N N_S \, . 
\end{equation}
The other superpotential parameters are rotated as
\begin{eqnarray}
 & \hat{Y}_\nu = V^\dagger_\nu Y^T_\nu U_{\nu^c} \, \hspace{0.5cm} \hat{\mu}_N = U^\dagger_N \mu_N U_N \, \hspace{0.5cm} \hat{Y}_{LS} = V^\dagger_\nu Y^T_{LS} U_N & 
\end{eqnarray}
while the soft-breaking terms read in the new basis
\begin{eqnarray}
 & \hat{T}_{\eta \nu^c} = U^\dagger_{\nu^c} T_{\eta \nu^c} U_{\nu^c} \, \hspace{1cm} \hat{T}_{IS}  =  U^\dagger_N T^T_{IS} U_{\nu^c}   & \\
 & \hat{T}_\nu = V^\dagger_\nu T^T_\nu U_{\nu^c} \,  \hspace{0.5cm} \hat{T}_{LS} = V^\dagger_\nu T^T_{LS} U_N & \\
 & \hat{B}_{N} = U^\dagger_N B_N U_N \, \\
 & \hat{m}_{\nu^c}^2 =  U^\dagger_{\nu^c} m^{2,T}_{\nu^c} U_{\nu^c} \, \hspace{0.5cm}  \hat{m}_{N}^2 =  U^\dagger_{N} m^{2,T}_{N} U_{N} &
\end{eqnarray}
Since the phase of the field $N_{S'}$ appearing in linear and inverse seesaw has no physical impact it is not necessary to fix it. 

As for $U(1)_R \times U(1)_{B-L}$, the neutrino Yukawa couplings can be choose in analogy to \eq{eq:Ynu},\eq{eq:YnuInv} and \eq{eq:YnuLin}.

\section{Tree-level mass matrices}
We give here the mass matrices of the different models which change in comparison to the MSSM. The equations are based on the \LaTeX\ output for the Mathematica package  {\tt SARAH} \cite{Staub:2008uz,Staub:2009bi,Staub:2010jh}. \\
For a given basis $\Phi$, the scalar mass matrix $m^2$ is defined
as
\begin{equation}
L = -  \Phi^\dagger m^2 \Phi
\end{equation}
while for fermions the conventions are for Majorana mass matrices $m_M$ and Dirac mass matrices $m_D$
\begin{align}
 L = & - \Psi^T m_M \Psi \\
 L = & - \Psi_1^T m_D \Psi_2
\end{align}
with basis vectors $\Psi$, $\Psi_1$ and $\Psi_2$ given in Weyl spinors.

\subsection{Linear and inverse seesaw}
We show here the mass matrices for an imaginary model which has all terms of the inverse and linear at once. The mass matrices for the physical relevant models, separated linear and inverse seesaw, can easily obtained by setting the unnecessary terms to zero. 
\subsubsection{Mass matrix for pseudo scalar sneutrinos}
Basis: \( \left(\sigma_L, \sigma_R, \sigma_S\right)\)
\begin{equation} 
m^2_{\nu^i} = \left( 
\begin{array}{ccc}
m_{11} &m^T_{21} &m^T_{31}\\ 
m_{21} &m_{22} &m^T_{32}\\ 
m_{31} &m_{32} &m_{33}\end{array} 
\right) 
\end{equation} 
\begin{align} 
m_{11} &= \frac{1}{8} \Big(2 v_{u}^{2} \Big(2 {\Re\Big({Y_{LN}^{T}  Y_{LN}^*}\Big)}  + 2 {\Re\Big({Y_{\nu}^{T}  Y_\nu^*}\Big)} \Big) + 8 {\Re\Big(m_l^2\Big)}  + \Big(g_{1}^{2} + g_{2}^{2}\Big){\bf 1} \Big(- v_{u}^{2}  + v_{d}^{2}\Big)\Big)\\ 
m_{21} &= \frac{1}{2} \frac{1}{\sqrt{2}} \Big(2 v_d {\Re\Big(\mu Y_\nu^* \Big)}  + v_u \Big(2 {\Re\Big({M_\nu  Y_{LN}^*}\Big)}  -2 {\Re\Big(T_\nu\Big)} \Big)\Big)\\ 
m_{22} &= \frac{1}{4} \Big(2 v_{u}^{2} {\Re\Big({Y_\nu  Y_{\nu}^{\dagger}}\Big)}  + 4 {\Re\Big(m^2_{\nu^c}\Big)}  + 4 {\Re\Big({M_\nu  M_{\nu}^{\dagger}}\Big)} \Big)\\ 
m_{31} &= \frac{1}{4} \frac{1}{\sqrt{2}} \Big(4 v_d {\Re\Big(\mu Y_{LN}^* \Big)}  + v_u \Big(4 {\Re\Big({M_{\nu}^{T}  Y_\nu^*}\Big)}  + 4 {\Re\Big({\mu_N  Y_{LN}^*}\Big)}  -4 {\Re\Big(T_{LN}\Big)} \Big)\Big)\\ 
m_{32} &= \frac{1}{4} \Big(2 m_{\nu^c s}^{2,T}  + 2 v_{u}^{2} {\Re\Big({Y_{LN}  Y_{\nu}^{\dagger}}\Big)}  -4 {\Re\Big(B_{\nu}^{T}\Big)}  + 4 {\Re\Big({\mu_N  M_{\nu}^{\dagger}}\Big)} \Big)\\ 
m_{33} &= \frac{1}{8} \Big(2 \Big(2 v_{u}^{2} {\Re\Big({Y_{LN}  Y_{LN}^{\dagger}}\Big)}  + 4 {\Re\Big({M_{\nu}^{T}  M_\nu^*}\Big)}  - 2 \Re(B_N) \Big) -4 {\Re\Big(B_N\Big)}  + 8 {\Re\Big(m^2_{N}\Big)}  + 8 {\Re\Big({\mu_N  \mu_N^*}\Big)} \Big)
\end{align} 
This matrix is diagonalized by \(Z^i\): 
\begin{equation} 
Z^i m^2_{\nu^i} Z^{i,\dagger} = m^{dia}_{2,\nu^i} 
\end{equation}

\subsubsection{Mass matrix for scalar sneutrinos}
Basis: \( \left(\phi_L, \phi_R, \phi_S\right)\)
\begin{equation} 
m^2_{\nu^R} = \left( 
\begin{array}{ccc}
m_{11} &m^T_{21} &m^T_{31}\\ 
m_{21} &m_{22} &m^T_{32}\\ 
m_{31} &m_{32} &m_{33}\end{array} 
\right) 
\end{equation} 
\begin{align} 
m_{11} &= \frac{1}{8} \Big(2 v_{u}^{2} \Big(2 {\Re\Big({Y_{LN}^{T}  Y_{LN}^*}\Big)}  + 2 {\Re\Big({Y_{\nu}^{T}  Y_\nu^*}\Big)} \Big) + 8 {\Re\Big(m_l^2\Big)}  + \Big(g_{1}^{2} + g_{2}^{2}\Big){\bf 1} \Big(- v_{u}^{2}  + v_{d}^{2}\Big)\Big)\\ 
m_{21} &= \frac{1}{2} \frac{1}{\sqrt{2}} \Big(-2 v_d {\Re\Big(\mu Y_\nu^* \Big)}  + v_u \Big(2 {\Re\Big({M_\nu  Y_{LN}^*}\Big)}  + 2 {\Re\Big(T_\nu\Big)} \Big)\Big)\\ 
m_{22} &= \frac{1}{4} \Big(2 v_{u}^{2} {\Re\Big({Y_\nu  Y_{\nu}^{\dagger}}\Big)}  + 4 {\Re\Big(m^2_{\nu^c}\Big)}  + 4 {\Re\Big({M_\nu  M_{\nu}^{\dagger}}\Big)} \Big)\\ 
m_{31} &= \frac{1}{4} \frac{1}{\sqrt{2}} \Big(-4 v_d {\Re\Big(\mu Y_{LN}^* \Big)}  + v_u \Big(4 {\Re\Big({M_{\nu}^{T}  Y_\nu^*}\Big)}  + 4 {\Re\Big({\mu_N  Y_{LN}^*}\Big)}  + 4 {\Re\Big(T_{LN}\Big)} \Big)\Big)\\ 
m_{32} &= \frac{1}{4} \Big(2 m_{\nu^c s}^{2,T}  + 2 v_{u}^{2} {\Re\Big({Y_{LN}  Y_{\nu}^{\dagger}}\Big)}  + 4 {\Re\Big(B_{\nu}^{T}\Big)}  + 4 {\Re\Big({\mu_N  M_{\nu}^{\dagger}}\Big)} \Big)\\ 
m_{33} &= \frac{1}{8} \Big(2 \Big(2 v_{u}^{2} {\Re\Big({Y_{LN}  Y_{LN}^{\dagger}}\Big)}  + 4 {\Re\Big({M_{\nu}^{T}  M_\nu^*}\Big)}  + 2 \Re(B_N) \Big) +  + 4 {\Re\Big(B_N\Big)}  + 8 {\Re\Big(m^2_{N}\Big)}  + 8 {\Re\Big({\mu_N  \mu_N^*}\Big)} \Big)
\end{align} 
This matrix is diagonalized by \(Z^R\): 
\begin{equation} 
Z^R m^2_{\nu^R} Z^{R,\dagger} = m^{dia}_{2,\nu^R} 
\end{equation} 

\subsubsection{Mass matrix for neutrinos}
Basis: \( \left(\nu_L, \nu_R, \tilde{S}\right)\) 
 
\begin{equation} 
m_{\nu} = \left( 
\begin{array}{ccc}
0 &\frac{1}{\sqrt{2}} v_u Y_{\nu}^{T}  &\frac{1}{\sqrt{2}} v_u Y_{LN}^{T} \\ 
\frac{1}{\sqrt{2}} v_u Y_\nu  &0 &M_\nu\\ 
\frac{1}{\sqrt{2}} v_u Y_{LN}  &M_{\nu}^{T} &\mu_N\end{array} 
\right) 
\end{equation} 
This matrix is diagonalized by \(U^V\): 
\begin{equation} 
U^{V,*} m_{\nu} U^{V,\dagger} = m^{dia}_{\nu} 
\end{equation} 

\subsection{\Seesaw\ in models with \URxUBL\ gauge sector}
We use in the following
\begin{equation*}
 v^2_{X_R} =  \left\{\begin{array}{c} 2 v^2_{\delta_R} \, \text{minimal \seesaw} \\
                                    v^2_{\xi_R} \, \text{linear and inverse \seesaw} 
                        \end{array} \right. \, , \hspace{0.5cm}
 v^2_{\bar{X}_R} =  \left\{\begin{array}{c} 2 v^2_{\bar{\delta}_R} \, \text{minimal \seesaw} \\
                                    v^2_{\bar{\xi}_R} \, \text{linear and inverse \seesaw} 
                        \end{array} \right. 
\end{equation*}

\subsubsection{Mass matrix for down squarks}
Basis: \( \left(\tilde{d}_{L,{{\alpha}}}, \tilde{d}_{R,{{\alpha}}}\right)\)
\begin{equation} 
m^2_{\tilde{d}} = \left( 
\begin{array}{cc}
m_{11} &\frac{1}{\sqrt{2}}  \Big(v_d T_d^\dagger  - v_u Y_d^\dagger \mu \Big)\\ 
\frac{1}{\sqrt{2}} \Big(v_d T_{d}  - v_u \mu^* Y_{d} \Big) &m_{22}\end{array} 
\right) 
\end{equation} 
\begin{align} 
m_{11} &= \frac{1}{24}  \Big({\bf 1} \Big(3 g_{L}^{2} \Big(v_{u}^{2} -  v_{d}^{2}\Big) + g_{BL}^{2} \Big(v_{\bar{X}_R}^{2}- v_{X_R}^{2} \Big) + g_{BL} \tilde{g} \Big(- v_{\bar{X}_R}^{2}  - v_{d}^{2}  + v_{X_R}^{2} + v_{u}^{2}\Big)\Big)+24 m_q^{2} \nonumber \\ 
 &+12 v_{d}^{2} {Y_d^\dagger  Y_{d}} \Big)\\ 
m_{22} &= \frac{1}{24} \Big({\bf 1} \Big(3 \Big(g_{R}^{2} + \tilde{g}^{2}\Big)\Big(- v_{\bar{X}_R}^{2}  - v_{d}^{2}  + v_{X_R}^{2} + v_{u}^{2}\Big) + g_{BL}^{2} \Big(v_{X_R}^{2} - v_{\bar{X}_R}^{2} \Big) + g_{BL} \tilde{g} \Big(4 v_{\bar{X}_R}^{2}  -4 v_{X_R}^{2}  - v_{u}^{2}  + v_{d}^{2}\Big)\Big)\nonumber \\ 
 &+24 m_d^{2} +12 v_{d}^{2} {Y_{d}  Y_d^{\dagger}} \Big)
\end{align} 
This matrix is diagonalized by \(Z^D\): 
\begin{equation} 
Z^D m^2_{\tilde{d}} Z^{D,\dagger} = m^{dia}_{2,\tilde{d}} 
\end{equation}

\subsubsection{Mass matrix for pseudo scalar sneutrinos}
\paragraph{minimal seesaw}
Basis: \( \left(\sigma_L, \sigma_R\right)\)
\begin{equation} 
m^2_{\nu^i} = \left( 
\begin{array}{cc}
m_{11} &m^T_{21} \\ 
m_{21} &m_{22}  \end{array} 
\right) 
\end{equation} 
\begin{align} 
m_{11} &= \frac{1}{8} \Big({\bf 1} \Big(2 g_{BL}^{2} \Big(- v_{\bar{\delta}_R}^{2}  + v_{\delta_R}^{2}\Big) + g_{BL} \tilde{g} \Big(2 v_{\bar{\delta}_R}^{2}  -2 v_{\delta_R}^{2}  - v_{u}^{2}  + v_{d}^{2}\Big) + g_{L}^{2} \Big(- v_{u}^{2}  + v_{d}^{2}\Big)\Big)+8 {\Re\Big(m_l^2\Big)} \nonumber \\ 
 &+4 v_{u}^{2} {\Re\Big({Y_{v}^{T}  Y_v^*}\Big)} \Big)\\ 
m_{21} &= \frac{1}{4} \Big(-2 \sqrt{2} v_d {\Re\Big(\mu Y_v^* \Big)}  + v_u \Big(2 \sqrt{2} {\Re\Big(T_\nu\Big)}  - v_{\delta_R} \Big(2 {\Re\Big({Y_{M}^{T}  Y_v^*}\Big)}  + 2 {\Re\Big({Y_{M}  Y_v^*}\Big)} \Big)\Big)\Big)\\ 
m_{22} &= \frac{1}{8} \Big({\bf 1} \Big(2 g_{BL}^{2} v_{\bar{\delta}_R}^{2}  + g_{BL} \tilde{g} \Big(-4 v_{\bar{\delta}_R}^{2}  - v_{d}^{2}  + v_{u}^{2}\Big) + \Big(g_{R}^{2} + \tilde{g}^{2}\Big)\Big(2 v_{\bar{\delta}_R}^{2}  - v_{u}^{2}  + v_{d}^{2}\Big)\Big)+2 \sqrt{2} \mu_{\delta} v_{\bar{\delta}_R} Y_{M}^{\dagger} \nonumber \\ 
 &+2 \sqrt{2} \mu_{\delta} v_{\bar{\delta}_R} Y_{M}^* +8 {\Re\Big(m_{\nu}^2\Big)} +4 v_{u}^{2} {\Re\Big({Y_v  Y_{v}^{\dagger}}\Big)} \nonumber \\ 
 &+2 v_{\delta_R}^{2} \Big(2 {\Re\Big({Y_{M}^{T}  Y_{M}^*}\Big)}  + 2 {\Re\Big({Y_{M}^{T}  Y_{M}^{\dagger}}\Big)}  + 2 {\Re\Big({Y_{M}  Y_{M}^*}\Big)}  + 2 {\Re\Big({Y_{M}  Y_{M}^{\dagger}}\Big)}  - \Big(g_{R}^{2} + \Big(- \tilde{g}  + g_{BL}\Big)^{2}\Big){\bf 1} \Big)\nonumber \\ 
 &-2 \sqrt{2} v_{\delta_R} \Big(2 {\Re\Big(T_{M}\Big)}  + 2 {\Re\Big(T_{M}^{T}\Big)} \Big)+2 \sqrt{2} v_{\bar{\delta}_R} \mu_{\delta}^* \Big(Y_{M} + Y_{M}^{T}\Big)\Big)
\end{align} 

\paragraph{linear and inverse seesaw}
Basis: \( \left(\sigma_L, \sigma_R, \sigma_S\right)\)
\begin{equation} 
m^2_{\nu^i} = \left( 
\begin{array}{ccc}
m_{11} &m^T_{21} &\frac{1}{2} v_{\xi_R} v_u {\Re\Big({Y_{v}^{T}  Y_{N \nu^c}^{\dagger}}\Big)} \\ 
m_{21} &m_{22} &m^T_{32}\\ 
\frac{1}{2} v_{\xi_R} v_u {\Re\Big({Y_{N \nu^c}  Y_v^*}\Big)}  &m_{32} &m_{33}\end{array} 
\right) 
\end{equation} 
\begin{align} 
m_{11} &= \frac{1}{8} \Big(4 v_{u}^{2} {\Re\Big({Y_{v}^{T}  Y_v^*}\Big)}  + 8 {\Re\Big(m_l^2\Big)}  + {\bf 1} \Big(g_{BL}^{2} \Big(- v_{\bar{\xi}_R}^{2}  + v_{\xi_R}^{2}\Big) + g_{BL} \tilde{g} \Big(- v_{u}^{2}  - v_{\xi_R}^{2}  + v_{\bar{\xi}_R}^{2} + v_{d}^{2}\Big) + g_{L}^{2} \Big(- v_{u}^{2}  + v_{d}^{2}\Big)\Big)\Big)\\ 
m_{21} &= -\frac{1}{2} \frac{1}{\sqrt{2}} \Big(2 v_d {\Re\Big(\mu Y_v^* \Big)}  -2 v_u {\Re\Big(T_\nu\Big)} \Big)\\ 
m_{22} &= \frac{1}{8} \Big({\bf 1} \Big(g_{BL}^{2} \Big(- v_{\xi_R}^{2}  + v_{\bar{\xi}_R}^{2}\Big) + g_{BL} \tilde{g} \Big(-2 v_{\bar{\xi}_R}^{2}  + 2 v_{\xi_R}^{2}  - v_{d}^{2}  + v_{u}^{2}\Big) - \Big(g_{R}^{2} + \tilde{g}^{2}\Big)\Big(- v_{\bar{\xi}_R}^{2}  - v_{d}^{2}  + v_{\xi_R}^{2} + v_{u}^{2}\Big)\Big)\nonumber \\ 
 &+8 {\Re\Big(m_{\nu}^2\Big)} +2 \Big(2 v_{u}^{2} {\Re\Big({Y_v  Y_{v}^{\dagger}}\Big)}  + 2 v_{\xi_R}^{2} {\Re\Big({Y_{N \nu^c}^{T}  Y_{N \nu^{c} *}}\Big)} \Big)\Big)\\ 
m_{32} &= -\frac{1}{2} \frac{1}{\sqrt{2}} \Big(2 v_{\bar{\xi}_R} {\Re\Big(\mu_{\xi} Y_{N \nu^{c},*} \Big)}  + v_{\xi_R} \Big(-2 {\Re\Big(T_{N \nu^c}\Big)}  + 4 {\Re\Big({\mu_N  Y_{N \nu^{c} *}}\Big)} \Big)\Big)\\ 
m_{33} &= \frac{1}{8} \Big(4 \Big(8 {\Re\Big({\mu_N  \mu_N^*}\Big)}  - B_N \Big) -4 B_N^*  -8 {\Re\Big(B_N\Big)}  + 8 {\Re\Big(m_{N}^2\Big)}  + v_{\xi_R}^{2} \Big(2 v_{d}^{2} {\Re\Big({\tilde{W}_{LS}^{T}  \tilde{W}_{LS}^*}\Big)}  + 4 {\Re\Big({Y_{N \nu^c}  Y_{N \nu^c}^{\dagger}}\Big)} \Big)\Big)
\end{align} 
This matrix is diagonalized by \(Z^i\): 
\begin{equation} 
Z^i m^2_{\nu^i} Z^{i,\dagger} = m^{dia}_{2,\nu^i} 
\end{equation}

\subsubsection{Mass matrix for scalar sneutrinos}
\paragraph{minimal seesaw}
Basis: \( \left(\phi_L, \phi_R\right)\)
\begin{equation} 
m^2_{\nu^R} = \left( 
\begin{array}{cc}
m_{11} &m^T_{21} \\ 
m_{21} &m_{22} \end{array} 
\right) 
\end{equation} 
\begin{align} 
m_{11} &= \frac{1}{8} \Big({\bf 1} \Big(2 g_{BL}^{2} \Big(- v_{\bar{\delta}_R}^{2}  + v_{\delta_R}^{2}\Big) + g_{BL} \tilde{g} \Big(2 v_{\bar{\delta}_R}^{2}  -2 v_{\delta_R}^{2}  - v_{u}^{2}  + v_{d}^{2}\Big) + g_{L}^{2} \Big(- v_{u}^{2}  + v_{d}^{2}\Big)\Big)+8 {\Re\Big(m_l^2\Big)} \nonumber \\ 
 &+4 v_{u}^{2} {\Re\Big({Y_{v}^{T}  Y_v^*}\Big)} \Big)\\ 
m_{21} &= \frac{1}{4} \Big(-2 \sqrt{2} v_d {\Re\Big(\mu Y_v^* \Big)}  + v_u \Big(2 \sqrt{2} {\Re\Big(T_\nu\Big)}  + v_{\delta_R} \Big(2 {\Re\Big({Y_{M}^{T}  Y_v^*}\Big)}  + 2 {\Re\Big({Y_{M}  Y_v^*}\Big)} \Big)\Big)\Big)\\ 
m_{22} &= \frac{1}{8} \Big({\bf 1} \Big(2 g_{BL}^{2} v_{\bar{\delta}_R}^{2}  + g_{BL} \tilde{g} \Big(-4 v_{\bar{\delta}_R}^{2}  - v_{d}^{2}  + v_{u}^{2}\Big) + \Big(g_{R}^{2} + \tilde{g}^{2}\Big)\Big(2 v_{\bar{\delta}_R}^{2}  - v_{u}^{2}  + v_{d}^{2}\Big)\Big)-2 \sqrt{2} \mu_{\delta} v_{\bar{\delta}_R} Y_{M}^{\dagger} \nonumber \\ 
 &-2 \sqrt{2} \mu_{\delta} v_{\bar{\delta}_R} Y_{M}^* +8 {\Re\Big(m_{\nu}^2\Big)} +4 v_{u}^{2} {\Re\Big({Y_v  Y_{v}^{\dagger}}\Big)} \nonumber \\ 
 &+2 v_{\delta_R}^{2} \Big(2 {\Re\Big({Y_{M}^{T}  Y_{M}^*}\Big)}  + 2 {\Re\Big({Y_{M}^{T}  Y_{M}^{\dagger}}\Big)}  + 2 {\Re\Big({Y_{M}  Y_{M}^*}\Big)}  + 2 {\Re\Big({Y_{M}  Y_{M}^{\dagger}}\Big)}  - \Big(g_{R}^{2} + \Big(- \tilde{g}  + g_{BL}\Big)^{2}\Big){\bf 1} \Big)\nonumber \\ 
 &+2 \sqrt{2} v_{\delta_R} \Big(2 {\Re\Big(T_{M}\Big)}  + 2 {\Re\Big(T_{M}^{T}\Big)} \Big)-2 \sqrt{2} v_{\bar{\delta}_R} \mu_{\delta}^* \Big(Y_{M} + Y_{M}^{T}\Big)\Big)
\end{align} 
\paragraph{linear and inverse seesaw}
Basis: \( \left(\phi_L, \phi_R, \phi_S\right)\)
\begin{equation} 
m^2_{\nu^R} = \left( 
\begin{array}{ccc}
m_{11} &m^T_{21} &\frac{1}{2} v_{\xi_R} v_u {\Re\Big({Y_{v}^{T}  Y_{N \nu^c}^{\dagger}}\Big)} \\ 
m_{21} &m_{22} &m^T_{32}\\ 
\frac{1}{2} v_{\xi_R} v_u {\Re\Big({Y_{N \nu^c}  Y_v^*}\Big)}  &m_{32} &m_{33}\end{array} 
\right) 
\end{equation} 
\begin{align} 
m_{11} &= \frac{1}{8} \Big(4 v_{u}^{2} {\Re\Big({Y_{v}^{T}  Y_v^*}\Big)}  + 8 {\Re\Big(m_l^2\Big)}  + {\bf 1} \Big(g_{BL}^{2} \Big(- v_{\bar{\xi}_R}^{2}  + v_{\xi_R}^{2}\Big) + g_{BL} \tilde{g} \Big(- v_{u}^{2}  - v_{\xi_R}^{2}  + v_{\bar{\xi}_R}^{2} + v_{d}^{2}\Big) + g_{L}^{2} \Big(- v_{u}^{2}  + v_{d}^{2}\Big)\Big)\Big)\\ 
m_{21} &= -\frac{1}{2} \frac{1}{\sqrt{2}} \Big(2 v_d {\Re\Big(\mu Y_v^* \Big)}  -2 v_u {\Re\Big(T_\nu\Big)} \Big)\\ 
m_{22} &= \frac{1}{8} \Big({\bf 1} \Big(g_{BL}^{2} \Big(- v_{\xi_R}^{2}  + v_{\bar{\xi}_R}^{2}\Big) + g_{BL} \tilde{g} \Big(-2 v_{\bar{\xi}_R}^{2}  + 2 v_{\xi_R}^{2}  - v_{d}^{2}  + v_{u}^{2}\Big) - \Big(g_{R}^{2} + \tilde{g}^{2}\Big)\Big(- v_{\bar{\xi}_R}^{2}  - v_{d}^{2}  + v_{\xi_R}^{2} + v_{u}^{2}\Big)\Big)\nonumber \\ 
 &+8 {\Re\Big(m_{\nu}^2\Big)} +2 \Big(2 v_{u}^{2} {\Re\Big({Y_v  Y_{v}^{\dagger}}\Big)}  + 2 v_{\xi_R}^{2} {\Re\Big({Y_{N \nu^c}^{T}  Y_{N \nu^{c} *}}\Big)} \Big)\Big)\\ 
m_{32} &= \frac{1}{2} \frac{1}{\sqrt{2}} \Big(-2 v_{\bar{\xi}_R} {\Re\Big(\mu_{\xi} Y_{N \nu^{c},*} \Big)}  + v_{\xi_R} \Big(2 {\Re\Big(T_{N \nu^c}\Big)}  + 4 {\Re\Big({\mu_N  Y_{N \nu^{c} *}}\Big)} \Big)\Big)\\ 
m_{33} &= \frac{1}{8} \Big(4 \Big(8 {\Re\Big({\mu_N  \mu_N^*}\Big)}  + B_N\Big) + 4 B_N^*  + 8 {\Re\Big(B_N\Big)}  + 8 {\Re\Big(m_{N}^2\Big)}  + v_{\xi_R}^{2} \Big(2 v_{d}^{2} {\Re\Big({\tilde{W}_{LS}^{T}  \tilde{W}_{LS}^*}\Big)}  + 4 {\Re\Big({Y_{N \nu^c}  Y_{N \nu^c}^{\dagger}}\Big)} \Big)\Big)
\end{align} 
This matrix is diagonalized by \(Z^R\): 
\begin{equation} 
Z^R m^2_{\nu^R} Z^{R,\dagger} = m^{dia}_{2,\nu^R} 
\end{equation} 

\subsubsection{Mass matrix for up squarks}
Basis: \( \left(\tilde{u}_{L,{{\alpha}}}, \tilde{u}_{R,{{\alpha}}}\right)\)
\begin{equation} 
m^2_{\tilde{u}} = \left( 
\begin{array}{cc}
m_{11} &\frac{1}{\sqrt{2}} \Big(- v_d Y_u^\dagger \mu  + v_u T^\dagger_u \Big)\\ 
\frac{1}{\sqrt{2}} \Big(- v_d \mu^* Y_{u}  + v_u T_{u} \Big)  &m_{22}\end{array} 
\right) 
\end{equation} 
\begin{align} 
m_{11} &= \frac{1}{24}  \Big({\bf 1} \Big(3 g_{L}^{2} \Big(v_{d}^{2} - v_{u}^{2} \Big) + g_{BL}^{2} \Big(v_{\bar{X}_R}^{2}- v_{X_R}^{2} \Big) + g_{BL} \tilde{g} \Big(- v_{\bar{X}_R}^{2}  - v_{d}^{2}  + v_{X_R}^{2} + v_{u}^{2}\Big)\Big)+24 m_q^{2} \nonumber \\ 
 &+12 v_{u}^{2} {Y^{\dagger}_u  Y_{u}} \Big)\\ 
m_{22} &= \frac{1}{24}  \Big({\bf 1} \Big(3 \Big(g_{R}^{2} + \tilde{g}^{2}\Big)\Big(- v_{X_R}^{2}  - v_{u}^{2}  + v_{\bar{X}_R}^{2} + v_{d}^{2}\Big) + g_{BL}^{2} \Big(v_{X_R}^{2} - v_{\bar{X}_R}^{2} \Big) + g_{BL} \tilde{g} \Big(-2 v_{\bar{X}_R}^{2}  + 2 v_{X_R}^{2}  - v_{u}^{2}  + v_{d}^{2}\Big)\Big)\nonumber \\ 
 &+24 m_u^{2} +12 v_{u}^{2} {Y_{u}  Y^{\dagger}_u} \Big)
\end{align} 
This matrix is diagonalized by \(Z^U\): 
\begin{equation} 
Z^U m^2_{\tilde{u}} Z^{U,\dagger} = m^{dia}_{2,\tilde{u}} 
\end{equation} 

\subsubsection{Mass matrix charged sleptons}
Basis: \( \left(\tilde{e}_L, \tilde{e}_R\right)\)
\begin{equation} 
m^2_{\tilde{e}} = \left( 
\begin{array}{cc}
m_{11} &\frac{1}{\sqrt{2}} \Big(v_d T^{\dagger}_e  - v_u Y^{\dagger}_e \mu \Big)\\ 
\frac{1}{\sqrt{2}} \Big(v_d T_{e}  - v_u \mu^* Y_{e} \Big) &m_{22}\end{array} 
\right) 
\end{equation} 
\begin{align} 
m_{11} &= \frac{1}{8} \Big(4 v_{d}^{2} Y^{\dagger}_e  Y_{e}  + 8 m_l^{2}  + {\bf 1} \Big(g_{BL}^{2} \Big(v_{X_R}^{2} - v_{\bar{X}_R}^{2} \Big) + g_{BL} \tilde{g} \Big(- v_{X_R}^{2}  - v_{u}^{2}  + v_{\bar{X}_R}^{2} + v_{d}^{2}\Big) + g_{L}^{2} \Big(v_{u}^{2} -  v_{d}^{2}\Big)\Big)\Big)\\ 
m_{22} &= \frac{1}{8} \Big({\bf 1} \Big(g_{BL}^{2} \Big(v_{\bar{X}_R}^{2}- v_{X_R}^{2} \Big) + g_{BL} \tilde{g} \Big(v_{u}^{2} -  v_{d}^{2}\Big) + \Big(g_{R}^{2} + \tilde{g}^{2}\Big)\Big(- v_{\bar{X}_R}^{2}  - v_{d}^{2}  + v_{X_R}^{2} + v_{u}^{2}\Big)\Big)+8 m_e^{2} \nonumber \\ 
 &+4 v_{d}^{2} Y_{e}  Y^{\dagger}_e \Big)
\end{align} 
This matrix is diagonalized by \(Z^E\): 
\begin{equation} 
Z^E m^2_{\tilde{e}} Z^{E,\dagger} = m^{dia}_{2,\tilde{e}} 
\end{equation} 

\subsubsection{Mass matrix for scalar Higgs}
\paragraph{minimal seesaw}
Basis: \( \left(\sigma_{d}, \sigma_{u}, \sigma_{\delta}, {\sigma}_{\bar{\delta}}\right)\)
\begin{equation} 
m^2_{h} = \left( 
\begin{array}{cccc}
m_{11} &m_{21} &m_{31} &m_{41}\\ 
m_{21} &m_{22} &m_{32} &m_{42}\\ 
m_{31} &m_{32} &m_{33} &m_{43}\\ 
m_{41} &m_{42} &m_{43} &m_{44}\end{array} 
\right) 
\end{equation} 
\begin{align} 
m_{11} &= \frac{1}{8} \Big(8 m_{H_d}^2 +g_{L}^{2} \Big(3 v_{d}^{2}  - v_{u}^{2} \Big)- g_{R}^{2} \Big(-2 v_{\bar{\delta}_R}^{2}  + 2 v_{\delta_R}^{2}  -3 v_{d}^{2}  + v_{u}^{2}\Big)\nonumber \\ 
 &+\tilde{g} \Big(2 g_{BL} \Big(- v_{\bar{\delta}_R}^{2}  + v_{\delta_R}^{2}\Big) - \tilde{g} \Big(-2 v_{\bar{\delta}_R}^{2}  + 2 v_{\delta_R}^{2}  -3 v_{d}^{2}  + v_{u}^{2}\Big)\Big)+8 |\mu|^2 \Big)\\ 
m_{21} &= -\frac{1}{4} \Big(g_{L}^{2} + g_{R}^{2} + \tilde{g}^{2}\Big)v_d v_u  - {\Re\Big(B_{\mu}\Big)} \\ 
m_{22} &= \frac{1}{8} \Big(8 m_{H_u}^2 - g_{L}^{2} \Big(-3 v_{u}^{2}  + v_{d}^{2}\Big)+g_{R}^{2} \Big(-2 v_{\bar{\delta}_R}^{2}  + 2 v_{\delta_R}^{2}  + 3 v_{u}^{2}  - v_{d}^{2} \Big)\nonumber \\ 
 &+\tilde{g} \Big(-2 g_{BL} \Big(- v_{\bar{\delta}_R}^{2}  + v_{\delta_R}^{2}\Big) + \tilde{g} \Big(-2 v_{\bar{\delta}_R}^{2}  + 2 v_{\delta_R}^{2}  + 3 v_{u}^{2}  - v_{d}^{2} \Big)\Big)+8 |\mu|^2 \Big)\\ 
m_{31} &= -\frac{1}{2} \Big(\tilde{g} \Big(- g_{BL}  + \tilde{g}\Big) + g_{R}^{2}\Big)v_{\delta_R} v_d \\ 
m_{32} &= \frac{1}{2} \Big(\tilde{g} \Big(- g_{BL}  + \tilde{g}\Big) + g_{R}^{2}\Big)v_{\delta_R} v_u \\ 
m_{33} &= \frac{1}{4} \Big(4 m_{\delta}^2 +g_{BL}^{2} \Big(-2 v_{\bar{\delta}_R}^{2}  + 6 v_{\delta_R}^{2} \Big)+g_{BL} \tilde{g} \Big(-12 v_{\delta_R}^{2}  + 4 v_{\bar{\delta}_R}^{2}  - v_{u}^{2}  + v_{d}^{2}\Big)\nonumber \\ 
 &+\Big(g_{R}^{2} + \tilde{g}^{2}\Big)\Big(-2 v_{\bar{\delta}_R}^{2}  + 6 v_{\delta_R}^{2}  - v_{d}^{2}  + v_{u}^{2}\Big)+4 |\mu_{\delta}|^2 \Big)\\ 
m_{41} &= \frac{1}{2} \Big(\tilde{g} \Big(- g_{BL}  + \tilde{g}\Big) + g_{R}^{2}\Big)v_{\bar{\delta}_R} v_d \\ 
m_{42} &= -\frac{1}{2} \Big(\tilde{g} \Big(- g_{BL}  + \tilde{g}\Big) + g_{R}^{2}\Big)v_{\bar{\delta}_R} v_u \\ 
m_{43} &= - \Big(g_{R}^{2} + \Big(- \tilde{g}  + g_{BL}\Big)^{2}\Big)v_{\delta_R} v_{\bar{\delta}_R}  - {\Re\Big(B_{\delta}\Big)} \\ 
m_{44} &= \frac{1}{4} \Big(4 m_{\bar{\delta}}^2 -2 g_{BL}^{2} \Big(-3 v_{\bar{\delta}_R}^{2}  + v_{\delta_R}^{2}\Big)+g_{BL} \tilde{g} \Big(-12 v_{\bar{\delta}_R}^{2}  + 4 v_{\delta_R}^{2}  - v_{d}^{2}  + v_{u}^{2}\Big)\nonumber \\ 
 &- \Big(g_{R}^{2} + \tilde{g}^{2}\Big)\Big(2 v_{\delta_R}^{2}  -6 v_{\bar{\delta}_R}^{2}  - v_{d}^{2}  + v_{u}^{2}\Big)+4 |\mu_{\delta}|^2 \Big)
\end{align} 
\paragraph{linear and inverse seesaw}
Basis: \( \left(\sigma_{d}, \sigma_{u}, \sigma_{\bar{\xi}}, {\sigma}_{\xi}\right)\)
\begin{equation} 
m^2_{h} = \left( 
\begin{array}{cccc}
m_{11} &m_{21} &m_{31} &m_{41}\\ 
m_{21} &m_{22} &m_{32} &m_{42}\\ 
m_{31} &m_{32} &m_{33} &m_{43}\\ 
m_{41} &m_{42} &m_{43} &m_{44}\end{array} 
\right) 
\end{equation} 
\begin{align} 
m_{11} &= \frac{1}{8} \Big(8 m_{H_d}^2 +g_{L}^{2} \Big(3 v_{d}^{2}  - v_{u}^{2} \Big)- g_{R}^{2} \Big(-3 v_{d}^{2}  - v_{\bar{\xi}_R}^{2}  + v_{\xi_R}^{2} + v_{u}^{2}\Big)\nonumber \\ 
 &+\tilde{g} \Big(g_{BL} \Big(- v_{\bar{\xi}_R}^{2}  + v_{\xi_R}^{2}\Big) - \tilde{g} \Big(-3 v_{d}^{2}  - v_{\bar{\xi}_R}^{2}  + v_{\xi_R}^{2} + v_{u}^{2}\Big)\Big)+8 |\mu|^2 \Big)\\ 
m_{21} &= -\frac{1}{4} \Big(g_{L}^{2} + g_{R}^{2} + \tilde{g}^{2}\Big)v_d v_u  - {\Re\Big(B_{\mu}\Big)} \\ 
m_{22} &= \frac{1}{8} \Big(8 m_{H_u}^2 - g_{L}^{2} \Big(-3 v_{u}^{2}  + v_{d}^{2}\Big)+g_{R}^{2} \Big(3 v_{u}^{2}  - v_{\bar{\xi}_R}^{2}  - v_{d}^{2}  + v_{\xi_R}^{2}\Big)\nonumber \\ 
 &+\tilde{g} \Big(g_{BL} \Big(- v_{\xi_R}^{2}  + v_{\bar{\xi}_R}^{2}\Big) + \tilde{g} \Big(3 v_{u}^{2}  - v_{\bar{\xi}_R}^{2}  - v_{d}^{2}  + v_{\xi_R}^{2}\Big)\Big)+8 |\mu|^2 \Big)\\ 
m_{31} &= -\frac{1}{4} \Big(\tilde{g} \Big(- g_{BL}  + \tilde{g}\Big) + g_{R}^{2}\Big)v_{\xi_R} v_d \\ 
m_{32} &= \frac{1}{4} \Big(\tilde{g} \Big(- g_{BL}  + \tilde{g}\Big) + g_{R}^{2}\Big)v_{\xi_R} v_u \\ 
m_{33} &= \frac{1}{8} \Big(8 m_{\xi}^2 +g_{BL}^{2} \Big(3 v_{\xi_R}^{2}  - v_{\bar{\xi}_R}^{2} \Big)+g_{BL} \tilde{g} \Big(2 v_{\bar{\xi}_R}^{2}  -6 v_{\xi_R}^{2}  - v_{u}^{2}  + v_{d}^{2}\Big)\nonumber \\ 
 &+\Big(g_{R}^{2} + \tilde{g}^{2}\Big)\Big(3 v_{\xi_R}^{2}  - v_{\bar{\xi}_R}^{2}  - v_{d}^{2}  + v_{u}^{2}\Big)+8 |\mu_{\xi}|^2 \Big)\\ 
m_{41} &= \frac{1}{4} \Big(\tilde{g} \Big(- g_{BL}  + \tilde{g}\Big) + g_{R}^{2}\Big)v_{\bar{\xi}_R} v_d \\ 
m_{42} &= -\frac{1}{4} \Big(\tilde{g} \Big(- g_{BL}  + \tilde{g}\Big) + g_{R}^{2}\Big)v_{\bar{\xi}_R} v_u \\ 
m_{43} &= \frac{1}{4} \Big(-4 {\Re\Big(B_{\xi}\Big)}  - \Big(g_{R}^{2} + \Big(- \tilde{g}  + g_{BL}\Big)^{2}\Big)v_{\xi_R} v_{\bar{\xi}_R} \Big)\\ 
m_{44} &= \frac{1}{8} \Big(8 m_{\bar{\xi}}^2 - g_{BL}^{2} \Big(-3 v_{\bar{\xi}_R}^{2}  + v_{\xi_R}^{2}\Big)+g_{BL} \tilde{g} \Big(2 v_{\xi_R}^{2}  -6 v_{\bar{\xi}_R}^{2}  - v_{d}^{2}  + v_{u}^{2}\Big)\nonumber \\ 
 &- \Big(g_{R}^{2} + \tilde{g}^{2}\Big)\Big(-3 v_{\bar{\xi}_R}^{2}  - v_{d}^{2}  + v_{\xi_R}^{2} + v_{u}^{2}\Big)+8 |\mu_{\xi}|^2 \Big)
\end{align} 
This matrix is diagonalized by \(Z^H\): 
\begin{equation} 
Z^H m^2_{h} Z^{H,\dagger} = m^{dia}_{2,h} 
\end{equation}

\subsubsection{Mass matrix for pseudo scalar Higgs}
\paragraph{minimal seesaw}
Basis: \( \left(\phi_{d}, \phi_{u}, \phi_{\delta}, \phi_{\bar{\delta}}\right)\) 
\begin{equation} 
m^2_{A_0} = \left( 
\begin{array}{cccc}
m_{11} &{\Re\Big(B_{\mu}\Big)} &0 &0\\ 
{\Re\Big(B_{\mu}\Big)} &m_{22} &0 &0\\ 
0 &0 &m_{33} &{\Re\Big(B_{\delta}\Big)}\\ 
0 &0 &{\Re\Big(B_{\delta}\Big)} &m_{44}\end{array} 
\right) 
\end{equation} 
\begin{align} 
m_{11} &= \frac{1}{8} \Big(8 m_{H_d}^2 +g_{L}^{2} \Big(- v_{u}^{2}  + v_{d}^{2}\Big)+g_{R}^{2} \Big(2 v_{\bar{\delta}_R}^{2}  -2 v_{\delta_R}^{2}  - v_{u}^{2}  + v_{d}^{2}\Big)\nonumber \\ 
 &+\tilde{g} \Big(2 g_{BL} \Big(- v_{\bar{\delta}_R}^{2}  + v_{\delta_R}^{2}\Big) + \tilde{g} \Big(2 v_{\bar{\delta}_R}^{2}  -2 v_{\delta_R}^{2}  - v_{u}^{2}  + v_{d}^{2}\Big)\Big)+8 |\mu|^2 \Big)\\ 
m_{22} &= \frac{1}{8} \Big(8 m_{H_u}^2 +g_{L}^{2} \Big(- v_{d}^{2}  + v_{u}^{2}\Big)+g_{R}^{2} \Big(-2 v_{\bar{\delta}_R}^{2}  + 2 v_{\delta_R}^{2}  - v_{d}^{2}  + v_{u}^{2}\Big)\nonumber \\ 
 &+\tilde{g} \Big(-2 g_{BL} \Big(- v_{\bar{\delta}_R}^{2}  + v_{\delta_R}^{2}\Big) + \tilde{g} \Big(-2 v_{\bar{\delta}_R}^{2}  + 2 v_{\delta_R}^{2}  - v_{d}^{2}  + v_{u}^{2}\Big)\Big)+8 |\mu|^2 \Big)\\ 
m_{33} &= \frac{1}{4} \Big(4 m_{\delta}^2 +2 g_{BL}^{2} \Big(- v_{\bar{\delta}_R}^{2}  + v_{\delta_R}^{2}\Big)+g_{BL} \tilde{g} \Big(4 v_{\bar{\delta}_R}^{2}  -4 v_{\delta_R}^{2}  - v_{u}^{2}  + v_{d}^{2}\Big)+\Big(g_{R}^{2} + \tilde{g}^{2}\Big)\Big(-2 v_{\bar{\delta}_R}^{2}  + 2 v_{\delta_R}^{2}  - v_{d}^{2}  + v_{u}^{2}\Big)\nonumber \\ 
 &+4 |\mu_{\delta}|^2 \Big)\\ 
m_{44} &= \frac{1}{4} \Big(4 m_{\bar{\delta}}^2 -2 g_{BL}^{2} \Big(- v_{\bar{\delta}_R}^{2}  + v_{\delta_R}^{2}\Big)+g_{BL} \tilde{g} \Big(-4 v_{\bar{\delta}_R}^{2}  + 4 v_{\delta_R}^{2}  - v_{d}^{2}  + v_{u}^{2}\Big)\nonumber \\ 
 &- \Big(g_{R}^{2} + \tilde{g}^{2}\Big)\Big(-2 v_{\bar{\delta}_R}^{2}  + 2 v_{\delta_R}^{2}  - v_{d}^{2}  + v_{u}^{2}\Big)+4 |\mu_{\delta}|^2 \Big)
\end{align} 
\paragraph{linear and inverse seesaw}
Basis: \( \left(\phi_{d}, \phi_{u}, \phi_{\bar{\xi}}, \phi_{\xi}\right)\) 
\begin{equation} 
m^2_{A_0} = \left( 
\begin{array}{cccc}
m_{11} &{\Re\Big(B_{\mu}\Big)} &0 &0\\ 
{\Re\Big(B_{\mu}\Big)} &m_{22} &0 &0\\ 
0 &0 &m_{33} &{\Re\Big(B_{\xi}\Big)}\\ 
0 &0 &{\Re\Big(B_{\xi}\Big)} &m_{44}\end{array} 
\right) 
\end{equation} 
\begin{align} 
m_{11} &= \frac{1}{8} \Big(8 m_{H_d}^2 +g_{L}^{2} \Big(- v_{u}^{2}  + v_{d}^{2}\Big)+g_{R}^{2} \Big(- v_{u}^{2}  - v_{\xi_R}^{2}  + v_{\bar{\xi}_R}^{2} + v_{d}^{2}\Big)\nonumber \\ 
 &+\tilde{g} \Big(g_{BL} \Big(- v_{\bar{\xi}_R}^{2}  + v_{\xi_R}^{2}\Big) + \tilde{g} \Big(- v_{u}^{2}  - v_{\xi_R}^{2}  + v_{\bar{\xi}_R}^{2} + v_{d}^{2}\Big)\Big)+8 |\mu|^2 \Big)\\ 
m_{22} &= \frac{1}{8} \Big(8 m_{H_u}^2 +g_{L}^{2} \Big(- v_{d}^{2}  + v_{u}^{2}\Big)+g_{R}^{2} \Big(- v_{\bar{\xi}_R}^{2}  - v_{d}^{2}  + v_{\xi_R}^{2} + v_{u}^{2}\Big)\nonumber \\ 
 &+\tilde{g} \Big(g_{BL} \Big(- v_{\xi_R}^{2}  + v_{\bar{\xi}_R}^{2}\Big) + \tilde{g} \Big(- v_{\bar{\xi}_R}^{2}  - v_{d}^{2}  + v_{\xi_R}^{2} + v_{u}^{2}\Big)\Big)+8 |\mu|^2 \Big)\\ 
m_{33} &= \frac{1}{8} \Big(8 m_{\xi}^2 +g_{BL}^{2} \Big(- v_{\bar{\xi}_R}^{2}  + v_{\xi_R}^{2}\Big)+g_{BL} \tilde{g} \Big(2 v_{\bar{\xi}_R}^{2}  -2 v_{\xi_R}^{2}  - v_{u}^{2}  + v_{d}^{2}\Big)+\Big(g_{R}^{2} + \tilde{g}^{2}\Big)\Big(- v_{\bar{\xi}_R}^{2}  - v_{d}^{2}  + v_{\xi_R}^{2} + v_{u}^{2}\Big)\nonumber \\ 
 &+8 |\mu_{\xi}|^2 \Big)\\ 
m_{44} &= \frac{1}{8} \Big(8 m_{\bar{\xi}}^2 +g_{BL}^{2} \Big(- v_{\xi_R}^{2}  + v_{\bar{\xi}_R}^{2}\Big)+g_{BL} \tilde{g} \Big(-2 v_{\bar{\xi}_R}^{2}  + 2 v_{\xi_R}^{2}  - v_{d}^{2}  + v_{u}^{2}\Big)- \Big(g_{R}^{2} + \tilde{g}^{2}\Big)\Big(- v_{\bar{\xi}_R}^{2}  - v_{d}^{2}  + v_{\xi_R}^{2} + v_{u}^{2}\Big)\nonumber \\ 
 &+8 |\mu_{\xi}|^2 \Big)
\end{align} 
This matrix is diagonalized by \(Z^A\): 
\begin{equation} 
Z^A m^2_{A_0} Z^{A,\dagger} = m^{dia}_{2,A_0} 
\end{equation} 
Using the solution of the tadpole equations the mass matrix can also be written as
\begin{equation}
m^2_{A^0} = \left(\begin{array}{cccc}
B_\mu \tan\beta & B_\mu & 0 & 0 \\
B_\mu  & B_\mu \cot\beta & 0 & 0 \\
 0 & 0  &B_{X} \tan\beta_R & B_{X} \\
 0 & 0  &B_{X} & B_{X} \cot\beta_R 
\end{array}
\right) \,.
\end{equation}
with $X=\xi,\delta$. This leads to following the masses of the physical states:
\begin{equation}
m^2_{A^0} = \frac{2 B_\mu}{\sin2\beta} \thickspace, \hspace{1cm} 
m^2_{A^0_X}  = \frac{2 B_{X}}{\sin2\beta_R} \thickspace.
\end{equation}

\subsubsection{Mass matrix for charged Higgs}
Basis: \( \left(H_d^-, H_u^{+,*}\right)\)
\begin{equation} 
m^2_{H^-} = \left( 
\begin{array}{cc}
m_{11} &\frac{1}{4} g_{L}^{2} v_d v_u  + B^*_{\mu}\\ 
\frac{1}{4} g_{L}^{2} v_d v_u  + B_{\mu} &m_{22}\end{array} 
\right) 
\end{equation} 
\begin{align} 
m_{11} &= \frac{1}{8} \Big(8 m_{H_d}^2 +g_{R}^{2} \Big(- v_{X_R}^{2}  - v_{u}^{2}  + v_{\bar{X}_R}^{2} + v_{d}^{2}\Big)+g_{L}^{2} \Big(v_{d}^{2} + v_{u}^{2}\Big)\nonumber \\ 
 &+\tilde{g} \Big(g_{BL} \Big(v_{X_R}^{2} - v_{\bar{X}_R}^{2} \Big) + \tilde{g} \Big(- v_{X_R}^{2}  - v_{u}^{2}  + v_{\bar{X}_R}^{2} + v_{d}^{2}\Big)\Big)+8 |\mu|^2 \Big)\\ 
m_{22} &= \frac{1}{8} \Big(8 m_{H_u}^2 +g_{R}^{2} \Big(- v_{\bar{X}_R}^{2}  - v_{d}^{2}  + v_{X_R}^{2} + v_{u}^{2}\Big)+g_{L}^{2} \Big(v_{d}^{2} + v_{u}^{2}\Big)\nonumber \\ 
 &+\tilde{g} \Big(g_{BL} \Big(v_{\bar{X}_R}^{2}- v_{X_R}^{2} \Big) + \tilde{g} \Big(- v_{\bar{X}_R}^{2}  - v_{d}^{2}  + v_{X_R}^{2} + v_{u}^{2}\Big)\Big)+8 |\mu|^2 \Big)
\end{align} 
This matrix is diagonalized by \(Z^+\): 
\begin{equation} 
Z^+ m^2_{H^-} Z^{+,\dagger} = m^{dia}_{2,H^-} 
\end{equation} 

\subsubsection{Mass matrix for neutralinos}
\paragraph{minimal seesaw}
Basis: \( \left(\lambda_{\tilde{B}}, \tilde{W}^0, \tilde{H}_d^0, \tilde{H}_u^0, \lambda_{R}, \tilde{\delta}, \tilde{\bar{\delta}}\right)\)
\begin{equation} 
m_{\tilde{\chi}^0} = \left( 
\begin{array}{ccccccc}
M_1 &0 &-\frac{1}{2} \tilde{g} v_d  &\frac{1}{2} \tilde{g} v_u  &\frac{1}{2} M_{B R}  &m_{61} &m_{71}\\ 
0 &M_2 &\frac{1}{2} g_L v_d  &-\frac{1}{2} g_L v_u  &0 &0 &0\\ 
-\frac{1}{2} \tilde{g} v_d  &\frac{1}{2} g_L v_d  &0 &- \mu  &-\frac{1}{2} g_R v_d  &0 &0\\ 
\frac{1}{2} \tilde{g} v_u  &-\frac{1}{2} g_L v_u  &- \mu  &0 &\frac{1}{2} g_R v_u  &0 &0\\ 
\frac{1}{2} M_{B R}  &0 &-\frac{1}{2} g_R v_d  &\frac{1}{2} g_R v_u  &M_4 &g_R v_{\delta_R}  &- g_R v_{\bar{\delta}_R} \\ 
m_{61} &0 &0 &0 &g_R v_{\delta_R}  &0 &- \mu_{\delta} \\ 
m_{71} &0 &0 &0 &- g_R v_{\bar{\delta}_R}  &- \mu_{\delta}  &0\end{array} 
\right) 
\end{equation} 
\begin{align} 
m_{61} &= \Big(- g_{BL}  + \tilde{g}\Big)v_{\delta_R} \\ 
m_{71} &= \Big(- \tilde{g}  + g_{BL}\Big)v_{\bar{\delta}_R} 
\end{align} 

\paragraph{linear and inverse seesaw}
Basis: \( \left(\lambda_{\tilde{B}}, \tilde{W}^0, \tilde{H}_d^0, \tilde{H}_u^0, \lambda_{R}, \tilde{\xi}, \tilde{\bar{\xi}}\right)\)
\begin{equation} 
m_{\tilde{\chi}^0} = \left( 
\begin{array}{ccccccc}
M_1 &0 &-\frac{1}{2} \tilde{g} v_d  &\frac{1}{2} \tilde{g} v_u  &\frac{1}{2} M_{B R}  &m_{61} &m_{71}\\ 
0 &M_2 &\frac{1}{2} g_L v_d  &-\frac{1}{2} g_L v_u  &0 &0 &0\\ 
-\frac{1}{2} \tilde{g} v_d  &\frac{1}{2} g_L v_d  &0 &- \mu  &-\frac{1}{2} g_R v_d  &0 &0\\ 
\frac{1}{2} \tilde{g} v_u  &-\frac{1}{2} g_L v_u  &- \mu  &0 &\frac{1}{2} g_R v_u  &0 &0\\ 
\frac{1}{2} M_{B R}  &0 &-\frac{1}{2} g_R v_d  &\frac{1}{2} g_R v_u  &M_4 &\frac{1}{2} g_R v_{\xi_R}  &-\frac{1}{2} g_R v_{\bar{\xi}_R} \\ 
m_{61} &0 &0 &0 &\frac{1}{2} g_R v_{\xi_R}  &0 &- \mu_{\xi} \\ 
m_{71} &0 &0 &0 &-\frac{1}{2} g_R v_{\bar{\xi}_R}  &- \mu_{\xi}  &0\end{array} 
\right) 
\end{equation} 
with
\begin{align} 
m_{61} &= \frac{1}{2} \Big(- g_{BL}  + \tilde{g}\Big)v_{\xi_R} \\ 
m_{71} &= \frac{1}{2} \Big(- \tilde{g}  + g_{BL}\Big)v_{\bar{\xi}_R} 
\end{align} 
This matrix is diagonalized by \(N\): 
\begin{equation} 
N^* m_{\tilde{\chi}^0} N^{\dagger} = m^{dia}_{\tilde{\chi}^0} 
\end{equation}

\subsubsection{Mass matrix for neutrinos}
\paragraph{minimal seesaw}
Basis: \( \left(\nu_L, \nu^c\right)\) 
\begin{equation} 
m_{\nu} = \left( 
\begin{array}{cc}
0 &\frac{1}{\sqrt{2}} v_u Y_{v}^{T}  \\ 
\frac{1}{\sqrt{2}} v_u Y_v  &\frac{2}{\sqrt{2}} v_{\delta_R} Y_{M} \end{array} 
\right) 
\end{equation} 

\paragraph{linear and inverse seesaw}
Basis: \( \left(\nu_L, \nu^c, N\right)\) 
\begin{equation} 
m_{\nu} = \left( 
\begin{array}{ccc}
0 &\frac{1}{\sqrt{2}} v_u Y_{v}^{T}  & \tilde{Y} v_d v_{\xi_R} \\ 
\frac{1}{\sqrt{2}} v_u Y_v  &0 &\frac{1}{\sqrt{2}} v_{\xi_R} Y_{N \nu^c}^{T} \\ 
\tilde{Y}^T v_d v_{\xi_R} &\frac{1}{\sqrt{2}} Y_{N \nu^c} v_{\xi_R}  &2 \mu_N \end{array} 
\right) 
\end{equation}
$\tilde{Y}$ is the running, effective operator $\sim \hat{L} \hat{N}_S \hat{H}_d \hat{\xi}_R$ obtained by integrating out $\xi_L$ and $\bar{\xi}_L$.
The mass matrix is diagonalized by \(Z^V_{\nu}\): 
\begin{equation} 
U^{V,*} m_{\nu} U^{V,\dagger} = m^{dia}_{\nu} 
\end{equation}

\subsection{\Seesaw\ in models with \UYxUBL\ gauge sector}
We give here the mass matrices for an imaginary model which would contain all terms in the Lagrangian of the minimal \UYxUBL\ model as well of the model with inverse seesaw. The mass matrices for the physical relevant models are given in the limit that the unwanted interactions vanish.
\subsubsection{Mass matrix for down squarks}
Basis: \( \left(\tilde{d}_{L,{{\alpha}}}, \tilde{d}_{R,{{\alpha}}}\right)\)
 
\begin{equation} 
m^2_{\tilde{d}} = \left( 
\begin{array}{cc}
m_{11} &\frac{1}{\sqrt{2}} \Big(v_d T_{d}^{\dagger}  - v_u \mu Y_{d}^{\dagger} \Big)\\ 
\frac{1}{\sqrt{2}} \Big(v_d T_d  - v_u \mu^* Y_d \Big)  &m_{22}\end{array} 
\right) 
\end{equation} 
\begin{align} 
m_{11} &= -\frac{1}{24}  \Big({\bf 1} \Big(2 \Big(g_{B}^{2} + \tilde{g}^{2}\Big)\Big(v_{\eta}^{2} -  v_{\bar{\eta}}^{2} \Big) + 3 g_{2}^{2} \Big( v_{d}^{2} -  v_{u}^{2} \Big) + g_{1}^{2} \Big( v_{d}^{2} -  v_{u}^{2} \Big) + g_1 \tilde{g} \Big(2 v_{\eta}^{2}  -2 v_{\bar{\eta}}^{2}  - v_{u}^{2}  + v_{d}^{2}\Big)\Big)\nonumber \\ 
 &-12 \Big(2 m_q^{2}  + v_{d}^{2} {Y_{d}^{\dagger}  Y_d} \Big)\Big)\\ 
m_{22} &= \frac{1}{24}  \Big(12 v_{d}^{2} {Y_d  Y_{d}^{\dagger}}  + 24 m_d^{2,*}  + {\bf 1} \Big(- \Big(2 g_1  - \tilde{g} \Big)\Big(2 \tilde{g} \Big(v_{\eta}^{2} -  v_{\bar{\eta}}^{2} \Big) + g_1 \Big( v_{d}^{2} -  v_{u}^{2} \Big)\Big) + 2 g_{B}^{2} \Big(v_{\eta}^{2} -  v_{\bar{\eta}}^{2} \Big)\Big)\Big)
\end{align} 
This matrix is diagonalized by \(Z^D\): 
\begin{equation} 
Z^D m^2_{\tilde{d}} Z^{D,\dagger} = m^{dia}_{2,\tilde{d}} 
\end{equation}

\subsubsection{Mass matrix for up-squarks}
Basis: \( \left(\tilde{u}_{L,{{\alpha}}}, \tilde{u}_{R,{{\alpha}}}\right)\)
 
\begin{equation} 
m^2_{\tilde{u}} = \left( 
\begin{array}{cc}
m_{11} &\frac{1}{\sqrt{2}} \Big(- v_d \mu Y_{u}^{\dagger}  + v_u T_{u}^{\dagger} \Big)\\ 
\frac{1}{\sqrt{2}} \Big(- v_d \mu^* Y_u  + v_u T_u \Big)  &m_{22}\end{array} 
\right) 
\end{equation} 
\begin{align} 
m_{11} &= \frac{1}{24}  \Big({\bf 1} \Big(-2 \Big(g_{B}^{2} + \tilde{g}^{2}\Big)\Big(v_{\eta}^{2} -  v_{\bar{\eta}}^{2} \Big) + 3 g_{2}^{2} \Big( v_{d}^{2} -  v_{u}^{2} \Big) + g_{1}^{2} \Big( v_{u}^{2} -  v_{d}^{2} \Big) + g_1 \tilde{g} \Big(-2 v_{\eta}^{2}  + 2 v_{\bar{\eta}}^{2}  - v_{d}^{2}  + v_{u}^{2}\Big)\Big)\nonumber \\ 
 &+24 m_q^{2} +12 v_{u}^{2} {Y_{u}^{\dagger}  Y_u} \Big)\\ 
m_{22} &= \frac{1}{24} \Big(12 v_{u}^{2} {Y_u  Y_{u}^{\dagger}}  + 24 m_u^{2}  + {\bf 1} \Big(2 g_{B}^{2} \Big(v_{\eta}^{2} -  v_{\bar{\eta}}^{2} \Big) + \Big(4 g_1  + \tilde{g}\Big)\Big(2 \tilde{g} \Big(v_{\eta}^{2} -  v_{\bar{\eta}}^{2} \Big) + g_1 \Big( v_{d}^{2} -  v_{u}^{2} \Big)\Big)\Big)\Big)
\end{align} 
This matrix is diagonalized by \(Z^U\): 
\begin{equation} 
Z^U m^2_{\tilde{u}} Z^{U,\dagger} = m^{dia}_{2,\tilde{u}} 
\end{equation} 
with 

\subsubsection{Mass matrix for charged sleptons}
Basis: \( \left(\tilde{e}_L, \tilde{e}_R\right)\)
 
\begin{equation} 
m^2_{\tilde{e}} = \left( 
\begin{array}{cc}
m_{11} &\frac{1}{\sqrt{2}} \Big(v_d T_{e}^{\dagger}  - v_u \mu Y_{e}^{\dagger} \Big)\\ 
\frac{1}{\sqrt{2}} \Big(v_d T_e  - v_u \mu^* Y_e \Big) &m_{22}\end{array} 
\right) 
\end{equation} 
\begin{align} 
m_{11} &= \frac{1}{8} \Big({\bf 1} \Big(2 \Big(g_{B}^{2} + \tilde{g}^{2}\Big)\Big(v_{\eta}^{2} -  v_{\bar{\eta}}^{2} \Big) + g_{1}^{2} \Big( v_{d}^{2} -  v_{u}^{2} \Big) + g_1 \tilde{g} \Big(2 v_{\eta}^{2}  -2 v_{\bar{\eta}}^{2}  - v_{u}^{2}  + v_{d}^{2}\Big) + g_{2}^{2} \Big( v_{u}^{2} -  v_{d}^{2} \Big)\Big)\nonumber \\ 
 &+8 m_l^{2} +4 v_{d}^{2} {Y_{e}^{\dagger}  Y_e} \Big)\\ 
m_{22} &= \frac{1}{8} \Big(4 v_{d}^{2} {Y_e  Y_{e}^{\dagger}}  + 8 m_e^{2}  - {\bf 1} \Big(\Big(2 g_1  + \tilde{g}\Big)\Big(2 \tilde{g} \Big(v_{\eta}^{2} -  v_{\bar{\eta}}^{2} \Big) + g_1 \Big( v_{d}^{2} -  v_{u}^{2} \Big)\Big) + 2 g_{B}^{2} \Big(v_{\eta}^{2} -  v_{\bar{\eta}}^{2} \Big)\Big)\Big)
\end{align} 
This matrix is diagonalized by \(Z^E\): 
\begin{equation} 
Z^E m^2_{\tilde{e}} Z^{E,\dagger} = m^{dia}_{2,\tilde{e}} 
\end{equation} 

\subsubsection{Mass matrix for pseudo scalar sneutrinos}
\( \left(\phi_L, \phi_R, \phi_S\right)\)
\begin{equation} 
m^2_{\nu^i} = \left( 
\begin{array}{ccc}
m_{11} &m^T_{21} &\frac{1}{2} v_u v_{\eta} {\Re\Big({Y^\dagger_\nu  Y_{IS}}\Big)} \\ 
m_{21} &m_{22} &m^T_{32}\\ 
\frac{1}{2} v_u v_{\eta} {\Re\Big({Y_{IS}  Y_{\nu}^{*}}\Big)}  &m_{32} &m_{33}\end{array} 
\right) 
\end{equation} 
\begin{align} 
m_{11} &= \frac{1}{8} \Big({\bf 1} \Big(2 \Big(g_{B}^{2} + \tilde{g}^{2}\Big)\Big(- v_{\bar{\eta}}^{2}  + v_{\eta}^{2}\Big) + g_{1}^{2} \Big(- v_{u}^{2}  + v_{d}^{2}\Big) + g_1 \tilde{g} \Big(-2 v_{\bar{\eta}}^{2}  + 2 v_{\eta}^{2}  - v_{u}^{2}  + v_{d}^{2}\Big) + g_{2}^{2} \Big(- v_{u}^{2}  + v_{d}^{2}\Big)\Big)\nonumber \\ 
 &+8 {\Re\Big(m_l^2\Big)} +4 v_{u}^{2} {\Re\Big({Y^\dagger_\nu  Y_{\nu}}\Big)} \Big)\\ 
m_{21} &= \frac{1}{4} \Big(-2 \sqrt{2} v_d {\Re\Big(\mu^* Y_{\nu}^{\dagger} \Big)}  + v_u \Big(2 \sqrt{2} {\Re\Big(T_{\nu}\Big)}  -4 v_{\eta} {\Re\Big({Y_{\eta \nu^c}  Y_{\nu}^{*}}\Big)} \Big)\Big)\\ 
m_{22} &= \frac{1}{8} \Big(g_1 \tilde{g} {\bf 1} \Big(- v_{d}^{2}  + v_{u}^{2}\Big)+8 {\Re\Big({m_\nu^2}\Big)} +2 v_{\bar{\eta}} \Big(4 \sqrt{2} {\Re\Big(Y_{\eta \nu^c} \mu_{\eta}^* \Big)}  + \Big(g_{B}^{2} + \tilde{g}^{2}\Big){\bf 1} v_{\bar{\eta}} \Big)+4 v_{u}^{2} {\Re\Big({Y_{\nu}  Y_\nu^\dagger}\Big)} \nonumber \\ 
 &+2 v_{\eta} \Big(-4 \sqrt{2} {\Re\Big(T_{\eta \nu^c}\Big)}  + v_{\eta} \Big(2 {\Re\Big({Y_{IS}  Y_{IS}^*}\Big)}  + 8 {\Re\Big({Y_{\eta \nu^c}  Y^*_{\eta \nu^{c}}}\Big)}  - \Big(g_{B}^{2} + \tilde{g}^{2}\Big){\bf 1} \Big)\Big)\Big)\\ 
m_{32} &= \frac{1}{4} \Big(-2 \sqrt{2} v_{\bar{\eta}} {\Re\Big(Y_{IS} \mu_{\eta}^* \Big)}  + v_{\eta} \Big(2 \sqrt{2} {\Re\Big(T_{IS}\Big)}  -4 \sqrt{2} {\Re\Big({\mu_N  Y_{IS}^*}\Big)}  -4 v_{\eta} {\Re\Big({Y_{IS}  Y^*_{\eta \nu^{c}}}\Big)} \Big)\Big)\\ 
m_{33} &= \frac{1}{8} \Big({\bf 1} \Big(-2 \Big(g_{B}^{2} + \tilde{g}^{2}\Big)\Big(- v_{\bar{\eta}}^{2}  + v_{\eta}^{2}\Big) + g_1 \tilde{g} \Big(- v_{d}^{2}  + v_{u}^{2}\Big)\Big) \nonumber \\ 
 &+4 \Big(-2 {\Re\Big(B_N\Big)}  + 2 {\Re\Big(m_{N}^2\Big)}  + 8 {\Re\Big({\mu_N  \mu_N^*}\Big)}  - 2 \Re(B_N) \Big)+4 v_{\eta}^{2} {\Re\Big({Y_{IS}  Y_{IS}^*}\Big)} \Big)
\end{align} 
This matrix is diagonalized by \(Z^i\): 
\begin{equation} 
Z^i m^2_{\nu^i} Z^{i,\dagger} = m^{dia}_{2,\nu^i} 
\end{equation} 

\subsubsection{Mass matrix for scalar sneutrinos}
Basis: \( \left(\sigma_L, \sigma_R, \sigma_S\right)\)
\begin{equation} 
m^2_{\nu^R} = \left( 
\begin{array}{ccc}
m_{11} &m^T_{21} &\frac{1}{2} v_u v_{\eta} {\Re\Big({Y^\dagger_\nu  Y_{IS}}\Big)} \\ 
m_{21} &m_{22} &m^T_{32}\\ 
\frac{1}{2} v_u v_{\eta} {\Re\Big({Y_{IS}  Y_{\nu}^{*}}\Big)}  &m_{32} &m_{33}\end{array} 
\right) 
\end{equation} 
\begin{align} 
m_{11} &= \frac{1}{8} \Big({\bf 1} \Big(2 \Big(g_{B}^{2} + \tilde{g}^{2}\Big)\Big(- v_{\bar{\eta}}^{2}  + v_{\eta}^{2}\Big) + g_{1}^{2} \Big(- v_{u}^{2}  + v_{d}^{2}\Big) + g_1 \tilde{g} \Big(-2 v_{\bar{\eta}}^{2}  + 2 v_{\eta}^{2}  - v_{u}^{2}  + v_{d}^{2}\Big) + g_{2}^{2} \Big(- v_{u}^{2}  + v_{d}^{2}\Big)\Big)\nonumber \\ 
 &+8 {\Re\Big(m_l^2\Big)} +4 v_{u}^{2} {\Re\Big({Y^\dagger_\nu  Y_{\nu}}\Big)} \Big)\\ 
m_{21} &= \frac{1}{4} \Big(-2 \sqrt{2} v_d {\Re\Big(\mu^* Y_{\nu} \Big)}  + v_u \Big(2 \sqrt{2} {\Re\Big(T_{\nu}\Big)}  + 4 v_{\eta} {\Re\Big({Y_{\eta \nu^c}  Y_{\nu}^{*}}\Big)} \Big)\Big)\\ 
m_{22} &= \frac{1}{8} \Big(g_1 \tilde{g} {\bf 1} \Big(- v_{d}^{2}  + v_{u}^{2}\Big)+8 {\Re\Big({m_\nu^2}\Big)} +2 v_{\bar{\eta}} \Big(-4 \sqrt{2} {\Re\Big(Y_{\eta \nu^c} \mu_{\eta}^* \Big)}  + \Big(g_{B}^{2} + \tilde{g}^{2}\Big){\bf 1} v_{\bar{\eta}} \Big)+4 v_{u}^{2} {\Re\Big({Y_{\nu}  Y_\nu^\dagger}\Big)} \nonumber \\ 
 &+2 v_{\eta} \Big(4 \sqrt{2} {\Re\Big(T_{\eta \nu^c}\Big)}  + v_{\eta} \Big(2 {\Re\Big({Y_{IS}  Y_{IS}^*}\Big)}  + 8 {\Re\Big({Y_{\eta \nu^c}  Y^*_{\eta \nu^{c}}}\Big)}  - \Big(g_{B}^{2} + \tilde{g}^{2}\Big){\bf 1} \Big)\Big)\Big)\\ 
m_{32} &= \frac{1}{4} \Big(-2 \sqrt{2} v_{\bar{\eta}} {\Re\Big(Y_{IS} \mu_{\eta}^* \Big)}  + v_{\eta} \Big(4 v_{\eta} {\Re\Big({Y_{IS}  Y^*_{\eta \nu^{c}}}\Big)}  + \sqrt{2} \Big(2 {\Re\Big(T_{IS}\Big)}  + 4 {\Re\Big({\mu_N  Y_{IS}^*}\Big)} \Big)\Big)\Big)\\ 
m_{33} &= \frac{1}{8} \Big(4 \Big(2 {\Re\Big(B_N\Big)}  + 2 {\Re\Big(m_{N}^2\Big)}  + 8 {\Re\Big({\mu_N  \mu_N^*}\Big)}  + 2 \Re(B_N)\Big)  + 4 v_{\eta}^{2} {\Re\Big({Y_{IS}  Y_{IS}^*}\Big)}  \nonumber \\ 
 & + {\bf 1} \Big(-2 \Big(g_{B}^{2} + \tilde{g}^{2}\Big)\Big(- v_{\bar{\eta}}^{2}  + v_{\eta}^{2}\Big) + g_1 \tilde{g} \Big(- v_{d}^{2}  + v_{u}^{2}\Big)\Big)\Big)
\end{align} 
This matrix is diagonalized by \(Z^R\): 
\begin{equation} 
 Z^R m^2_{\nu^R} Z^{R,\dagger} = m^{dia}_{2,\nu^R} 
 \end{equation} 

\subsubsection{Mass matrix for scalar Higgs}
Basis: \( \left(\phi_{d}, \phi_{u}, \phi_{\eta}, \phi_{\bar{\eta}}\right)\)
 \begin{equation} 
m^2_{h} = \left( 
\begin{array}{cccc}
m_{11} &m_{21} &\frac{1}{2} g_1 \tilde{g} v_d v_{\eta}  &-\frac{1}{2} g_1 \tilde{g} v_d v_{\bar{\eta}} \\ 
m_{21} &m_{22} &-\frac{1}{2} g_1 \tilde{g} v_u v_{\eta}  &\frac{1}{2} g_1 \tilde{g} v_u v_{\bar{\eta}} \\ 
\frac{1}{2} g_1 \tilde{g} v_d v_{\eta}  &-\frac{1}{2} g_1 \tilde{g} v_u v_{\eta}  &m_{33} &m_{43}\\ 
-\frac{1}{2} g_1 \tilde{g} v_d v_{\bar{\eta}}  &\frac{1}{2} g_1 \tilde{g} v_u v_{\bar{\eta}}  &m_{43} &m_{44}\end{array} 
\right) 
\end{equation} 
\begin{align} 
m_{11} &= \frac{1}{8} \Big(2 g_1 \tilde{g} \Big(v_{\eta}^{2} -  v_{\bar{\eta}}^{2} \Big) + 8 m_{H_d}^2  + 8 |\mu|^2  + g_{1}^{2} \Big(3 v_{d}^{2}  - v_{u}^{2} \Big) + g_{2}^{2} \Big(3 v_{d}^{2}  - v_{u}^{2} \Big)\Big)\\ 
m_{21} &= -\frac{1}{4} \Big(g_{1}^{2} + g_{2}^{2}\Big)v_d v_u  - {\Re\Big(B_{\mu}\Big)} \\ 
m_{22} &= \frac{1}{8} \Big(2 g_1 \tilde{g} \Big(- v_{\eta}^{2}  + v_{\bar{\eta}}^{2}\Big) + 8 m_{H_u}^2  + 8 |\mu|^2  - g_{1}^{2} \Big(-3 v_{u}^{2}  + v_{d}^{2}\Big) - g_{2}^{2} \Big(-3 v_{u}^{2}  + v_{d}^{2}\Big)\Big)\\ 
m_{33} &= \frac{1}{4} \Big(2 \Big(g_{B}^{2} + \tilde{g}^{2}\Big)\Big(3 v_{\eta}^{2}  - v_{\bar{\eta}}^{2} \Big) + 4 m^2_{\eta}  + 4 |\mu_{\eta}|^2  + g_1 \tilde{g} \Big( v_{d}^{2} -  v_{u}^{2} \Big)\Big)\\ 
m_{43} &= - \Big(g_{B}^{2} + \tilde{g}^{2}\Big)v_{\eta} v_{\bar{\eta}}  - {\Re\Big(B_{\eta}\Big)} \\ 
m_{44} &= \frac{1}{4} \Big(-2 \Big(g_{B}^{2} + \tilde{g}^{2}\Big)\Big(-3 v_{\bar{\eta}}^{2}  + v_{\eta}^{2}\Big) + 4 m^2_{\bar{\eta}}  + 4 |\mu_{\eta}|^2  + g_1 \tilde{g} \Big( v_{u}^{2} -  v_{d}^{2} \Big)\Big)
\end{align} 
This matrix is diagonalized by \(Z^H\): 
\begin{equation} 
Z^H m^2_{h} Z^{H,\dagger} = m^{dia}_{2,h} 
\end{equation} 

\subsubsection{Mass matrix for pseudo scalar Higgs}
Basis: \( \left(\sigma_{d}, \sigma_{u}, \sigma_{\eta}, \sigma_{\bar{\eta}}\right)\)
 
\begin{equation} 
m^2_{A^0} = \left( 
\begin{array}{cccc}
m_{11} &{\Re\Big(B_{\mu}\Big)} &0 &0\\ 
{\Re\Big(B_{\mu}\Big)} &m_{22} &0 &0\\ 
0 &0 &m_{33} &{\Re\Big(B_{\eta}\Big)}\\ 
0 &0 &{\Re\Big(B_{\eta}\Big)} &m_{44}\end{array} 
\right) 
\end{equation} 
\begin{align} 
m_{11} &= \frac{1}{8} \Big(2 g_1 \tilde{g} \Big(v_{\eta}^{2} -  v_{\bar{\eta}}^{2} \Big) + 8 m_{H_d}^2  + 8 |\mu|^2  + g_{1}^{2} \Big( v_{d}^{2} -  v_{u}^{2} \Big) + g_{2}^{2} \Big( v_{d}^{2} -  v_{u}^{2} \Big)\Big)\\ 
m_{22} &= \frac{1}{8} \Big(2 g_1 \tilde{g} \Big(- v_{\eta}^{2}  + v_{\bar{\eta}}^{2}\Big) + 8 m_{H_u}^2  + 8 |\mu|^2  + g_{1}^{2} \Big( v_{u}^{2} -  v_{d}^{2} \Big) + g_{2}^{2} \Big( v_{u}^{2} -  v_{d}^{2} \Big)\Big)\\ 
m_{33} &= \frac{1}{4} \Big(2 \Big(g_{B}^{2} + \tilde{g}^{2}\Big)\Big(v_{\eta}^{2} -  v_{\bar{\eta}}^{2} \Big) + 4 m^2_{\eta}  + 4 |\mu_{\eta}|^2  + g_1 \tilde{g} \Big( v_{d}^{2} -  v_{u}^{2} \Big)\Big)\\ 
m_{44} &= \frac{1}{4} \Big(-2 \Big(g_{B}^{2} + \tilde{g}^{2}\Big)\Big(v_{\eta}^{2} -  v_{\bar{\eta}}^{2} \Big) + 4 m^2_{\bar{\eta}}  + 4 |\mu_{\eta}|^2  + g_1 \tilde{g} \Big( v_{u}^{2} -  v_{d}^{2} \Big)\Big)
\end{align} 
This matrix is diagonalized by \(ZA\): 
\begin{equation} 
Z^A m^2_{A^0} Z^{A,\dagger} = m^{dia}_{2,A^0} 
\end{equation} 
Using the solution of the tadpole equations the mass matrix can also be written as
\begin{equation}
m^2_{A^0} = \left(\begin{array}{cccc}
B_\mu \tan\beta & B_\mu & 0 & 0 \\
B_\mu  & B_\mu \cot\beta & 0 & 0 \\
 0 & 0  &B_{\mu'} \tan\beta' & B_{\mu'} \\
 0 & 0  &B_{\mu'} & B_{\mu'} \cot\beta' 
\end{array}
\right) \,.
\end{equation}
what gives the masses of the physical states:
\begin{equation}
m^2_{A^0} = \frac{2 B_\mu}{\sin2\beta} \thickspace, \hspace{1cm} 
m^2_{A^0_\eta}  = \frac{2 B_{\mu'}}{\sin2\beta'} \thickspace.
\end{equation}
\subsubsection{Mass matrix for charged Higgs}
Basis: \( \left(H_d^-, H_u^{+,*}\right)\)
 
\begin{equation} 
m^2_{H^-} = \left( 
\begin{array}{cc}
m_{11} &\frac{1}{4} g_{2}^{2} v_d v_u  + B^*_{\mu}\\ 
\frac{1}{4} g_{2}^{2} v_d v_u  + B_{\mu} &m_{22}\end{array} 
\right) 
\end{equation} 
\begin{align} 
m_{11} &= \frac{1}{8} \Big(2 g_1 \tilde{g} \Big(v_{\eta}^{2} -  v_{\bar{\eta}}^{2} \Big) + 8 m_{H_d}^2  + 8 |\mu|^2  + g_{1}^{2} \Big( v_{d}^{2} -  v_{u}^{2} \Big) + g_{2}^{2} \Big(v_{d}^{2} + v_{u}^{2}\Big)\Big)\\ 
m_{22} &= \frac{1}{8} \Big(2 g_1 \tilde{g} \Big(- v_{\eta}^{2}  + v_{\bar{\eta}}^{2}\Big) + 8 m_{H_u}^2  + 8 |\mu|^2  + g_{1}^{2} \Big( v_{u}^{2} -  v_{d}^{2} \Big) + g_{2}^{2} \Big(v_{d}^{2} + v_{u}^{2}\Big)\Big)
\end{align} 
This matrix is diagonalized by \(Z^+\): 
\begin{equation} 
Z^+ m^2_{H^-} Z^{+,\dagger} = m^{dia}_{2,H^-} 
\end{equation} 

\subsubsection{Mass matrix for neutralinos}
Basis: \( \left(\lambda_{\tilde{B}}, \tilde{W}^0, \tilde{H}_d^0, \tilde{H}_u^0, {\tilde{B}{}'}, \tilde{\eta}, \tilde{\bar{\eta}}\right)\)
 
\begin{equation} 
m_{\tilde{\chi}^0} = \left( 
\begin{array}{ccccccc}
M_1 &0 &-\frac{1}{2} g_1 v_d  &\frac{1}{2} g_1 v_u  &\frac{1}{2} M_{B B'}  &- \tilde{g} v_{\eta}  &\tilde{g} v_{\bar{\eta}} \\ 
0 &M_2 &\frac{1}{2} g_2 v_d  &-\frac{1}{2} g_2 v_u  &0 &0 &0\\ 
-\frac{1}{2} g_1 v_d  &\frac{1}{2} g_2 v_d  &0 &- \mu  &0 &0 &0\\ 
\frac{1}{2} g_1 v_u  &-\frac{1}{2} g_2 v_u  &- \mu  &0 &0 &0 &0\\ 
\frac{1}{2} M_{B B'}  &0 &0 &0 &M_{B-L} &- g_{B} v_{\eta}  &g_{B} v_{\bar{\eta}} \\ 
- \tilde{g} v_{\eta}  &0 &0 &0 &- g_{B} v_{\eta}  &0 &- \mu_{\eta} \\ 
\tilde{g} v_{\bar{\eta}}  &0 &0 &0 &g_{B} v_{\bar{\eta}}  &- \mu_{\eta}  &0\end{array} 
\right) 
\end{equation} 
This matrix is diagonalized by \(N\): 
\begin{equation} 
N^* m_{\tilde{\chi}^0} N^{\dagger} = m^{dia}_{\tilde{\chi}^0} 
\end{equation}

\subsubsection{Mass matrix for neutrinos}
Basis: \( \left(\nu_L, \nu^c, N\right) \) 
 
\begin{equation} 
m_{\nu} = \left( 
\begin{array}{ccc}
0 &\frac{1}{\sqrt{2}} v_u Y_\nu  & \frac{\tilde{Y}}{2} v_d v_\eta\\ 
\frac{1}{\sqrt{2}} v_u Y^T_{\nu}  &\frac{1}{\sqrt{2}} v_{\eta} \Big(Y_{\eta \nu^c} + Y_{\eta \nu^c}^{T}\Big) &\frac{1}{\sqrt{2}} v_{\eta} Y_iS \\ 
\frac{\tilde{Y}^T}{2} v_d v_\eta &\frac{1}{\sqrt{2}} v_{\eta} Y_{IS}^{T}  &2 \mu_N \end{array} 
\right) 
\end{equation} 
$\tilde{Y}$ is the running, effective operator $\sim \hat{L} \hat{N}_S \hat{H}_d \hat{\eta}$ obtained by integrating out $\rho$ and $\bar{\rho}_L$.
This matrix is diagonalized by \(U^V\): 
\begin{equation} 
U^{V,*} m_{\nu} U^{V,\dagger} = m^{dia}_{\nu} 
\end{equation}

\bibliography{SLHAbib}

\end{document}